\documentclass[apjl]{emulateapj}

\pdfoutput=1

\usepackage{graphicx,natbib}
\usepackage{amsmath}
\usepackage{apjfonts}

\newcommand\Ha{$\textrm{H}\alpha \,$}
\newcommand{\s}{$\sim$}

\begin{document}

\author{
Richard Beare\altaffilmark{1, 3},
Michael J. I. Brown\altaffilmark{1},
Kevin Pimbblet\altaffilmark{2, 1},
}

\altaffiltext{1}{Monash Centre for Astrophysics, Monash University, Clayton, Victoria 3800, Australia. }
\altaffiltext{2}{Department of Physics and Mathematics, University of Hull, Cottingham Road, Kingston-upon-Hull, HU6 7RX, UK.}
\altaffiltext{3}{Email: richard@beares.net}

\slugcomment{Accepted for publication in the Astrophysical Journal}

\date{\today}

\title{An accurate new method of calculating absolute magnitudes and K-corrections applied to the Sloan filter set}

\shorttitle{Absolute magnitudes and K-corrections}

\shortauthors{Beare \it {et al.}}

\begin{abstract}

We describe an accurate new method for determining absolute magnitudes, and hence also K-corrections, which is simpler than most previous methods, being based on a quadratic function of just one suitably chosen observed color. The method relies on the extensive and accurate new set of 129 empirical galaxy template SEDs from \citet {brown14}. A key advantage of our method is that we can reliably estimate random errors in computed absolute magnitudes due to galaxy diversity, photometric error and redshift error. We derive K-corrections for the five Sloan Digital Sky Survey filters and provide parameter tables for use by the astronomical community. Using the New York Value-Added Galaxy Catalog we compare our K-corrections with those from \textit{kcorrect}. Our K-corrections produce absolute magnitudes that are generally in good agreement with \textit{kcorrect}. Absolute $griz$ magnitudes differ by less than 0.02 mag, and those in the $u$-band by $\sim 0.04$ mag. The evolution of rest-frame colors as a function of redshift is better behaved using our method, with relatively few galaxies being assigned anomalously red colors and a tight red sequence being observed across the whole $0.0 <  z < 0.5$  redshift range.

\end{abstract}


\section{INTRODUCTION}
\label{sec:intro}

Measurements of luminosity evolution are key to understanding how galaxies have evolved, but account must be taken of the fact that light emitted by a distant galaxy in a given waveband is redshifted to longer wavelengths by the time that it arrives at a telescope. A \textit{K-correction} \citep[e.g.][]{oke68, hogg02} enables the restframe absolute magnitude $M_W$ in a waveband  $W$ to be determined from an observed apparent magnitude $m_Q$ in a (possibly different) observed waveband $Q$:

\begin{equation} \label{eq:KWQ} 
	M_W = m_Q - D_{\rm{M}} + K_{WQ}.
\end{equation}

where ${D_{\rm{M}} = 5 \log_{10} (d_L / \rm{10 \, pc})}$ is the distance modulus and $d_L$ the luminosity distance.

The K-correction for a particular galaxy and redshift depends on its spectral energy density or SED (defined as the flux emitted per unit wavelength or per unit frequency as a function of wavelength or frequency). It also depends on the transmission functions of the $W$ and $Q$ filters (taking account of the quantum efficiency of the detector and the transmission function of the telescope optics in the case of $Q$).  To determine the K-correction for a galaxy with unknown SED it must be matched to template galaxies with known SEDs.  Different methods have been used to do this but all rely on having a set of representative template SEDs. 

Template galaxy SEDs can be drawn from a representative range of real galaxies  \citep[e.g.][]{colem80, kinne96, brown14} or based on synthetic SEDs derived from stellar population synthesis (SPS) models \citep[e.g.][]{bruzu03, fioc97}.  Both types of template SED can suffer from the deficiency that they do not fully span the entire range of galaxy spectral types. Models have sometimes been used to extend observed optical and ultraviolet spectra of galaxies into the infrared when infrared spectroscopy has not been available.

A major benefit of using templates based on stellar population synthesis models is that additional physical quantities such as stellar mass and star formation rate can be derived from the models. However, the star formation histories (SFH) needed to match broadband photometry are not always unique and this can result in uncertainty in the values of derived quantities such as stellar mass. For example, \citet{simha14} show that the exponentially declining star formation rates often used in SPS models are less satisfactory than other parameterisations of SFH for modelling the results of smooth particle hydrodynamic simulations of galaxy formation and evolution.

In principle, observationally based templates should have the advantage that they are based on observed galaxy spectra.  However, the ones that have been most widely used historically, the four templates from \citep{colem80} and the 43 from \citet{kinne96}, are both largely the spectra of galactic nuclei rather than entire galaxies, so may produce systematic errors. Both sets of templates were based on the best data available at the time and are still widely used today. However their wavelength coverage is limited to the ultraviolet and optical, so that they need extending into the infrared using stellar population synthesis models.  

The most straightforward way of calculating the K-correction for a galaxy is to assume that it is a simple function of redshift and galaxy type only. For example, \citet{colem80} calculated K-corrections and colors out to $z = 2$ for four empirically based template SEDs representing E, Sbc, Scd and Irr galaxies. This broad brush approach cannot take account of the true diversity of galaxy spectral properties, but it has the virtue of simplicity.

Maximum likelihood fits of stellar population synthesis models to observed photometry have commonly been used to determine K-corrections \citep[e.g.][]{bell04, brown07}. Such models provide a reasonable approximation to the observed SEDs of red galaxies. However, the situation is more difficult for blue galaxies which have a wide variety of complex star formation histories, dust obscuration and nebular emission lines, and therefore lack the tightly constrained color-color relationships found in red sequence galaxies.

\citet{blant03b} introduced a method for calculating K-corrections based on matching broadband photometric observations of galaxies to template SEDs generated from combinations of \citet{bruzu03} SPS models with dust obscuration and nebular emission lines included. Their key innovation was to use multiple components to reproduce galaxy SEDs. \textit{kcorrect} v\_4\_2 \citep{blant07}, for example, uses a set of five template SEDs, each derived from non-negative linear combinations of 450 individual  \citet{bruzu03} models with instantaneous bursts of star formation and varying metallicities, together with 35 models of emission from ionized gas \citep{kewle01}.  \textit{kcorrect} uses the nonnegative matrix factorisation method \citep[NMF, ][]{lee99, blant07} which in some respects is similar to principal component analysis  \citep[e.g.][]{conno95}. Using NMF the five templates are optimised to span the space of observed galaxy spectra and broadband optical and near infrared photometry for a training set of a few thousand galaxies at various redshifts. This is done by minimising $\chi^2$ for the offsets between (a) spectroscopic and photometric data for the training set and (b) linear combinations of the five templates.  Once these five templates have been generated, NMF is  used to match broadband photometry for sample galaxies to linear combinations of the five templates, so producing estimates of their SEDs. From these, K-corrections, stellar mass to light ratios and other physical parameters, can be estimated.  Training set spectra were obtained from SDSS \citep{york00} and photometry from SDSS, 2MASS \citep{skrut06}, GALEX \citep{marti05}, DEEP2 \citep{davis03, willm06} and GOODS \citep{giava04}.  \textit{kcorrect} has proved very useful and has been widely used in the literature. A key advantage is that it enables physical properties of sample galaxies to be estimated, such as stellar mass to light ratios and star formation rates and histories, although there are concerns as already indicated that these may not be uniquely defined by the photometry. 

\citet{chili10} showed that K-corrections computed using both \textit{kcorrect} and the fitting of PEGASE models can be approximated at redshifts below 0.5 by polynomials involving the redshift $z$ and just one observed color. Their polynomials  are of fifth degree in observed color and third degree in redshift, and provide a readily accessible way of determining K-corrections, e.g.:

\begin{equation}		
	\label{eq:chilingarian}
	K_{gg} = \sum_{j = 0}^{5}  \sum_{k = 0}^{3}  a_{jk} z^j (g - r)^k.
\end{equation}

\citet{chili10} provide tables of polynomial coefficients and an on-line K-correction calculator and associated code, and these tools make it easy to calculate K-corrections for a wide range of observed colors. Their tables provide AB-based K-corrections for the $ugriz$ SDSS bands and the $PJHK$ bands used by the United Kingdom Infrared Telescope (UKIRT) Wide Field Camera, as well as Vega-based K-corrections for the Johnson-Cousins $UBVR_cI_c$ bands and the Two Micron All Sky Survey (2MASS) $JHK_s$ bands. \citet{omill11} derived a simpler, but more restricted, linear relation between $g$ and $r$-band K-corrections and redshift and $(g-r)$ color, based on K-corrections determined using \textit{kcorrect} v\_4\_2 for a sample consisting of the SDSS Main Galaxy Sample, the SDSS Luminous Red Galaxies sample and SDSS galaxies with AGN.

\citet{roche09} and \citet{westr10} have both produced empirical K-corrections for the $g$ and $r$-bands based on large samples of flux-calibrated spectra. \citet{roche09} determined $g$ and $r$-band K-corrections up to $z=0.36$ for $\sim70\,000$ E/S0 galaxies with SDSS spectra by integrating their SEDs and compared these with values obtained from \textit{kcorrect} v\_4\_2 \citep{blant07}, finding good agreement.  They then used these to study the evolution of the color-magnitude relation for E/S0 galaxies. \citet{westr10} used $\sim15\,000$ spectra from the Smithsonian Hectospec Lensing Survey \citep{gelle05} to derive functions giving $g$ and $r$-band K-corrections (to $z\sim0.68$ and $z\sim0.33$ respectively) as third order polynomials in the redshift and the quantity $D_n4000$ which measures the strength of the 4000\AA \, break. They point out that their method has a number of advantages for a spectroscopic survey, notably that $D_n4000$ is redshift-independent, barely affected by attenuation, and does not need flux-calibrated spectra. Their method gives similar scatter to methods based on $(g-r)$ color and shows no systematic bias.

\citet[][]{rudni03} developed a method of calculating K-corrections based on interpolating between computed $K_{WQ} = M_W - m_Q + D_{\rm{M}}$ values for the two templates whose $(m_P - m_Q)$ colors are closest to the observed color. One key advantage of this method is that neither observed waveband $P$ or $Q$ need be the same as the restframe waveband $W$. The redshifted SED can be observed in wavebands $P$ and $Q$ which are close to the redshifted restframe waveband $W$, so avoiding errors that arise at higher redshifts when the computed K-correction $K_{WW}$ refers to the same restframe and observed wavebands, but very different parts of the SED. Another key advantage of the method of \citet{rudni03} is that the difference between apparent and absolute magnitudes is determined as a function of galaxy color, i.e. magnitude differences, so that errors in the  template SEDs largely cancel out when the $P$ and $Q$ wavebands and the redshifted restframe $W$ waveband are all sampling coincident or adjacent sections of SED. Prior to \citet{rudni03}, \citet[][]{dokku96} and \citet[][]{dokku00} had used a similar method involving a linear relationship between absolute magnitude and two observed magnitudes to determine K-corrections for one specific restframe waveband, but both were studies of galaxies in one cluster.

\citet{taylo09} implemented the method of \citet{rudni03} in their publicly available \textit{InterRest} IDL code. This uses the six EAZY templates developed for use with the EAZY photometric redshift code of \citet{bramm08}. Five of these are derived from PEGASE \citep{fioc97} stellar population synthesis models using NMF, and the sixth is a dusty starburst model. This method shares the advantages of \citet{rudni03} while additionally providing an easy-to-use software tool.

\citet[][]{willm06} fitted second degree polynomials to to plots of $K_{BR}$ against both $(B - R)$ and $(R - I)$ color for 34 of the the 43 \citet{kinne96} templates for redshifts in the range $0 \leq z < 1.4$. The other 9 templates were rejected because they resulted in outliers. These polynomials were used to determine K-corrections for DEEP2 galaxies from observed colors, interpolating between the $(B - R)$ and $(R - I)$ derived quantities when the restframe $(M_U - M_B)$ color lay between the observed $(B - R)$ and $(R - I)$ colors, and otherwise using the closest color. This approach is similar to ours, and we discuss its strengths and limitations in the next section.

\subsection{Limitations of existing methods and motivation for the present work}
\label{motivation}

All methods used to determine K-corrections for galaxies depend on matching their properties to those of template or model galaxies at the redshift in question. Except where a simple classification by morphological type has been used, e.g. E, Sbc, Scd, Irr as in \citet{colem80}, the matching effectively depends on one or more observed colors. A small number of template SEDs may not be adequate to span the diversity of galaxy SEDs in a sample, even when the templates have been derived from a large number of component SEDs (e.g. the five \textit{kcorrect} templates which are combinations of 450 individual SPS models and 35 models of ionized gas emission). As a result, a small number of templates or template components may not adequately span the color space occupied by the full diversity of galaxies. Consequently, we expect that methods using only a small number of templates will result in many less ``typical'' galaxy SEDs being poorly matched to the templates, in turn resulting in inaccurate K-corrections. \citet{willm06} used a larger number of template SEDs, but as already noted they had to omit nine of the 43 Kinney templates as they produced outliers in $(B - R)$ versus $(R - I)$ color space (their Figure 16) and there is therefore concern that the omitted templates could represent valid examples of galaxies. Furthermore, the 35 templates they retained still exhibited considerable scatter in their plots (up to ~1 mag).

Many methods \citep[e.g. \textit{kcorrect} and ][]{chili10} calculate K-corrections $K_{WW}$ for identical restframe and observed wavebands. At $z\gtrsim 0.3$ this means the part of the SED observed through the $W$-band filter is very different from that emitted from the galaxy in the restframe $W$-band. Because of the diversity of galaxy SEDs considerable errors will therefore result for many galaxies, especially those with an SED that exhibits complex features between the restframe $W$ waveband and the shorter wavelengths that get redshifted into the observed $W$ waveband.

Another concern with some existing methods is the use of template SEDs based largely on the centres of bright galaxies rather than whole galaxies \citep[e.g.][]{colem80, kinne96}. For example, \citet{willm06} note that their use of the \citet{kinne96} template SEDs results in calculated $(U-B)$ colors for the reddest templates that are too red by \s0.08 mag when compared with values from the \textit{Third Reference Catalogue of Bright Galaxies} \citep{devau91} and state that the anomaly is to be expected given the direction of internal color gradients in galaxies.  Furthermore, \citet{willm06} do not detect any significant evolution in the $(M_U-M_B)$ color of the red sequence from $z =1.3$ to $z =0.3$ (their Figure 4) in contradiction to what we know must occur due to the passive evolution of red galaxies, \citep[e.g.,][who measure evolution of the strength of the 4000\AA \, break from $z = 1.42$ to $z=0.15$]{mores13}. 

\citet{taylo09} investigated the accuracy of their K-corrections in the redshift range $0 < z  < 1.2$, finding random scatter of \s0.05 mag and systematic offsets of a similar size when they compared computed and observed plots of $(R-I)$ against $(V-I)$, noting that for the reddest galaxies the systematic error could rise to \s0.1 mag. The accuracy was comparable whether they used their default EAZY templates, Kinney templates, or \citet{colem80} templates supplemented by a Kinney starburst SED. They found, however, that use of templates based on \citet{bruzu03} models could result in systematics as large as \s0.2 mag, while \textit{kcorrect} resulted in random and systematic errors at the 0.1 mag level (peak to peak).

The preceding discussion shows that there are a range of difficulties with existing methods for calculating K-corrections: K-corrections in the prior literature may not capture the full diversity of galaxy SEDs, do not use the observed photometry optimally, and can produce systematic errors in the observed colors of galaxies (and their evolution). The new atlas of 129 galaxy SEDs from \citet{brown14} provides the motivation for seeking a new method of determining K-corrections based on comparing observed colors with empirical models that have been fitted to all 129 templates in color-color space. Our method is similar to that of \citet{willm06} but we investigate in detail the criteria for choosing the optimum observed color at any particular redshift, and show that the nearest filters to the redshifted restframe waveband of interest are not always the best ones to use for the observed color.  We also provide reliable estimates of the random errors due to photometric error, redshift error and galaxy diversity. The resulting new K-correction method is both simpler than existing methods and in certain situations more accurate.

The structure of this paper is as follows. \S \ref{sec:method} first describes our data: the galaxy templates from \citet{brown14} and the subset of SDSS galaxies in the VAGC that we use to validate our method.  It then describes our method for determining K-corrections. In \S \ref{sec:results} we apply this method to calculation of K-corrections for the Sloan Digital Sky Survey $ugriz$ filter set, presenting the relevant model parameters in tabular form in an appendix, as well as electronically on the web. We then compare our K-corrections for VAGC galaxies with those from \textit{kcorrect}. Finally we summarise our findings in the appendix.

Our method in \S \ref{sec:method} is applicable for any cosmology and in any magnitude system but we use AB magnitudes and a cosmology with $\Omega_0 = 0.3$, $\Omega_{\rm{\Lambda}} = 0.7$ and $H_0 = 100h \, \textrm{km s}^{-1} \textrm{Mpc}^{-1}$ with $h = 1$ for the Sloan waveband comparisons with \textit{kcorrect} in \S\ref{sec:kcorrect}.

\begin{figure}
 	\centering
		\includegraphics[width=0.45\textwidth]{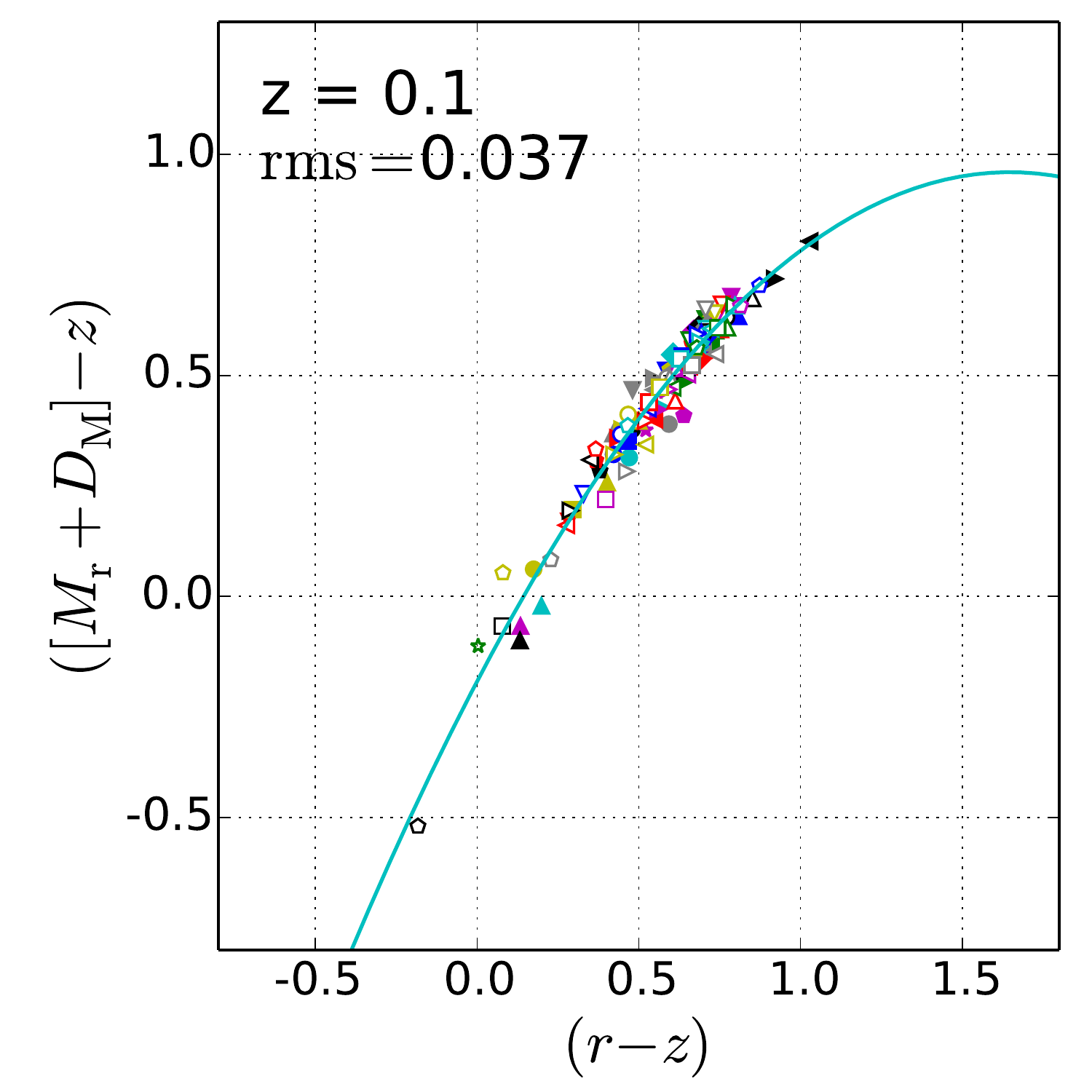}
		\caption{Determination of absolute $r$ magnitude from observed $(r-z)$ color at $z  = 0.1$. The colored markers represent values of ${K_{rz} \equiv (M_r + D_{\rm{M}}) -  z}$ plotted against ${(r - z)}$ computed at different redshifts for the 129 template SEDs in \citet{brown14}. The curve is the best fit second order polynomial to the template data points. Given the measured $(r-z)$ color of a galaxy on the $x$-axis, its absolute magnitude $M_u$ is equal to the corresponding $y$-value on the curve minus the distance modulus $ D_{\rm{M}}$ and plus the measured apparent magnitude $z$. Each template marker has a unique shape and color enabling it to be easily identified in the plot. Outliers more than 0.2 mag from the polynomials are excluded from the fit after three iterations of the best fit process.}
		\label{fig:r_calibration_example}
\end{figure}

\begin{figure*}
 	\centering
		\includegraphics[width=0.49\textwidth]{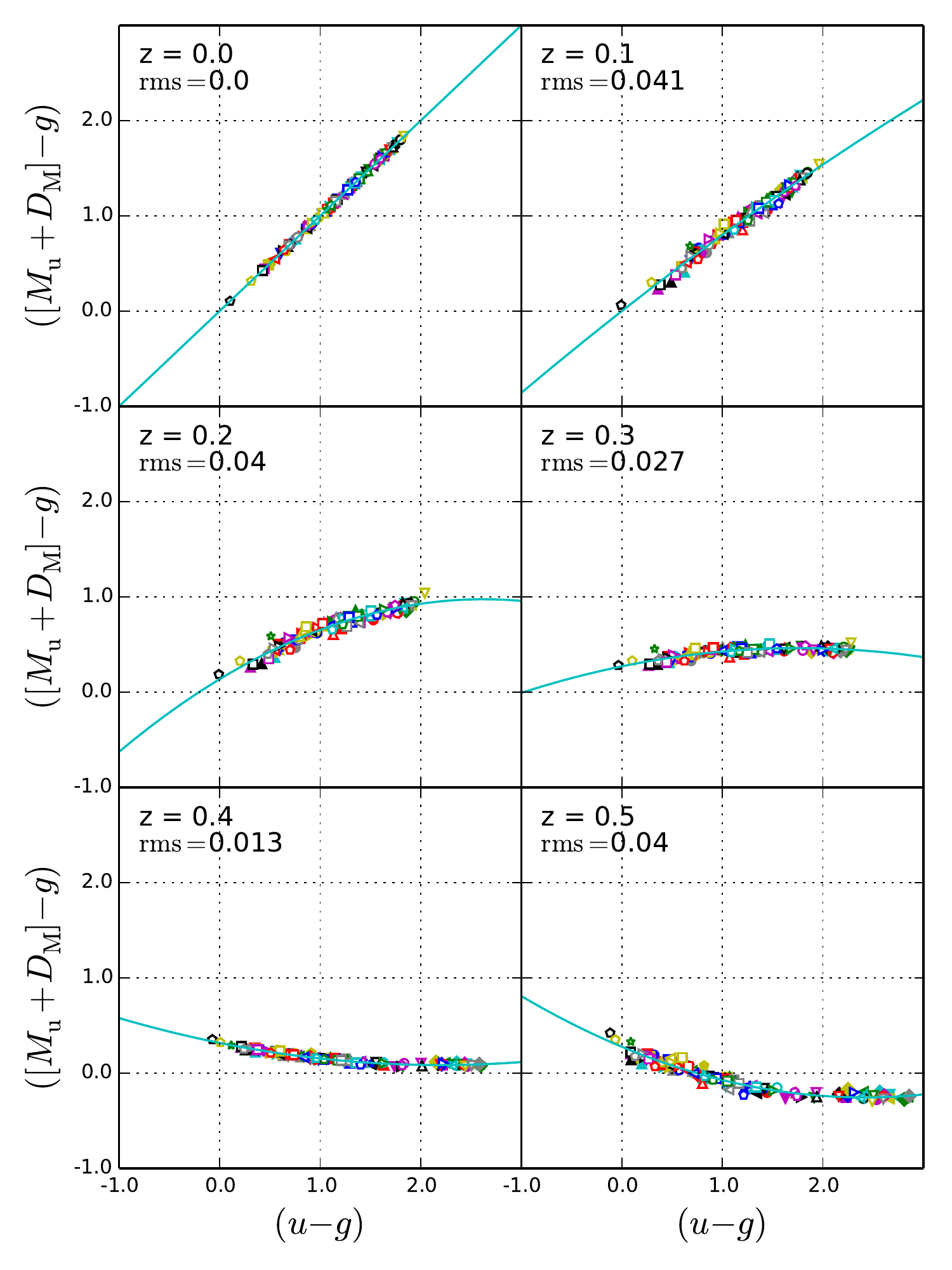} 
		\includegraphics[width=0.49\textwidth]{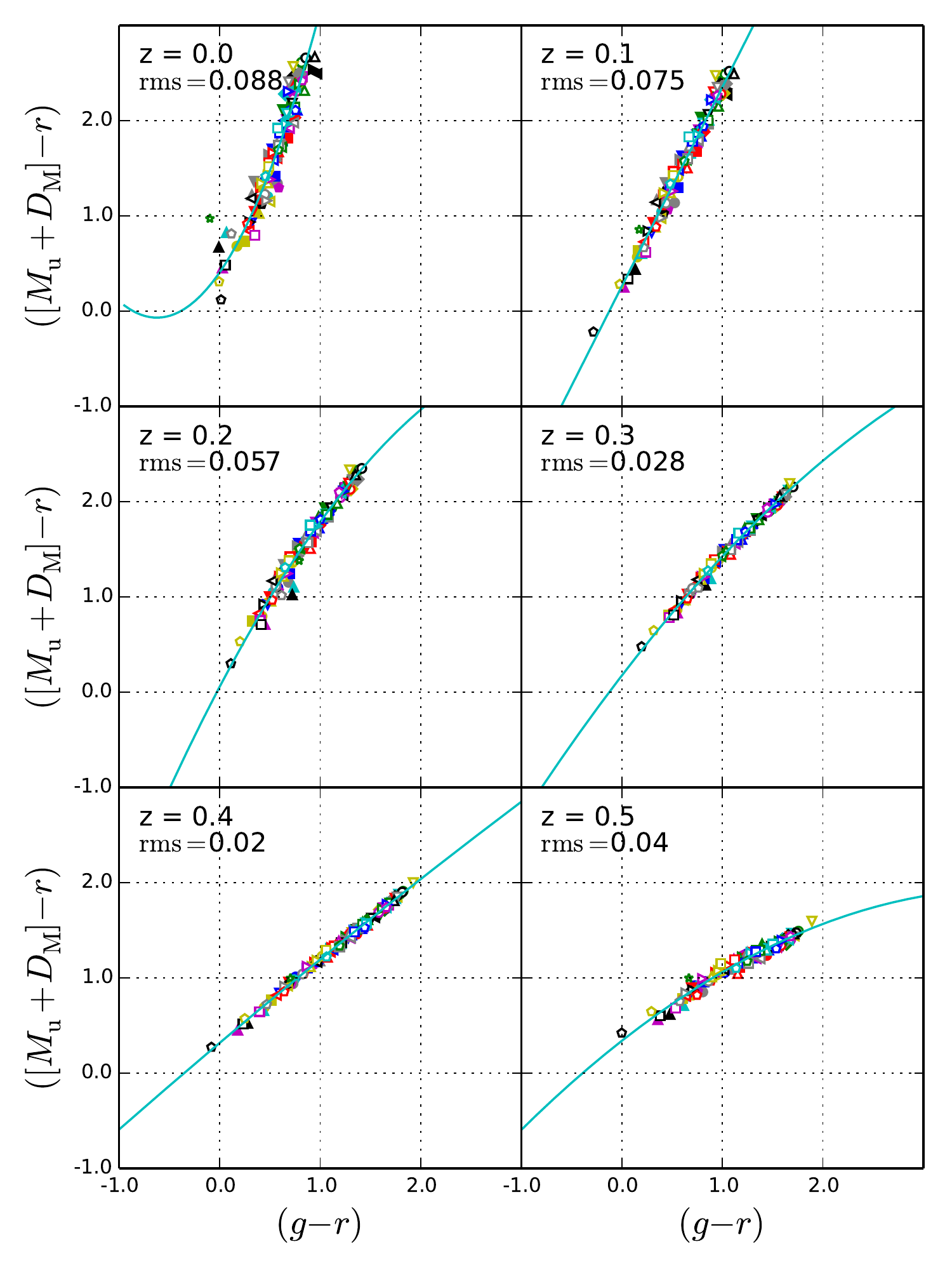}
		\caption{Determination of absolute Sloan $u$ magnitudes from observed $(u-g)$ or $(g-r)$ colors. The curves are the best fit second order polynomial fits to data points calculated for the 129 SED templates of \citet{brown14} at the redshifts in question. Given the measured color of a galaxy on the $x$-axis, its absolute magnitude $M_u$ is equal to the corresponding $y$-value on the curve minus the distance modulus $ D_{\rm{M}}$ and plus the appropriate apparent magnitude ($g$ on the left and $r$ on the right).  If good quality $u$-band data is available we prefer to use observed $(u-g)$ color as this results in smaller RMS offsets of the template SED points from the best fit polynomials (as given in the top left corner of each plot). If this is not the case we can still use $(g-r)$ colors. Each template marker has a unique shape and color enabling it to be easily identified in the plot.  Outliers more than 0.2 mag from the polynomials are excluded from the fit after three iterations of the best fit process.}
		\label{fig:u_calibration}
\end{figure*}

\begin{figure*}
 	\centering
		\includegraphics[width=0.49\textwidth]{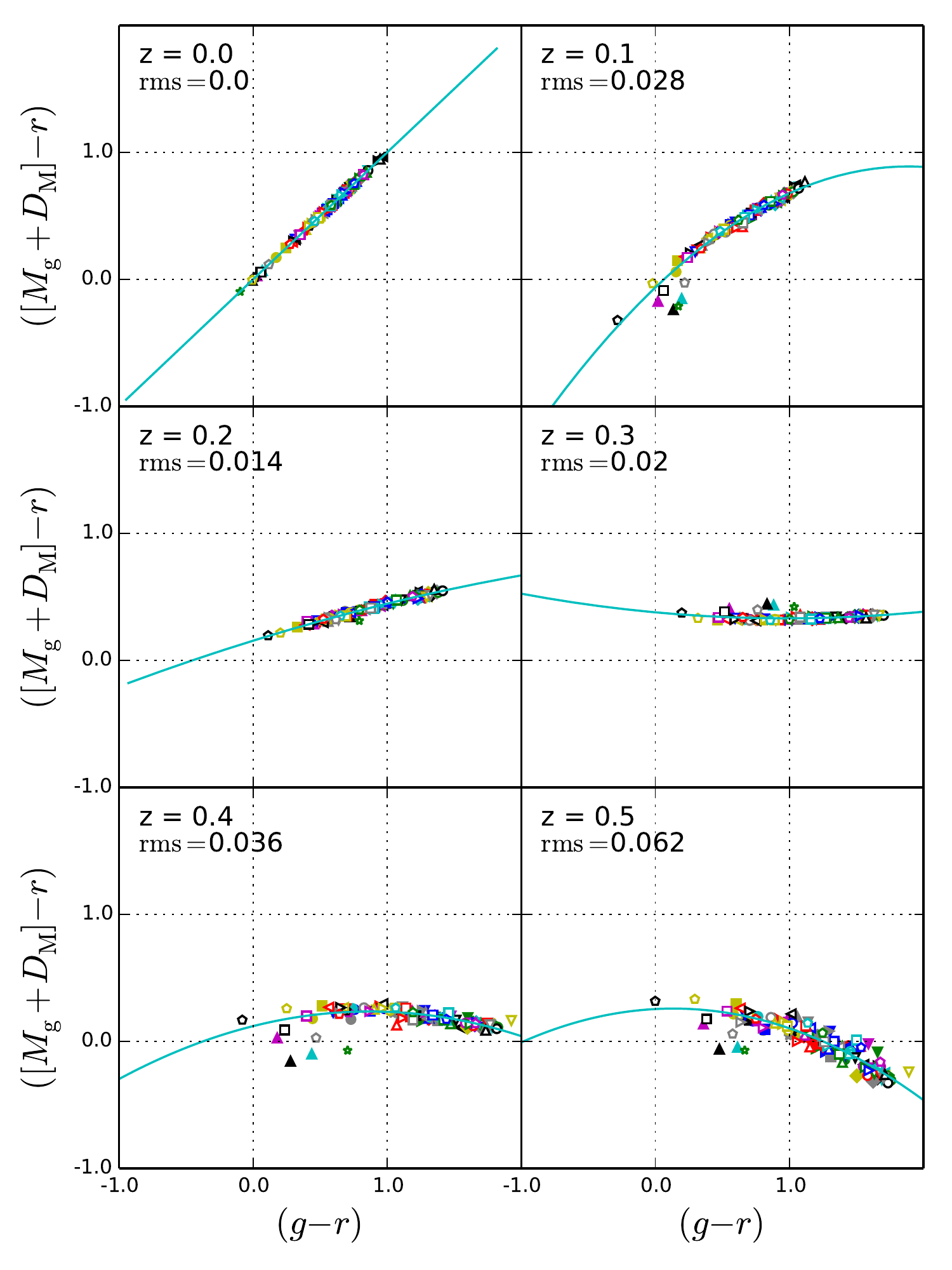}
		\includegraphics[width=0.49\textwidth]{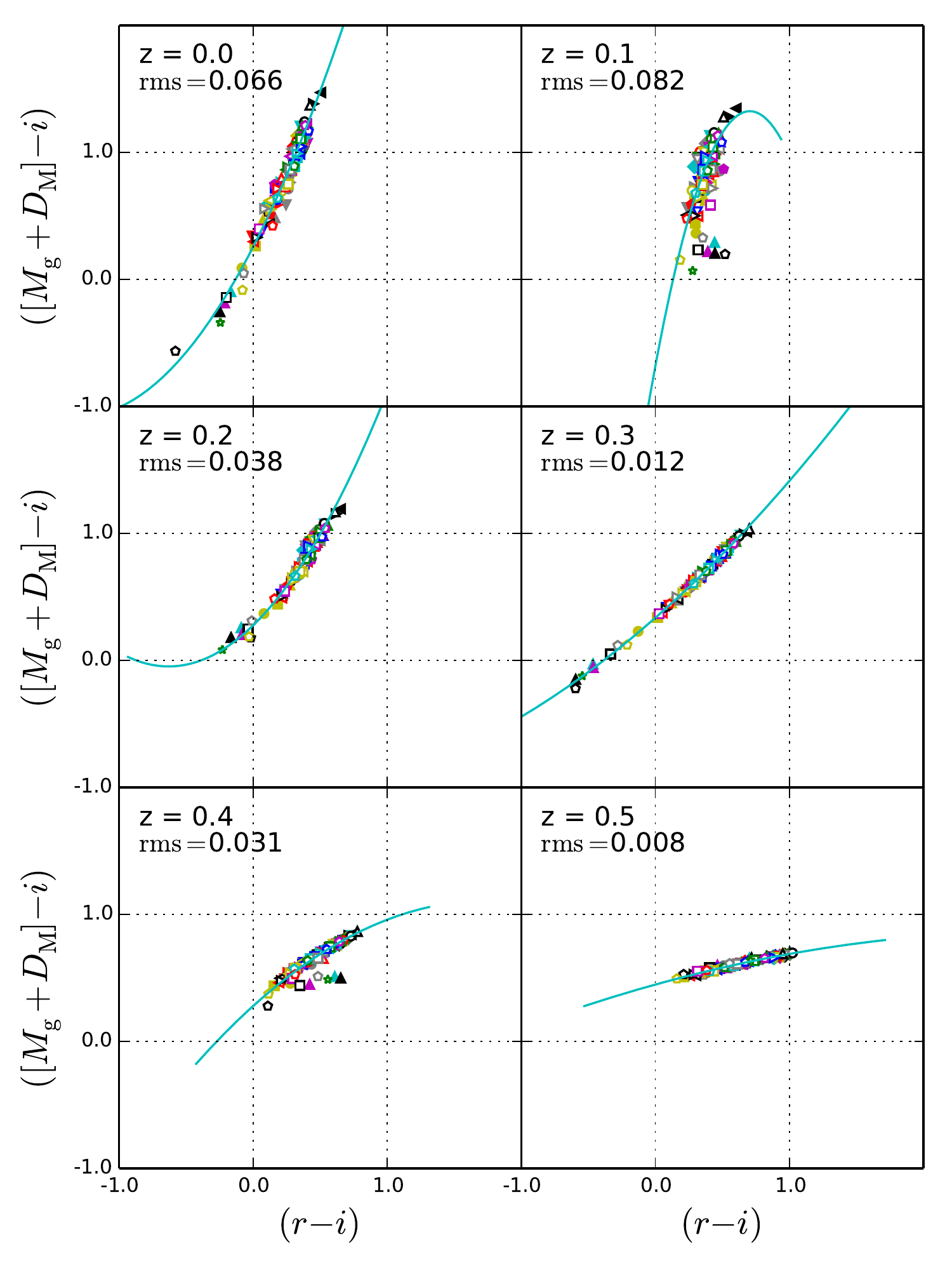}
		\caption{Determination of absolute Sloan $g$ magnitudes from observed $(g-r)$ or $(r-i)$ colors.  The curves are the best fit second order polynomial fits to data points calculated for the 129 SED templates of \citet{brown14} at the redshifts in question. Given the measured color of a galaxy on the $x$-axis, its absolute magnitude $M_g$ is equal to the corresponding $y$-value on the curve minus the distance modulus $ D_{\rm{M}}$ and plus the appropriate apparent magnitude ($r$ on the left and $i$ on the right). We prefer to use observed $(g-r)$ color from $z =0$ to $z =0.34$ and $(r-i)$ color from $z =0.34$ to $z =0.5$, as this minimises the RMS offsets of the template SED points from the best fit polynomials (as given in the top left corner of each plot).   Each template marker has a unique shape and color enabling it to be easily identified in the plot. The templates significantly offset from the polynomials at $z  \sim 0.1$ and  $z  \geq 0.4$ are: UM461, UGCA410 and UGCA6850 (compact blue galaxies), MRK930 (a starburst galaxy), and MRK1450 (a compact starburst galaxy).  Outliers more than 0.2 mag from the polynomials are excluded from the fit after three iterations of the best fit process.}
		\label{fig:g_calibration}
\end{figure*}

\begin{figure*}
 	\centering
		\includegraphics[width=0.49\textwidth]{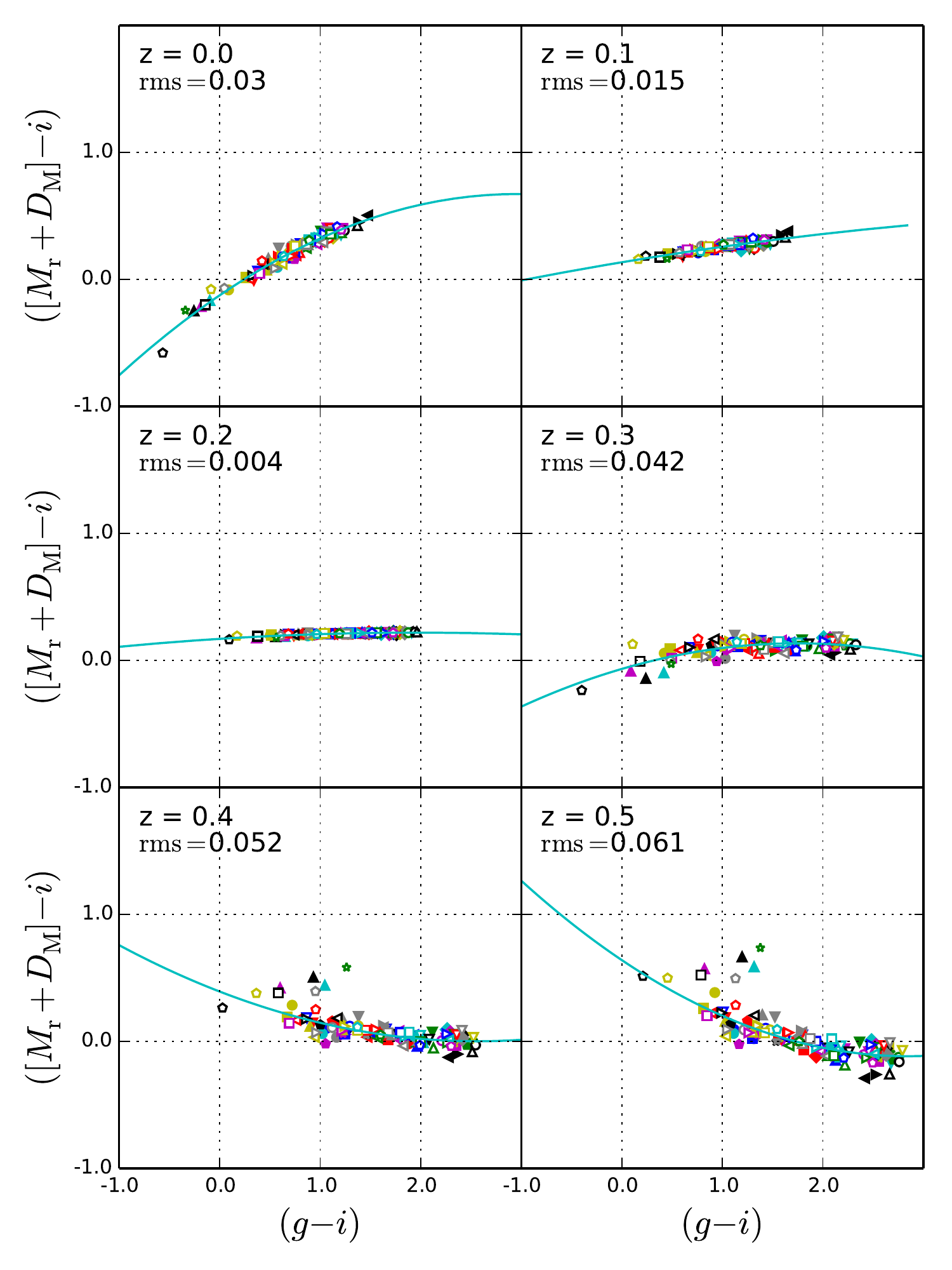} 
		\includegraphics[width=0.49\textwidth]{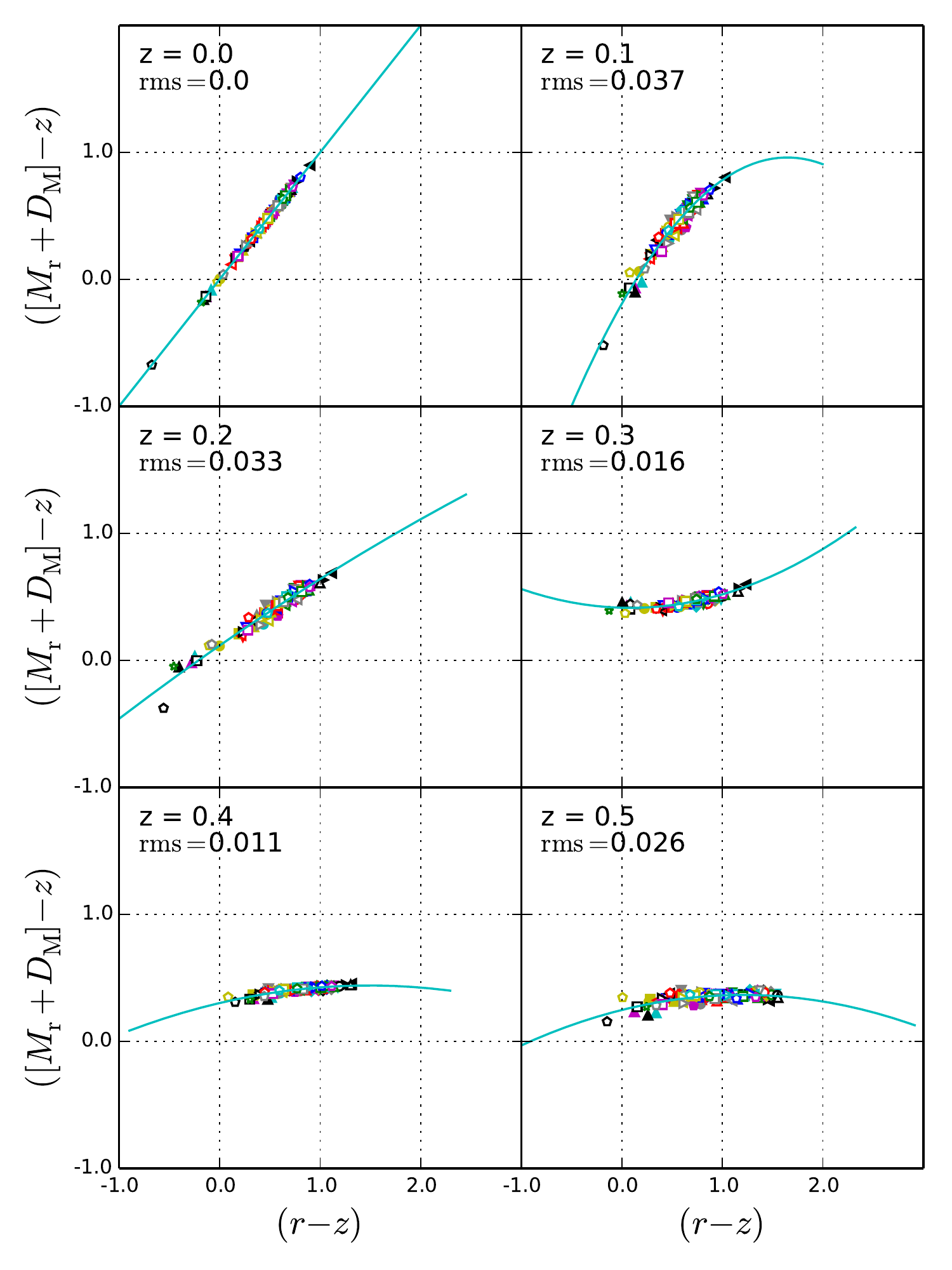}
		\caption{Determination of absolute Sloan $r$ magnitudes from observed $(g-i)$ or $(r-z)$ colors.  The curves are the best fit second order polynomial fits to data points calculated for the 129 SED templates of \citet{brown14} at the redshifts in question. Given the measured color of a galaxy on the $x$-axis, its absolute magnitude $M_r$ is equal to the corresponding $y$-value on the curve minus the distance modulus $ D_{\rm{M}}$ and plus the appropriate apparent magnitude ($i$ on the left and $z$ on the right). We prefer to use observed $(g-i)$ color from $z =0$ to $z =0.25$ and $(r-z)$ color from $z =0.25$ to $z =0.5$, as this minimises the RMS offsets of the template SED points from the best fit polynomials (as given in the top left corner of each plot).   Each template marker has a unique shape and color enabling it to be easily identified in the plot.  Outliers more than 0.2 mag from the polynomials are excluded from the fit after three iterations of the best fit process.}
		\label{fig:r_calibration}
\end{figure*}

\begin{figure*}
 	\centering
		\includegraphics[width=0.49\textwidth]{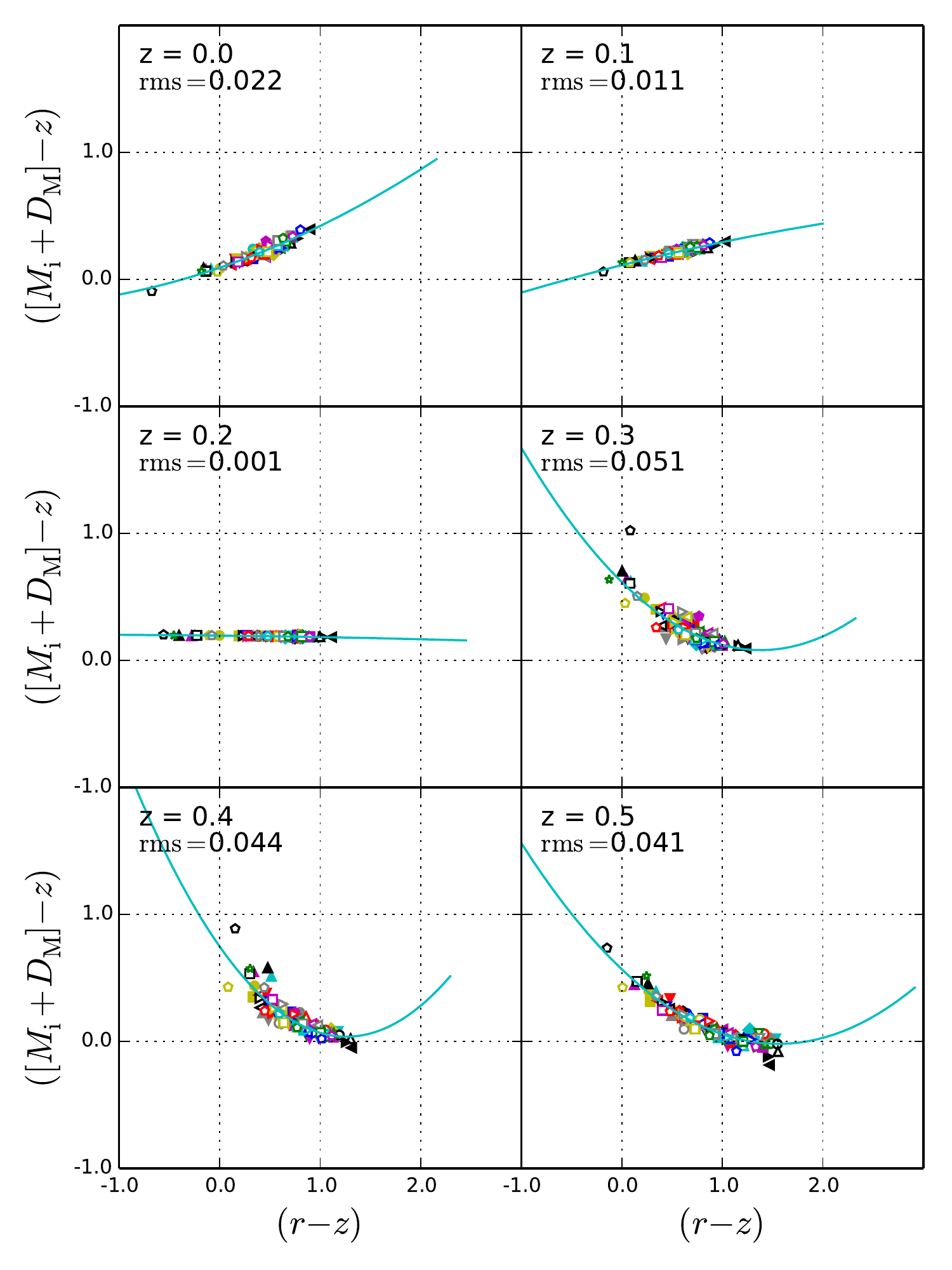} 
		\includegraphics[width=0.49\textwidth]{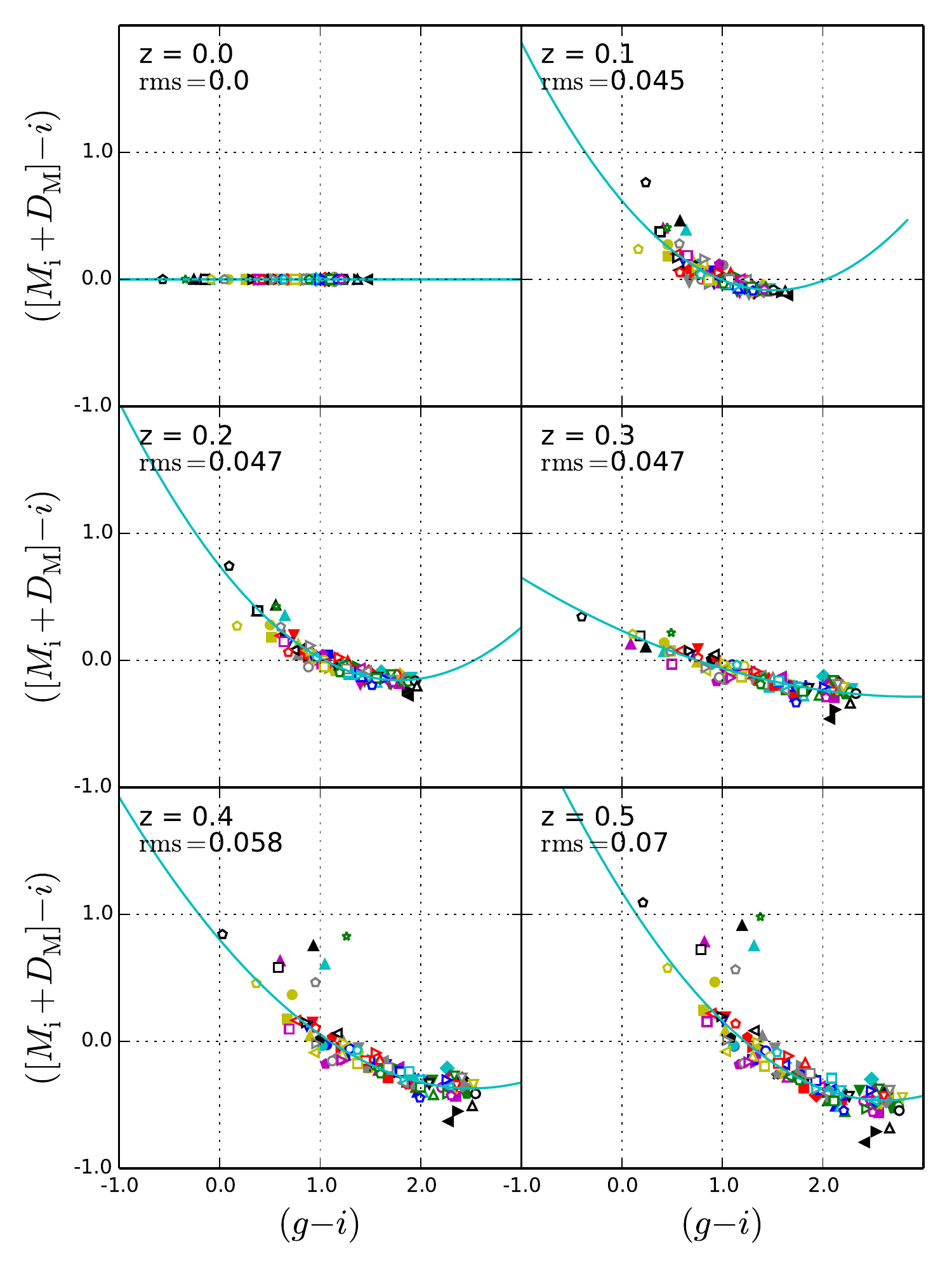}
		\caption{Determination of absolute Sloan $i$ magnitudes from observed $(r-z)$ or $(g-i)$ colors. The curves are the best fit second order polynomial fits to data points calculated for the 129 SED templates of \citet{brown14} at the redshifts in question. Given the measured color of a galaxy on the $x$-axis, its absolute magnitude $M_i$ is equal to the corresponding $y$-value on the curve minus the distance modulus $ D_{\rm{M}}$ and plus the appropriate apparent magnitude ($z$ on the left and $i$ on the right).  If good quality $z$-band data is available we prefer to use observed $(r-z)$ color as this results in smaller RMS offsets of the template SED points from the best fit polynomials (as given in the top left corner of each plot). If this is not the case we can still use $(g-i)$ colors. Each template marker has a unique shape and color enabling it to be easily identified in the plot. Outliers more than 0.2 mag from the polynomials are excluded from the fit after three iterations of the best fit process.}
		\label{fig:i_calibration}
\end{figure*}

\begin{figure*}
 	\centering
		\includegraphics[width=0.49\textwidth]{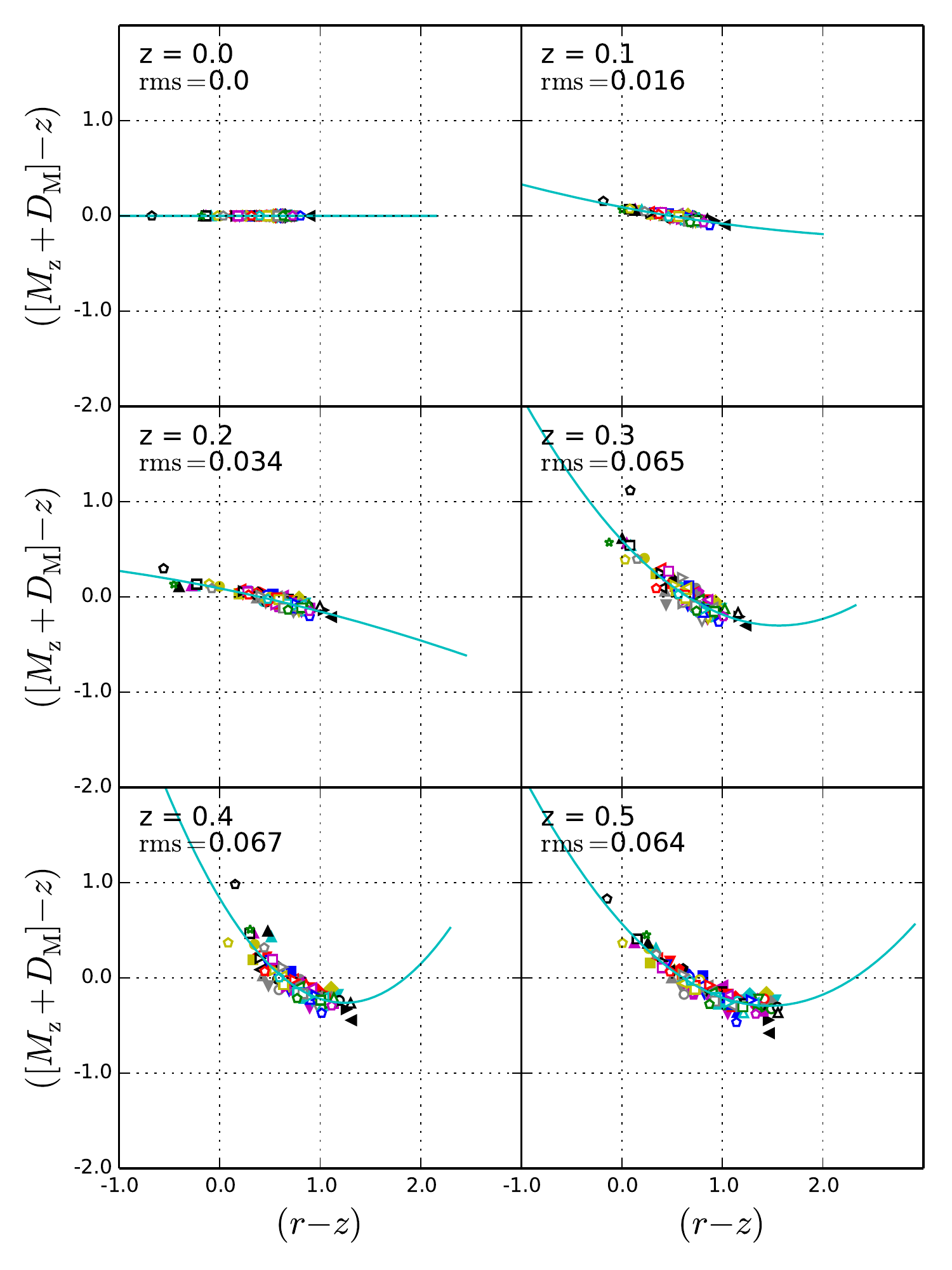} 
		\includegraphics[width=0.49\textwidth]{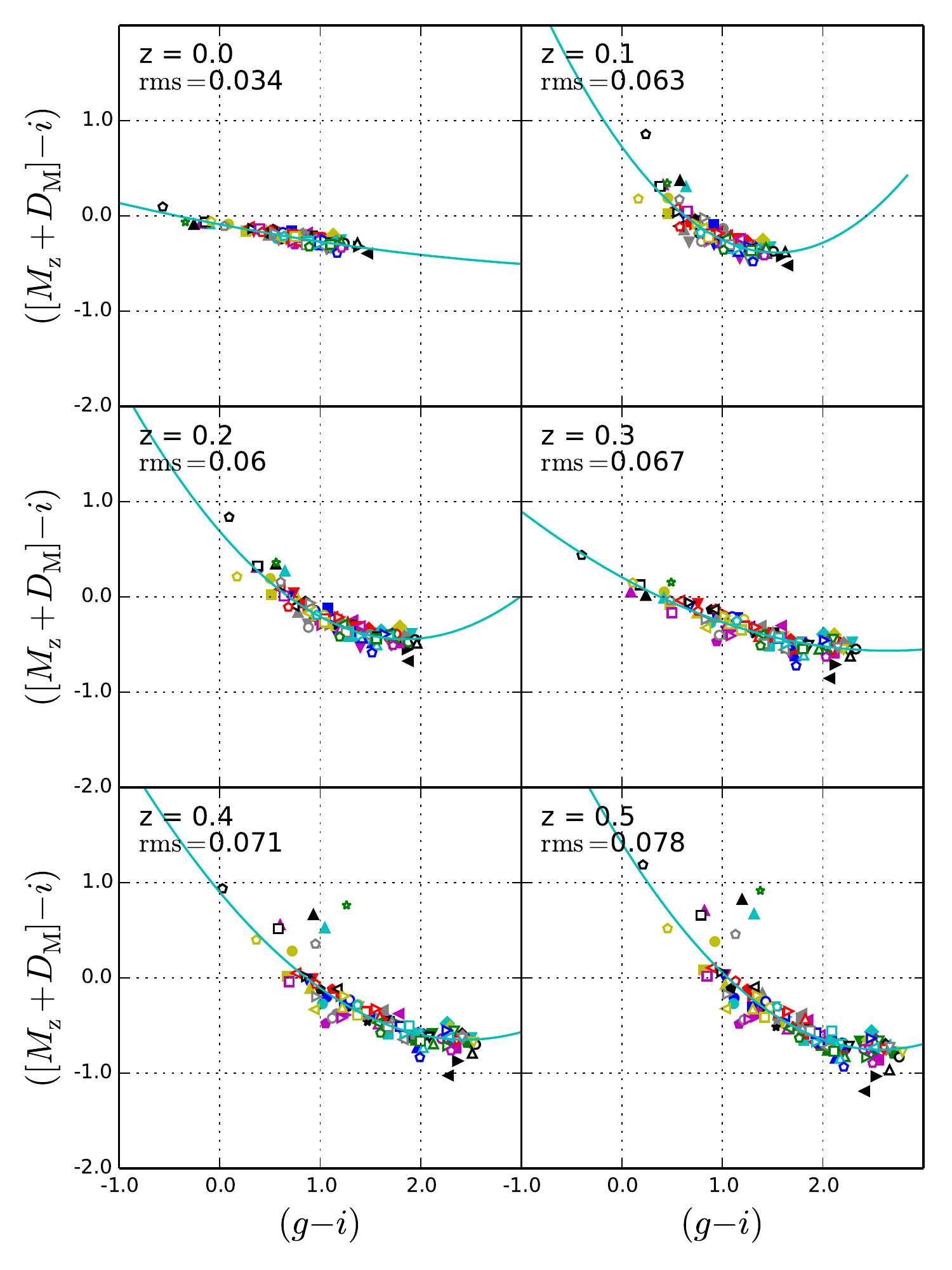}
		\caption{Determination of absolute Sloan $z$ magnitudes from observed $(r-z)$ or $(g-i)$ colors. The curves are the best fit second order polynomial fits to data points calculated for the 129 SED templates of \citet{brown14} at the redshifts in question. Given the measured color of a galaxy on the $x$-axis, its absolute magnitude $M_z$ is equal to the corresponding $y$-value on the curve minus the distance modulus $ D_{\rm{M}}$ and plus the appropriate apparent magnitude ($z$ on the left and $i$ on the right). If good quality $z$-band data is available we prefer to use observed $(r-z)$ color as this results in smaller RMS offsets of the template SED points from the best fit polynomials (as given in the top left corner of each plot). If this is not the case we can still use $(g-i)$ colors.  Each template marker has a unique shape and color enabling it to be easily identified in the plot. Outliers more than 0.2 mag from the polynomials are excluded from the fit after three iterations of the best fit process.}
		\label{fig:z_calibration}
\end{figure*}

\newpage

\section{DATA}
\label{sec:data}

\subsection{The templates}
\label{sec:templates}

We use the \citet {brown14} atlas of 129 ultraviolet to  mid-infrared spectral energy distributions (SEDs) of nearby galaxies. These combine ground-based and space-based observations in 26 photometric bands, gaps in spectral coverage being filled using MAGPHYS models \citep{cunha08}. The atlas spans a broad range of
 absolute magnitudes ($-14.4 < M_g < -22.3$) and colors ($0.1 < u-g < 1.9$). The systematic offsets and standard deviations for the residuals between between the actual observed magnitudes and the observed magnitudes predicted by integrating each SED over the filter transmission curves are all less than 0.03 mag in the $ugriz$ wavebands except for the $u$-band standard deviation which is 0.06. This provides a high degree of accuracy for our K-correction calculations. The broad spectral coverage enables our method to be applied to the determination of infrared K-corrections and absolute magnitudes, although we do not do so in this paper.

The atlas includes a large diversity of galaxy types, for example ellipticals, spirals, merging galaxies, blue compact dwarfs and luminous infrared galaxies (LIRGs). Within each type there is considerable diversity. For example, the atlas includes much of the diversity of ultraviolet SEDs observed in ellipticals, as well as ellipticals with significant star formation and dust emission in the mid-infrared. Our templates span the range of galaxy colors significantly better than previous SED libraries because (a) their large number enables a wider range of galaxy types to be represented, (b) they are derived from whole galaxy spectra rather than just the central regions of galaxies, and (c) they are based on accurate modern photometry. We refer the reader to \citet{brown14} for a fuller discussion of how the template SEDs compare with observed photometry.

\subsection{The New York University Value-Added Galaxy Catalog}
\label{sec:vagc}

We used SDSS $ugriz$ apparent AB magnitudes and extinction values from the New York University Value-Added Galaxy Catalogue \citep[VAGC,][]{blant05b} to compare absolute magnitudes that we calculated using our method with those that we calculated using the latest version of \textit{kcorrect} (v\_4\_2) \citep{blant07}.  As described in \citet{blant07}, we adjusted the apparent magnitudes to account for extinction and the small offsets between the ``SDSS natural system'' and the AB system (-0.036, 0.012, 0.010, 0.028, 0.040 respectively for $ugriz$). We chose to use the VAGC because it was used during the development of \textit{kcorrect} and therefore provides a good sample for benchmarking our approach relative to the prior literature. However we note that the majority of objects in this catalog have redshifts below $z\sim0.3$ so that we need to bear in mind that the sample will be biased above this redshift. The SDSS DR7 algorithm webpages \footnote{http://classic.sdss.org/dr7/algorithms/} recommend using Sloan "model magnitudes" for accurate determination of galaxy colors and we do this, rather than using the alternative Petrosian magnitudes. Model magnitudes are generated by fitting both exponential and de Vaucouleurs profiles and then choosing the profile which provides the best fit in the $r$-band.  The VAGC used independent and improved photometric calibration of the SDSS data.

\section{THE METHOD}
\label{sec:method}

\subsection{Calculation of K-corrections for the template SEDs}
\label{K_theory}

In order be able to derive absolute magnitudes and K-corrections for a sample of galaxies with photometry available in several wavebands, we need to be able to compare their photometry with that of the \citet{brown14} SED templates at the same redshift. We first determine how apparent and absolute magnitudes for the template SEDs depend on observed colours at a range of redshifts. \citet{hogg02} give a derivation of K-correction calculations for known SEDs, based on the original papers of \citet{humas56} and \citet{oke68}. Following similar derivations to theirs, we find the following formulae for the apparent magnitude in waveband $X$ and its absolute (restframe) magnitude in waveband $W$ for a galaxy with unnormalized flux density per unit wavelength $S(\lambda)$:

\begin{equation} \label{eq:mX_SED_em}
m_X =
 -2.5 \log_{10} 
 \left[
 \frac{
[1+z ]
}
{
4 \pi d_L^2
}
 \frac { 
 \int {
 \lambda_{em}
 S(\lambda_{em})
 T^X([1+z ]\lambda_{em})
 d \lambda_{em} 
  }
 } 
 { 
 \int {
\lambda
G(\lambda)
T^X(\lambda)
d \lambda 
 }
 } 
 \right]
\end{equation}

and

\begin{equation} \label{eq:MW_SED}
M_W =
 -2.5 \log_{10} 
\left[
 \frac
 {
1
}
{
4 \pi d_{10}^2
}
 \frac 
 { 
 \int {
 \lambda_{em}
 S(\lambda_{em})
 T^W(\lambda_{em})
 d \lambda_{em} 
  }
 } 
 { 
 \int {
\lambda
H(\lambda)
T^W(\lambda)
d \lambda 
 }
 } 
 \right].
\end{equation}

Here $T^X(\lambda)$, $T^W(\lambda)$ are the transmission functions of the $X$ and $W$ filters, i.e. the probabilities that a photon of wavelength $\lambda$ will get transmitted and counted. In the case of the observed filter $X$ this probability must in practise take account of the quantum efficiency of the CCD and any other relevant factors such as absorption by the telescope optics. $G(\lambda)$ and $H(\lambda)$ are the zero magnitude reference spectra used for the apparent and absolute magnitudes respectively (usually either both Vega or both AB). $d_L$ is the luminosity distance and $d_{10} = \rm{10 \, pc}$. For clarity we use the suffix ``\textit{em}'' in integrals where we are effectively integrating over emitted (restframe) wavelengths.

\subsection{Determination of absolute magnitudes from observed colors}
\label{sec:K_calibrations}

At any given redshift we use Equations \ref{eq:mX_SED_em} and \ref{eq:MW_SED} to produce a plot of $K_{WQ} \equiv (M_W + D_{\rm{M}}) - m_Q$ against an observed color $(m_P - m_Q)$ for the 129 template SEDs from \citet{brown14}. An example is shown in Figure \ref{fig:r_calibration_example}. We then determine the parameters $a(z), b(z), c(z)$ that provide the best second degree polynomial fits to this template plot over the range of redshifts $z$ of interest: 

\begin{equation}
	\label{eq:quadratic_K}
	K_{WQ}([m_P - m_Q], z) = a(z)(m_P - m_Q)^2 + b(z)(m_P - m_Q) + c(z).
\end{equation}

Then for any sample galaxy we can determine its absolute magnitude $M_W$  from its observed color $(m_P - m_Q)$ and its redshift $z$:

\begin{equation}
	\label{eq:quadratic}
	M_W = a(z)(m_P - m_Q)^2 + b(z)(m_P - m_Q) + c(z) - D_{\rm{M}}(z) + m_Q.
\end{equation}

As we show below, at any given redshift the observed wavebands $P$ and $Q$ can be chosen to minimise the RMS $y$-offset between the best fit quadratic and the individual template points. To determine the absolute magnitude $M_W$ of a sample galaxy we first measure the apparent magnitudes $m_P$, $m_Q$ in these two carefully selected wavebands. We then use the template based second degree polynomial model (Equation \ref{eq:quadratic}) for the same redshift to determine the absolute magnitude $M_W$.

Our method provides a quick and simple way of accurately determining the absolute magnitudes of galaxies using just one observed color and it requires no lengthy calculations or specially written software such as \textit{kcorrect} or \textit{InterRest}. Because the templates span almost the whole spectral range of observed galaxies, one can be confident in the model used and in the RMS error due to both galaxy diversity and random photometric errors in the observed observed color. Figures \ref{fig:u_calibration} to \ref{fig:z_calibration} show the polynomials we use to determine  absolute $ugriz$magnitudes.

\begin{figure}
 	\centering
		\includegraphics[width=0.47\textwidth]{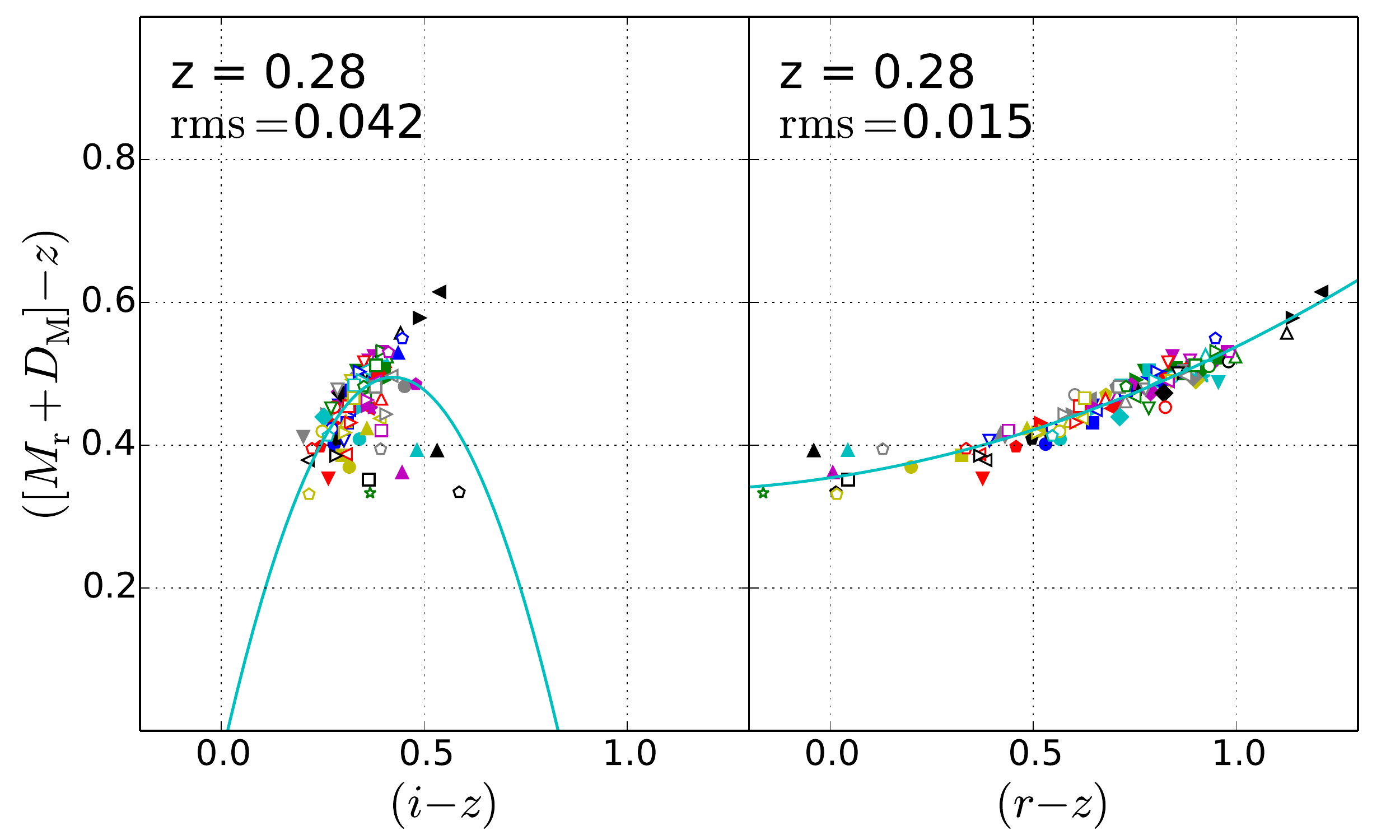}
		\caption{The nearest filters to the redshifted restframe waveband do not necessarily provide the best comparison color. Although the redshifted restrame $r$-band lies between the $i$ and $z$-bands at $z =0.28$, $(r-z)$ produces a much better defined template locus for determining $M_r$ than $(i-z)$, a much larger range of observed color, and no outliers.}
		\label{fig:r_calibration_iz}
\end{figure}

\begin{figure*}
 	\centering
		\includegraphics[width=0.75\textwidth]{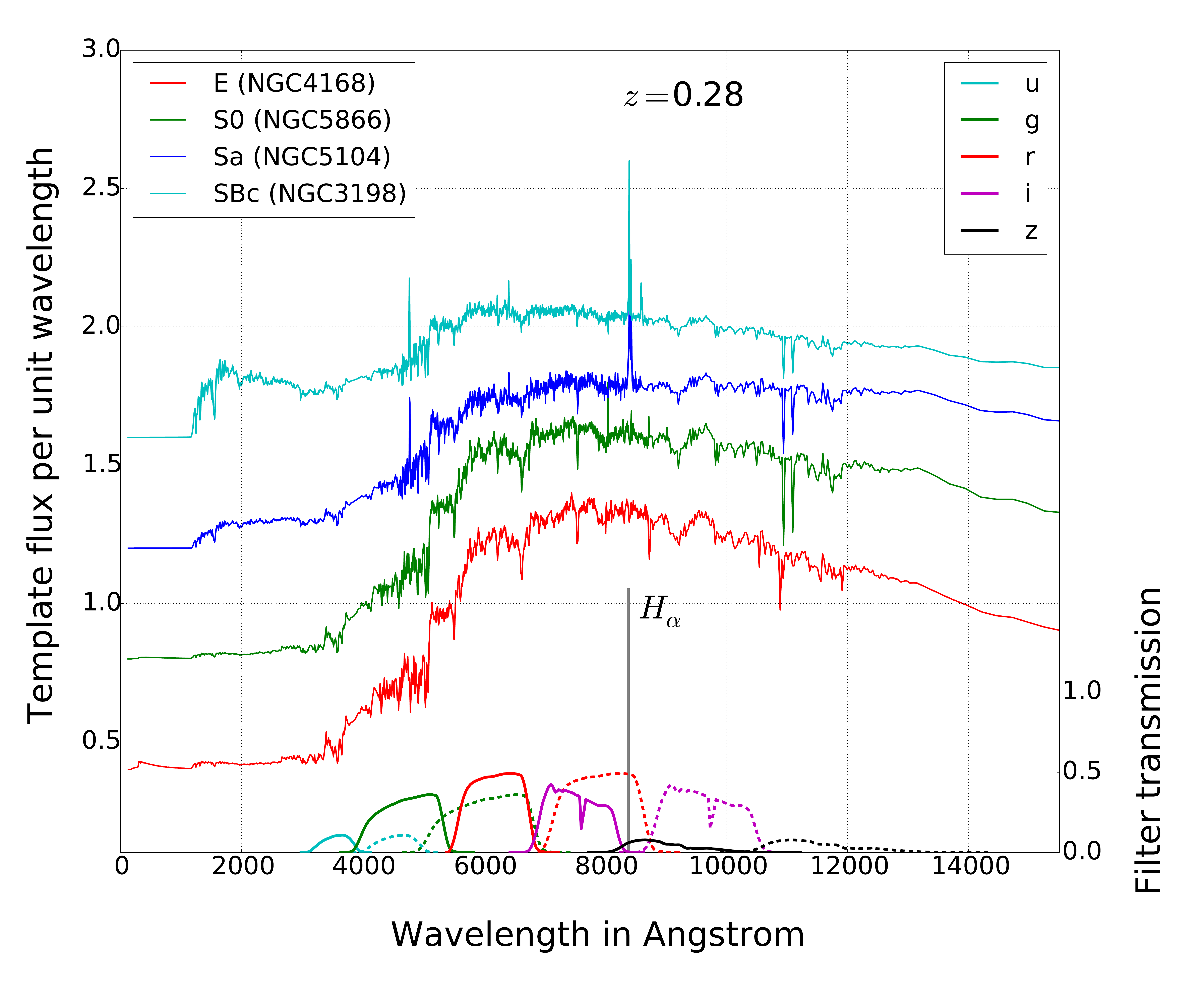}
		\caption{Observed SEDs of four representative galaxies at $z  = 0.28$ (offset for clarity). Also shown are transmission curves for the Sloan filter set (solid lines) and the Sloan filter set in the redshifted restframe of the galaxies (dashed lines). The observed SEDs are essentially parallel within the $i$ and $z$-bands and the \Ha line lies exactly between these two bands. Consequently there is very little spread in $(i-z)$ color at $z  = 0.28$ (Figure \ref{fig:r_calibration_iz}).}
		\label{fig:r_calibration_spectra}
\end{figure*}

\begin{figure*}
 	\centering
		\includegraphics[width=0.9\textwidth]{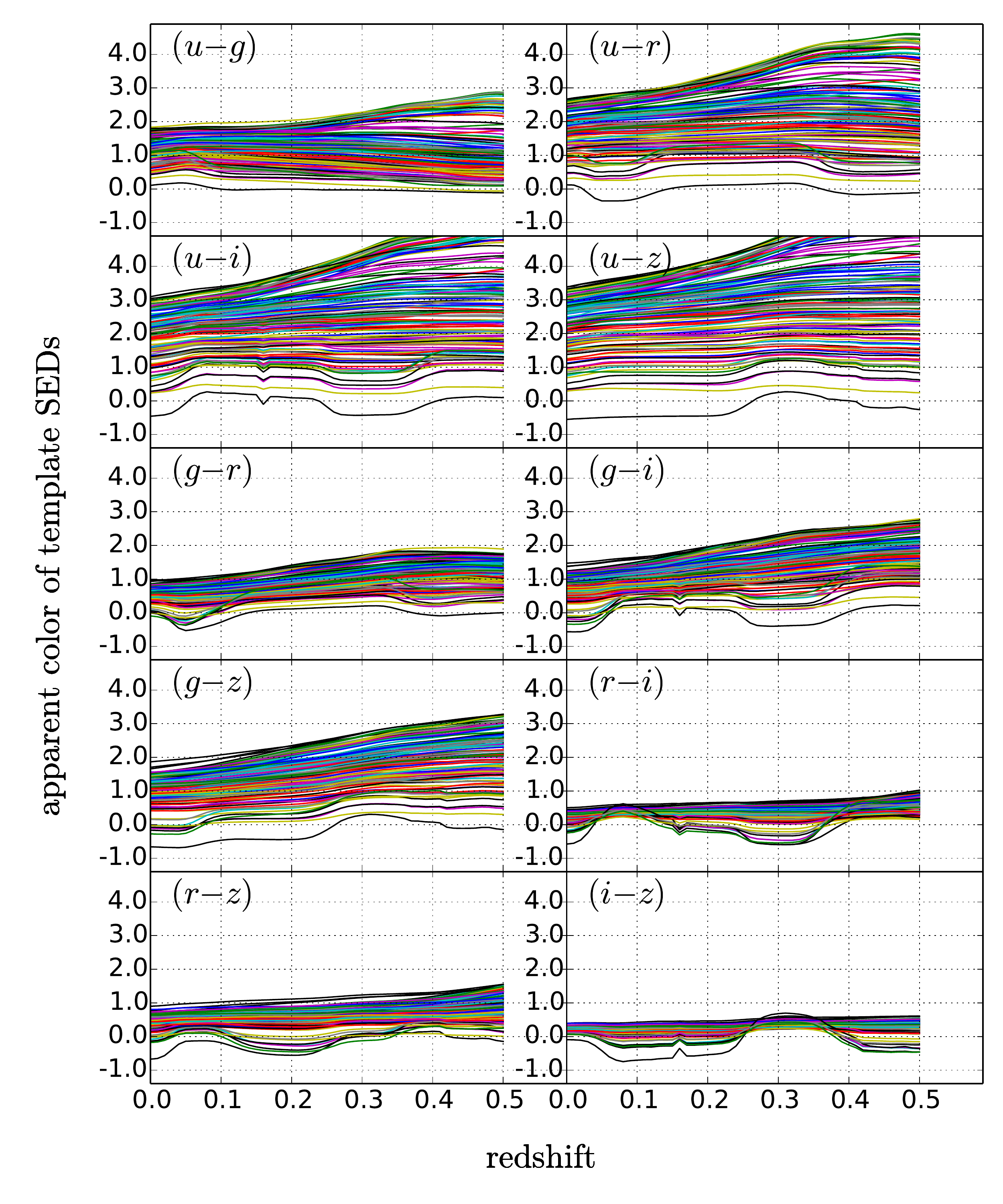}
		\caption{Variation with redshift of the ten possible observed colors for the $ugriz$ filter set. Colors are plotted for all 129 template SEDs. There is much less variation in observed color at longer wavelengths where there is little variation between SEDs, except for absorption and emission lines. A good range of observed color is necessary, otherwise the polynomial fits are not  properly constrained, as is the case for $(i - z)$ at $z = 0.28$ (see \S\ref{sec:color_choices} and the left hand plot of Figure \ref{fig:r_calibration_iz}).} 
		\label{fig:colour_evolution}
\end{figure*}

\newpage

\subsubsection{Errors}
\label{sec:errors}

A benefit of our method is that it enables the random errors in computed absolute magnitude values to be determined from the random errors in apparent magnitudes and redshifts, and the template scatter $\sigma_{\textrm{templates}}$ in plots such as that in Figure \ref{fig:r_calibration_example}.

Random and systematic errors in computed absolute magnitudes arising from random errors in the observed color can be analysed as follows.  Let $m_P$ and $m_Q$ denote the true values of the $P$ and $Q$ apparent magnitudes for a particular galaxy. Let the measured values be $m_P + \delta_P$ and $m_Q + \delta_Q$ where $\delta_P$ and $\delta_Q$ are random measurement errors with variances $\sigma_P^2$ and $\sigma_Q^2$.  Let $M$ be the true value of the absolute magnitude $M_W$. The value deduced from the measurements is then $M + \Delta$ where, from Equation \ref{eq:quadratic}:

\begin{multline}
	(M + \Delta)  + D_{\rm{M}} - (m_Q + \delta_Q) = \\
		\; {} a[(m_P+ \delta_P) - (m_Q + \delta_Q)]^2 +b[(m_P+ \delta_P) - (m_Q + \delta_Q)] + c. 
\end{multline}

Subtracting ${M + D_{\rm{M}} - m_Q = a(m_P - m_Q)^2 +b(m_P - m_Q) + c}$:

\begin{equation}\label{eq:E}
	\Delta = a(\delta_P - \delta_Q)^2 + 2a(m_P - m_Q)(\delta_P - \delta_Q) +  b(\delta_P - \delta_Q) + \delta_Q.
\end{equation}

Ignoring powers higher than the second in $\delta_P$ and/or $\delta_Q$ the variance $\sigma_{\textrm{phot}}^2 = {\overline{\Delta}}^2 - \overline{\Delta^2}$ over a large number repeated measurements is:

\begin{equation}\label{eq:E_var}
	\sigma_{\textrm{phot}}^2 = \beta^2 \sigma_P^2 +  (1 - \beta )^2 \sigma_Q^2 \quad \textrm{ where } \, \beta = 2a(m_P - m_Q) + b.
\end{equation}

Errors in redshift also give rise to errors in computed absolute magnitude. A fractional redshift error $\delta z/z$ produces an  error in $M_W = m_Q + K_{WQ} - D_{\rm{M}}$  of:

\begin{equation}\label{eq:E_redshift}
	\delta M_W = z \left( \frac{\partial K_{WQ}}{\partial z} - \frac{d D_{\rm{M}}}{dz}\right) \left(\frac{\delta z}{z}\right).
\end{equation},

where $K_{WQ}([m_P - m_Q], z)$ is given by Equation \ref{eq:quadratic_K}. We use fractional redshift errors $\delta z/z$ because they are a more meaningful measure than absolute errors $\delta z$ and because $z \, d D_{\rm{M}}/dz$ is bounded as $z \rightarrow 0$ (tending to the value $5/\ln10 = 2.1715$ as a result of the inverse square law). 

A fractional RMS error of $\sigma_z/z$ in redshift results in a an RMS absolute magnitude error of:

\begin{equation}\label{eq:sigma_redshift}
	\sigma_{\textrm{redshift}} = z \left( \frac{da}{dz}[m_P - m_Q]^2 + \frac{db}{dz}[m_P - m_Q] + \frac{dc}{dz} - \frac{d D_{\rm{M}}}{dz}\right) \frac{\sigma_z}{z}.
\end{equation}.

Combining the three sources of random error  in quadrature, we have the total variance in absolute magnitude for each galaxy in a sample:

\begin{equation}\label{eq_E_total}
	\sigma_{\textrm{total}}^2 =\sigma_{\textrm{templates}}^2 +  \sigma_{\textrm{phot}}^2 + \sigma_{\textrm{redshift}}^2.
\end{equation}

Values of $\sigma_{\textrm{templates}}$ are tabulated in Tables \ref{tab:u_parameters} to \ref{tab:z_parameters} alongside the polynomial coefficients $a$, $b$ and $c$. $\sigma_{\textrm{templates}}$ provides a measure of the range of polynomial offsets to be expected for a sample of real galaxies. It will be smaller than the RMS scatter for the templates (even with gross outliers excluded) because the template atlas includes several rare types of galaxy SED and does not attempt to provide a similar frequency distribution of SED types to that found in real galaxy samples.  $\sigma_{\textrm{phot}}$ values can be calculated from $a$, $b$ and $c$ using Equation \ref{eq:E_var}. Values for $\sigma_{\textrm{redshift}}$ can be determined by extracting values of $da/dz$, $db/dz$ and $dc/dz$ from the tables and using these in Equation \ref{eq:sigma_redshift}.

\subsection{Choice of the best observed color}
\label{sec:color_choices}

As the example in Figure \ref{fig:r_calibration_example} shows, the diversity of galaxy SEDs produces scatter in the template values of ${(M_W + D_{\rm{M}}) - m_Q}$ about the best fit polynomial.  Assuming that the 129 templates adequately  span the full range of galaxy colors, the template scatter about the best fit polynomial scatter for observed galaxies will be similar to that for the templates and so will also have variance $\sim \sigma_{\textrm{templates}}^2$. We aim to choose observed colors which minimise the RMS offset $\sigma_{\textrm{templates}}$.

Choosing observed colors which straddle or are close to the redshifted restframe waveband $W$, as in \citet{rudni03} and the $InterRest$ software \citep{taylo09}, does not always result in the smallest template scatter $\sigma_{\textrm{templates}}$. For example, at $z  = 0.28$, the redshifted restframe $r$ band lies mid-way between the observed $i$ and $z$-bands. One might expect, therefore, that the best observed color to use for $M_r$ at this redshift would be ${(i-z)}$. Figure \ref{fig:r_calibration_iz} shows that this is not the case and that $(r-z)$ is a better color as it produces a very well-defined sequence of points, while  ${(i-z)}$ gives rise to a small range of observed color, a tightly bunched set of points, a highly indeterminate polynomial fit, and a number of significant outliers.  The RMS offset $\sigma_{\textrm{templates}}$ is 0.042 for ${(i-z)}$ but only 0.015 for $(r-z)$. 

It is instructive to examine typical SEDs at $z  = 0.28$ to understand why there is such a small range in ${(i-z)}$ color. Figure \ref{fig:r_calibration_spectra} shows that the redshifted SEDs of representative galaxies are essentially parallel in the region covered by the $i$ and $z$ filters, while the \Ha line lies exactly between them so that variations in \Ha emission line strength between galaxies do not affect the ${(i-z)}$ color. It might be thought that using filters which do not sample \Ha emission would lead to significant errors (up to \s0.1 mag), as a result of this prominent spectral feature not being sampled. However, the strength of the \Ha line is correlated with the overall shape of the SED so that we do not find galaxies with otherwise similar SEDs but very different  \Ha equivalent line widths. This means that enough information is encoded in other parts of the SED to provide an accurate measurement of the \Ha line strength and hence absolute magnitude.  This is confirmed by the right hand plot in Figure \ref{fig:r_calibration_iz} which shows little scatter around the best fit polynomial.

In general, we note that the variations in color on both axes due to galaxy diversity depend on a complex interplay of factors, including the overall shape of the SED and specific features such as absorption and emission line strengths and the prominence of the Balmer jump and 4000\AA \, break. Figure \ref{fig:colour_evolution} shows how observed color varies with redshift for all 129 templates and all 10 possible combinations of observed $ugriz$ filters. There is a relatively small range of color values at longer wavelengths where there is little variation between SEDs apart from absorption and emission lines. The small range in $(i-z)$ color at $z \sim 0.28$ already alluded to can be seen clearly in the figure.

As well as variance in computed absolute magnitudes due to galaxy diversity, we also need to consider the variance $\sigma_{\textrm{phot}}^2$ in computed absolute magnitudes due to random photometric errors. Equation \ref{eq:E_var} shows that this depends on the quantity $\beta$, which is a function of the best fit parameters $a$ and $b$ and the color value $(m_P-m_Q)$.  Choosing an observed color which minimises $|a|$ and $|b|$ will in general result in a smaller value of $|\beta|$ and hence $\sigma_{\textrm{phot}}$. (The exception would be when $2a(m_P - m_Q)$ and $b$ cancel, but for any given polynomial this would only arise at specific values of the color $(m_P - m_Q)$, so it is not useful to consider this situation further.)

The gradient of the polynomial fit in Equation \ref{eq:quadratic} with respect to color is ${\beta = 2a(m_P - m_Q) + b}$. If the range of observed colors $(m_P - m_Q)$ on the $x$-axis  is small relative to the range of $(M_W + D_{\rm{M}}) - m_Q$ values on the $y$-axis, the slope $\beta$ will be large. This will mean that the error $\sigma_{\textrm{phot}}$ from Equation \ref{eq:E_var} will also be large. In practice we find that observed colors with a small range of values on the $x$-axis relative to the range on the $y$-axis also result in larger template scatter, and consequently larger values of $\sigma_{\textrm{templates}}$ as well, but we note that formally this need not necessarily be the case.  An example is shown in Figure \ref{fig:r_calibration_iz}, which also makes the point that a small range of observed colors will result in greater uncertainty in the best fit parameters, in the sense that they will be sensitive to small changes in the template SEDs.

If $a$ is small the gradient ${\beta = 2a(m_P - m_Q) + b}$ changes little with observed color and the polynomial is close to being linear.  This occurs when the restframe $W$-band coincides with the observed $Q$-band and the polynomial is then close to being a horizontal line (i.e. $a, b, \beta \approx 0$, as in the lefthand panel of Figure \ref{fig:g_calibration} at $z  = 0.3$). It also occurs when the restframe $W$-band coincides with the observed $P$-band and the polynomial is then close to being a line of slope +1 (i.e. $a \approx 0$ and $b, \beta \approx 1$, as in the righthand panel o Figure \ref{fig:g_calibration} at $z  = 0.3$). When the restframe $W$-band lies between the observed $P$ and $Q$-bands, we find $0 \lesssim \beta \lesssim 1$. 

It is very helpful to visually examine the polynomial plots of $(M_W + D_{\rm{M}}) - m_Q$ against  $(m_P - m_Q)$ for different observed colors. This gives an instant visual impression of how the best fit polynomials evolve with redshift and how large the RMS scatter $\sigma_{\textrm{templates}}$ is at different redshifts. It can also reveal unexpected polynomial behavior at certain redshifts, such as the small ${(i-z)}$ color range and large scatter in $(M_r + D_{\rm{M}}) - m_z$ at $z  = 0.28$ referred to above (Figure \ref{fig:r_calibration_iz}).

Small apparent magnitude errors can result in large absolute magnitude errors when the value of $a$ is large so that the polynomial is highly curved in the region of interest and possibly even has a sharp maximum or minimum. This is illustrated by the righthand $z =0.1$ panel in Figure \ref{fig:g_calibration}, (although we do in fact prefer not to use $(r-i)$ as observed color for determining $M_g$ at $z =0.1$). In this case $a=-5.04$ and $b=6.44$ so that $\beta \approx 3.4$ at $(r-i) = 0.3$ and ${\sigma_{\textrm{phot}}^2 = 11.6 \sigma_r^2 +  5.8 \sigma_i^2}$. Photometric errors of $\sigma_r = \sigma_i = 0.1$ will give rise to absolute magnitude errors $\sigma_{\textrm{phot}} \approx 0.4$, four times larger than either of the individual apparent magnitude errors. We note also that in this case the polynomial turns over rapidly at the red end of the observed color range so that galaxies a little redder than this would be expected to be assigned absolute magnitudes \s0.5 mag or more too bright.) Outliers are also revealed by visual inspection, as well as by the quantitative criterion we use that they are offset from the polynomial by more than 0.2. We discuss offsets further in \S \ref{sec:outliers}.

\subsubsection{Different observed colors}
\label{sec:alternatives}

\begin{figure}
 	\centering
		\includegraphics[width=0.45\textwidth]{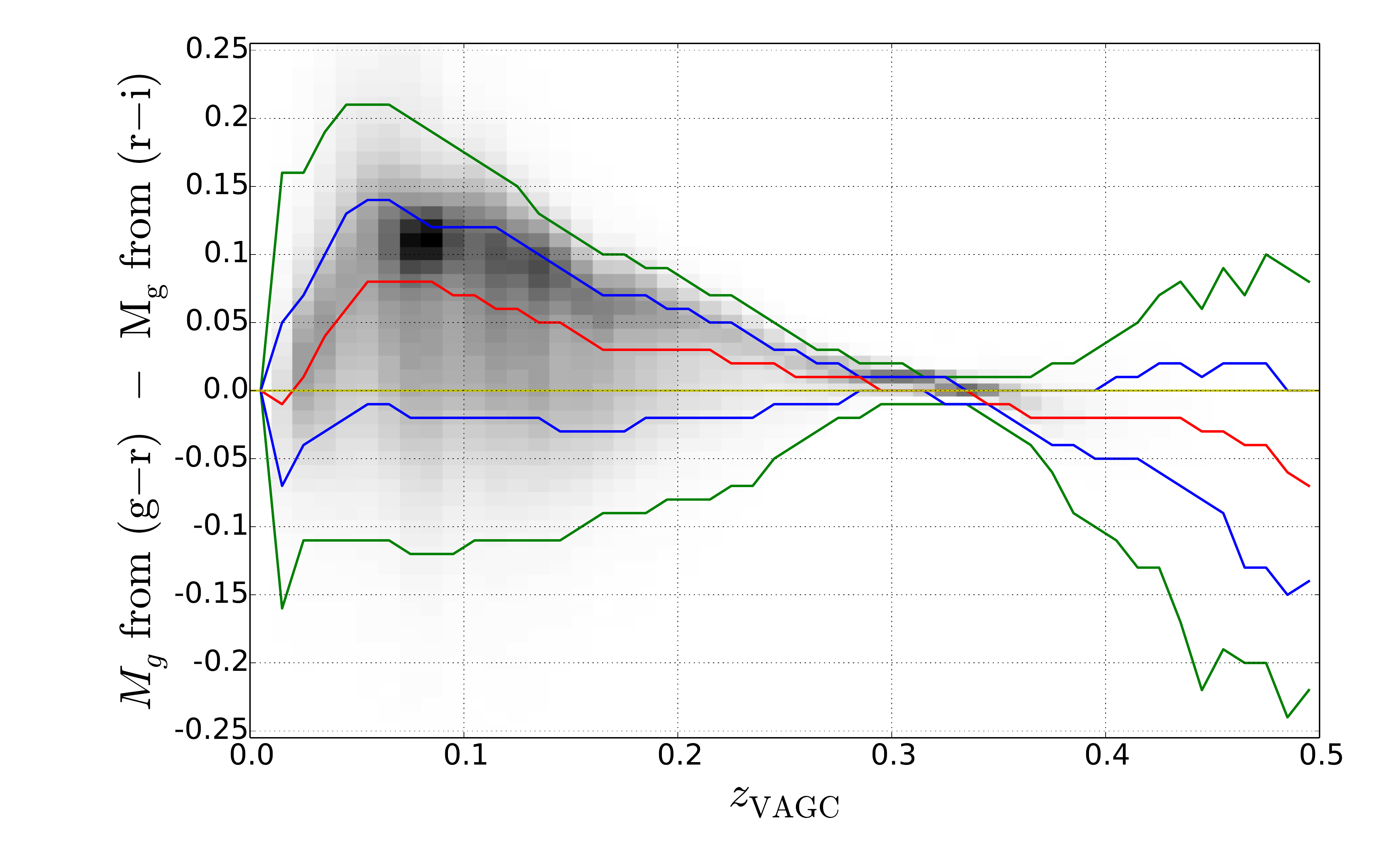}
		\caption{Showing that there is no discontinuity in calculated $M_g$ values on switching from $(g-r)$ to $(r-i)$ observed color at $z  = 0.28$. The absolute $g$-magnitudes compared are for galaxies in the VAGC calculated using our method. The red curve denotes the median of the $y$-axis distribution, blue denotes the 15.9 and 84.1 percentiles, and green the 2.3 and 9.7 percentiles.}
		\label{fig:g_continuity}
\end{figure}

To minimise absolute magnitude errors, the optimum observed color to use can be different in different redshift ranges, as summarised in Table \ref{tab:sloan_colors}.  This is the case for $M_g$ and $M_r$, but not $M_u$, $M_i$ and $M_z$ (unless the $u$ or $z$-band photometry normally used is poor or missing).  For determining $M_g$ we prefer $(g-r)$ from $z  = 0$ to $z  = 0.34$ and $(r-i)$ from $z  = 0.34$ to $z  = 0.5$. This strategy uses the observed color with the smallest RMS offset $\sigma_{\textrm{templates}}$ at every redshift.  The benefit of switching observed colors is significant: at $z  = 0.1$, $\sigma_{\textrm{phot}}$ is 3.1 times smaller for $(g-r)$ than for $(r-i)$, while at $z  = 0.5$ it is 10 times smaller for $(r-i)$  than for $(g-r)$.  Using data from the VAGC, we show in Figure \ref{fig:g_continuity} that switching observed color at $z  = 0.28$ does not result in a discontinuity in computed $M_g$ values. 

\begin{figure}
 	\centering
		\includegraphics[width=0.45\textwidth]{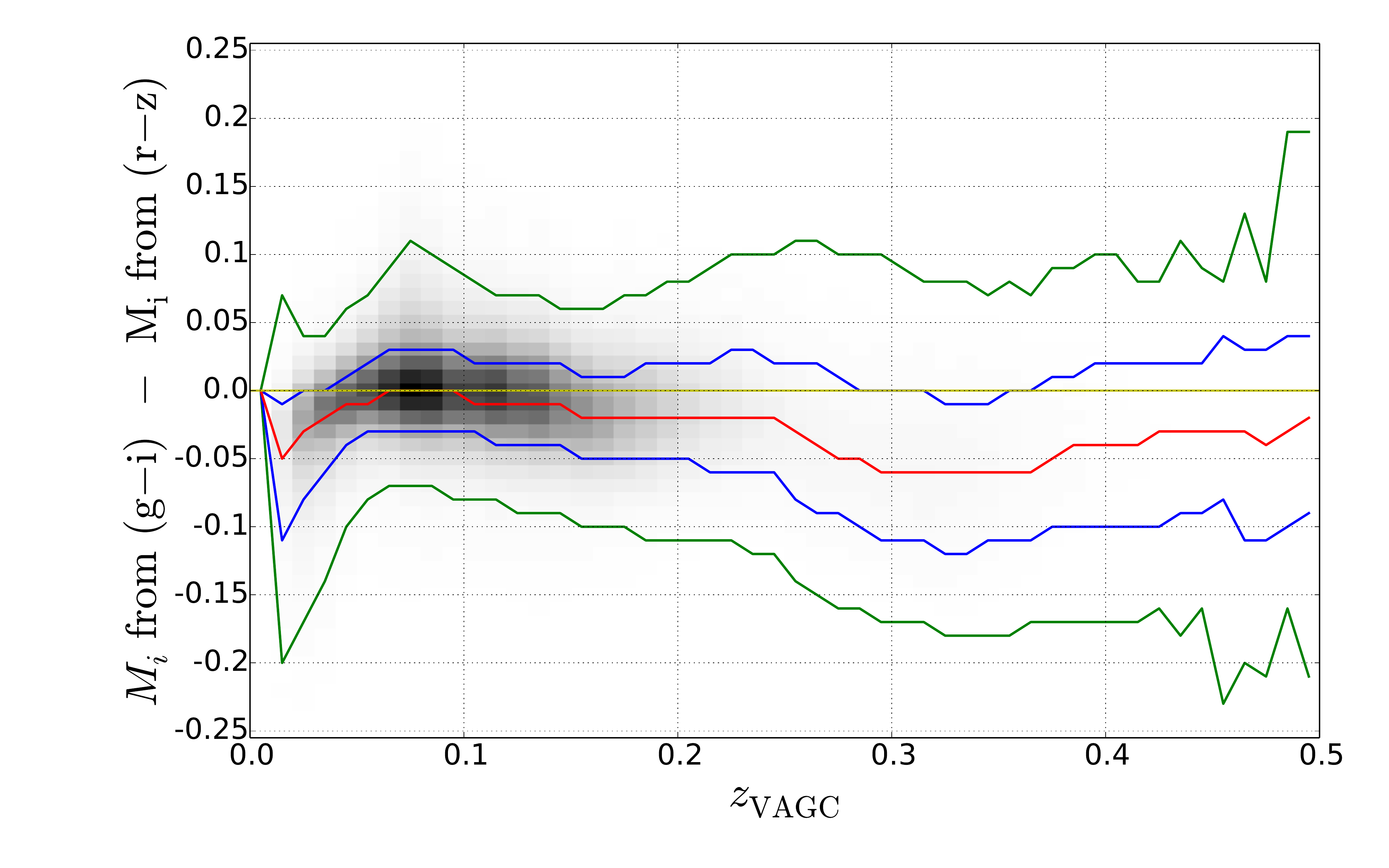}
		\caption{Showing that alternative observed colors can be used when measurements in particular wavebands are not available. The plot compares $M_i$ values calculated using $(g-i)$ and $(r-z)$ observed colors. $(r-z)$ is normally preferred as observed color beyond redshift 0.25. However, if $z$-band apparent magnitudes are unavailable or have large random error,  $(g-i)$ can be used as observed color instead. The absolute $i$-magnitudes compared are for galaxies in the VAGC calculated using our method. The red curve denotes the median of the $y$-axis distribution, blue denotes the 15.9 and 84.1 percentiles, and green the 2.3 and 9.7 percentiles.}
		\label{fig:z_missing-a}
\end{figure}

One can generally find an alternative to the preferred observed color if measurements are not available or are too noisy in a particular waveband. For example, using redshift and apparent magnitude data from the VAGC Figure \ref{fig:z_missing-a} shows that if no $z$-band magnitudes are available, $(g-i)$ can be used for determining $M_i$ at all redshifts instead of $(r-z)$. For redshifts less than \s0.25 the systematic offset between the two sets of measurements is less than \s0.01 mag, and for $0.25 < z  < 0.5$ it is less than \s0.03 mag. We recall from \S\ref{sec:color_choices} that the more obvious choice of $(i-z)$ as observed color is not satisfactory for determining either $M_i$ or $M_z$. 

Because $u$-band photometry generally has much larger errors than photometry in the $griz$ wavebands (\s20 times larger for faint SDSS galaxies) and may sometimes be unavailable, we also tabulate parameters for calculating $M_u$ using observed color $(g-r)$ instead of $(u-g)$ (Figure \ref{fig:u_calibration}). Similarly, we provide parameters for calculating $M_i$ and $M_z$ without the use of $z$-band photometry for situations where this is absent or of poor quality, (Figures \ref{fig:i_calibration} and \ref{fig:z_calibration}).

\subsubsection{Outliers}
\label{sec:outliers}

To prevent outliers from skewing our polynomial fits, we iteratively exclude objects that lie more than 0.2 magnitudes from the best fit polynomial until we achieve a stable solution. For any given observed color, outliers only become significant at certain redshifts. The majority of them result from faint starburst and compact blue galaxies. To assess the impact of outliers with faint absolute magnitudes, we determined the apparent magnitudes at which these galaxies would be visible in galaxy surveys. A detailed examination reveals that none of the gross outliers for our preferred observed colors would be visible in a survey with limiting magnitude $g = 21.5$. 

In a survey with limiting magnitude $g = 23.0$ only the following template galaxies would be visible and exhibit outliers at specific redshifts: $M_g$ for compact blue dwarf UGCA410 at $z  \sim 0.1$,  $M_i$ and $M_z$ for compact blue dwarf UGCA219  at $z  \sim 0.3$, $M_g$ for compact starburst galaxy MRK1450 at $z  \sim 0.1$,  $M_g$ for compact starburst galaxy MRK930 at $z  \sim 0.4$, and $M_z$ for star-forming galaxy NGC660 at $z  \sim 0.4-0.5$. The SED plots in \citet{brown14} show that all these galaxies have very strong emission lines that contribute significantly to their optical colors. Visual inspection of NGC660 reveals prominent tidal tails and two intersecting dust lanes indicating that it is in fact two spiral galaxies in the process of merging. It is unusual because the intersecting dust lanes are being viewed edge-on. We therefore expect galaxies with similar SEDs to comprise a negligible fraction of any galaxy sample at $z  \sim 0.4-0.5$. We conclude therefore that outliers do not significantly affect our calculation of K-corrections except for surveys deeper than $g = 21.5$ where some faint compact blue and starburst galaxies will be assigned absolute magnitudes that are 0.2 to 0.3 mag too bright at certain redshifts.

\begin{deluxetable}{ccccc}	
\tablewidth{0pt}
\tablecolumns{5}
\tabletypesize{\scriptsize}
\tablecaption {The preferred observed colors used to determine K-corrections.}
\tablehead{
\colhead{restframe} & \colhead{central} & \colhead{redshift} & \colhead{preferred} & \colhead{alternative}
\\
\colhead{waveband} & \colhead{wavelength} & \colhead{range} & \colhead{color} & \colhead{color}
 \\
\colhead{$M_W$} & \colhead{$\mu m$} & \colhead{ } & \colhead{$(m_P - m_Q)$} & \colhead{$(m_P - m_Q)$}
}
 \startdata
$u$ & 0.3551 & 0.0 to 0.5 & $(u-g)$ & $(g-r)$ \\
\\
$g$ & 0.4686 & 0.0 to 0.34 & $(g-r)$ & \\
$g$ & 0.4686 & 0.34 to 0.5 & $(r-i)$ & \\
\\
$r$ & 0.6166 & 0.0 to 0.25 & $(g-i)$ & \\
$r$ & 0.6166 & 0.25 to 0.5 & $(r-z)$ & \\
\\
$i$ & 0.7480 & 0.0 to 0.5 & $(r-z)$ & $(g-i)$ \\
\\
$z$ & 0.8932 & 0.0 to 0.5 & $(r-z)$ & $(g-i)$ \\
\enddata
\label{tab:sloan_colors}
\end{deluxetable}

\begin{figure*}
 	\centering
		\includegraphics[width=0.9\textwidth]{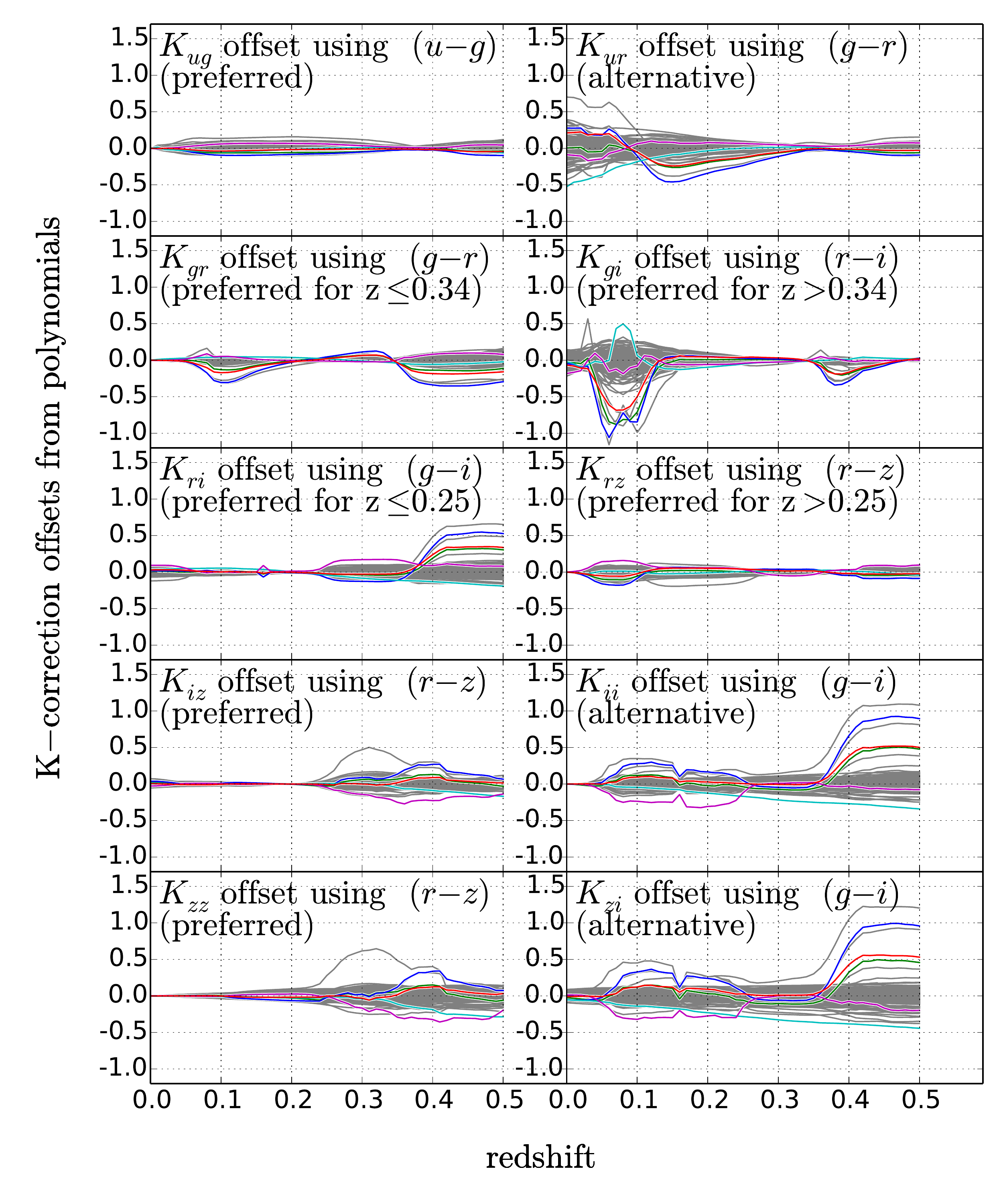}
		\caption{Variation with redshift of the differences between absolute magnitudes determined from the template SEDs and from the polynomials for our preferred observed colors.  For the bulk of galaxies (shown in gray) the preferred pairs of filters give maximum residuals in $ugriz$ of 0.05, 0.04, 0.05, 0.1 and 0.1 mag respectively. The RMS residual values  are smaller: 0.04, 0.03, 0.03, 0.06 and 0.07 mag respectively and these will be overestimates because the templates include a disproportionate number of rare galaxy types. The rare outlier galaxies referred to in the text are shown in color and are mainly faint compact blue and faint starburst galaxies. They only result in outliers at redshifts where they are visible in surveys with limiting magnitudes brighter than $g = 21.5$.} 
		\label{fig:offset_evolution}
\end{figure*}

\section{RESULTS AND DISCUSSION}
\label{sec:results}

\subsection{K-corrections for the Sloan filter set }
\label{sec:sloan}

For each Sloan filter we calculate the polynomial parameters $a$, $b$ and $c$ in Equation \ref{eq:quadratic} on a one-dimensional grid of 51 redshifts between $z  = 0.0$ and $z  = 0.5$ by minimising the sum of the squared $y$-offsets, performing three iterations in order to eliminate from the fit any templates with a $y$-offset of more than 0.2 mag.  The tables in the appendix list our results for the five Sloan filters $ugriz$ and indicate which are our preferred observed colors at different redshifts, and which are possible alternatives that avoid use of $u$ and $z$-band photometry. Also tabulated are the RMS $y$-offsets between the polynomial models and the template points.  Figure \ref{fig:offset_evolution} plots the differences between absolute magnitudes determined from the template SEDs and from the polynomials and how these differences vary with redshift. The preferred pairs of filters give maximum residuals in $ugriz$ of 0.05, 0.04, 0.05, 0.1 and 0.1 mag respectively, (except for the faint outlier galaxies discussed previously). The RMS residual values  are smaller: 0.04, 0.03, 0.03, 0.06 and 0.07 mag respectively, and these values may be overestimates because the \citet{brown14} SED atlas includes many rare types of galaxy and therefore produces a much broader distribution of offsets than any real galaxy survey would.

\subsection{Comparison with \textit{kcorrect}}
\label{sec:kcorrect}

\begin{figure*}
 	\centering
		\includegraphics[width=0.49\textwidth]{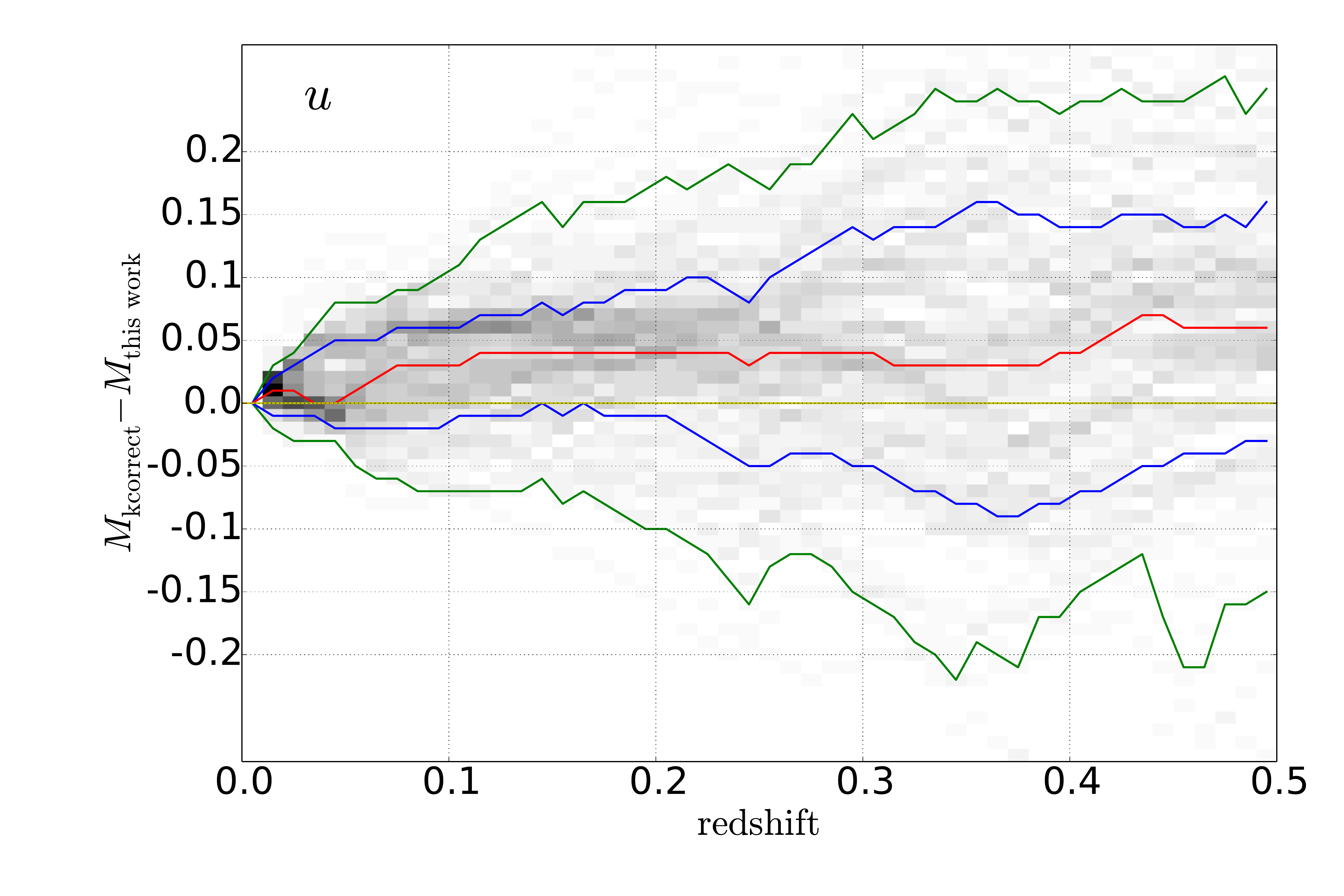}
		\includegraphics[width=0.49\textwidth]{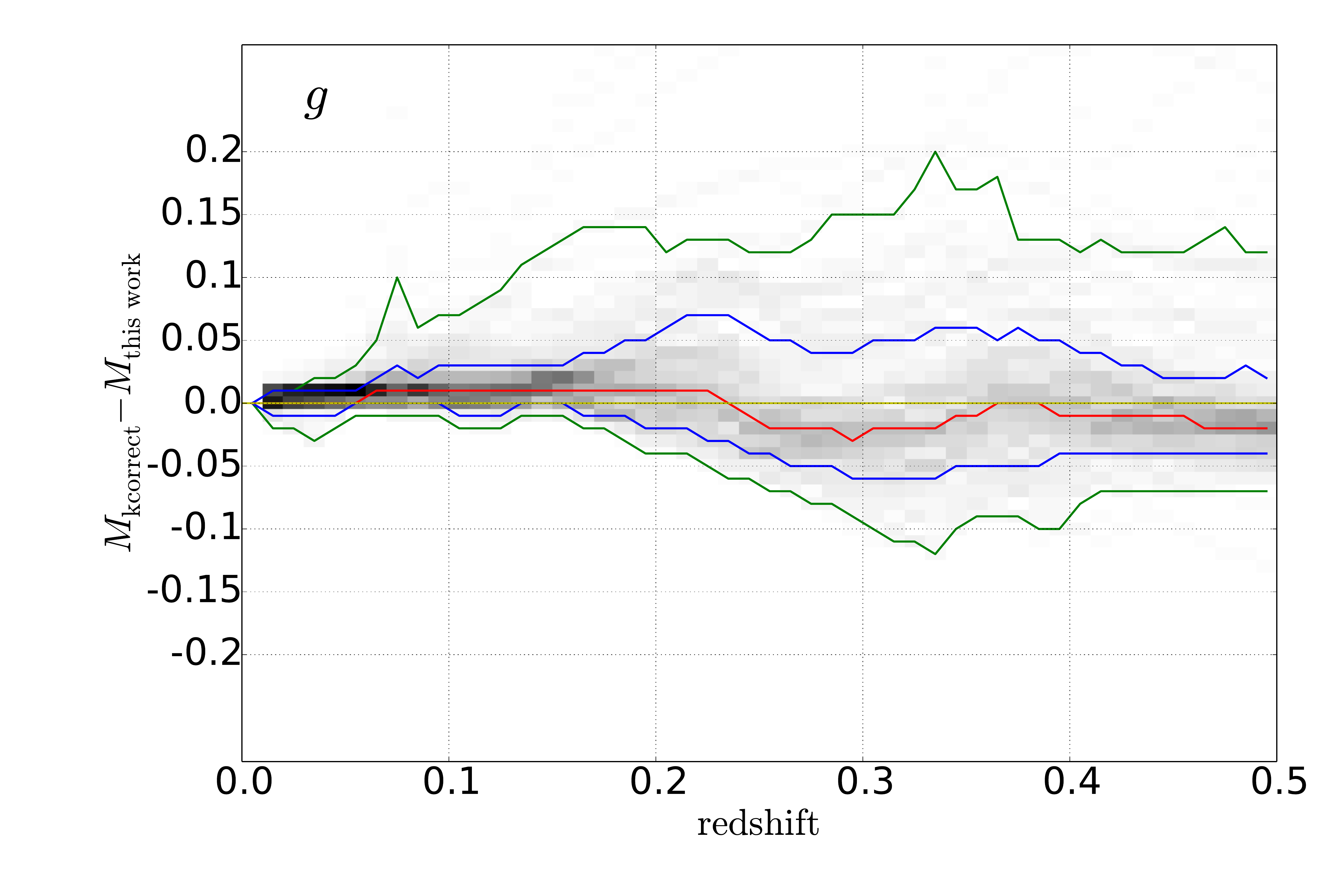}
		\caption{Binned plots comparing absolute magnitudes for synthetic photometry generated from the 129 \citet{brown14} templates; \textit{left:} calculated using our method, \textit{right:} calculated using the latest version of \textit{kcorrect} (v\_4\_2). $griz$ systematic offsets are less than 0.02. $u$-band absolute magnitudes are consistently $\sim 0.04$ fainter than ours, which we attribute to differences in the ultraviolet between our templates and the SPS templates used by \textit{kcorrect}.  The red line is the median difference and the green and blue lines are the 2.3, 15.9, 84.1 and 97.7 percentiles (corresponding to $1-\sigma$ and  $2-\sigma$ for a normal distribution).}
		\label {fig:ug_comparisons_templates}
\end{figure*}

In Figure \ref{fig:ug_comparisons_templates} we compare  $u$ and $g$-band absolute magnitudes calculated with our method and with \textit{kcorrect} v\_4\_2 for synthetic galaxies whose apparent magnitudes were derived from the 129 template SEDs of \citet{brown14}). For the rest frame $gri$ wavebands the systematic offset is less than $\sim$0.02,  while for $z$ it is less than $\sim$0.03. The $u$-band shows a relatively constant offset, with our method producing absolute magnitudes $\sim$0.04 mag brighter than those from \textit{kcorrect}. We expect this offset to be due to differences in the ultraviolet between our observational template SEDs and the SPS-based SEDs used by \textit{kcorrect}. We have used the latest version of \textit{kcorrect} for our comparisons (v\_4\_2) and note that the previous version (v\_4\_1, (as used for the published VAGC K-corrections and absolute magnitudes) results in a $u$-band offset which is twice as large ($\sim0.08$).

\begin{figure*}
 	\centering
		\includegraphics[width=0.49\textwidth]{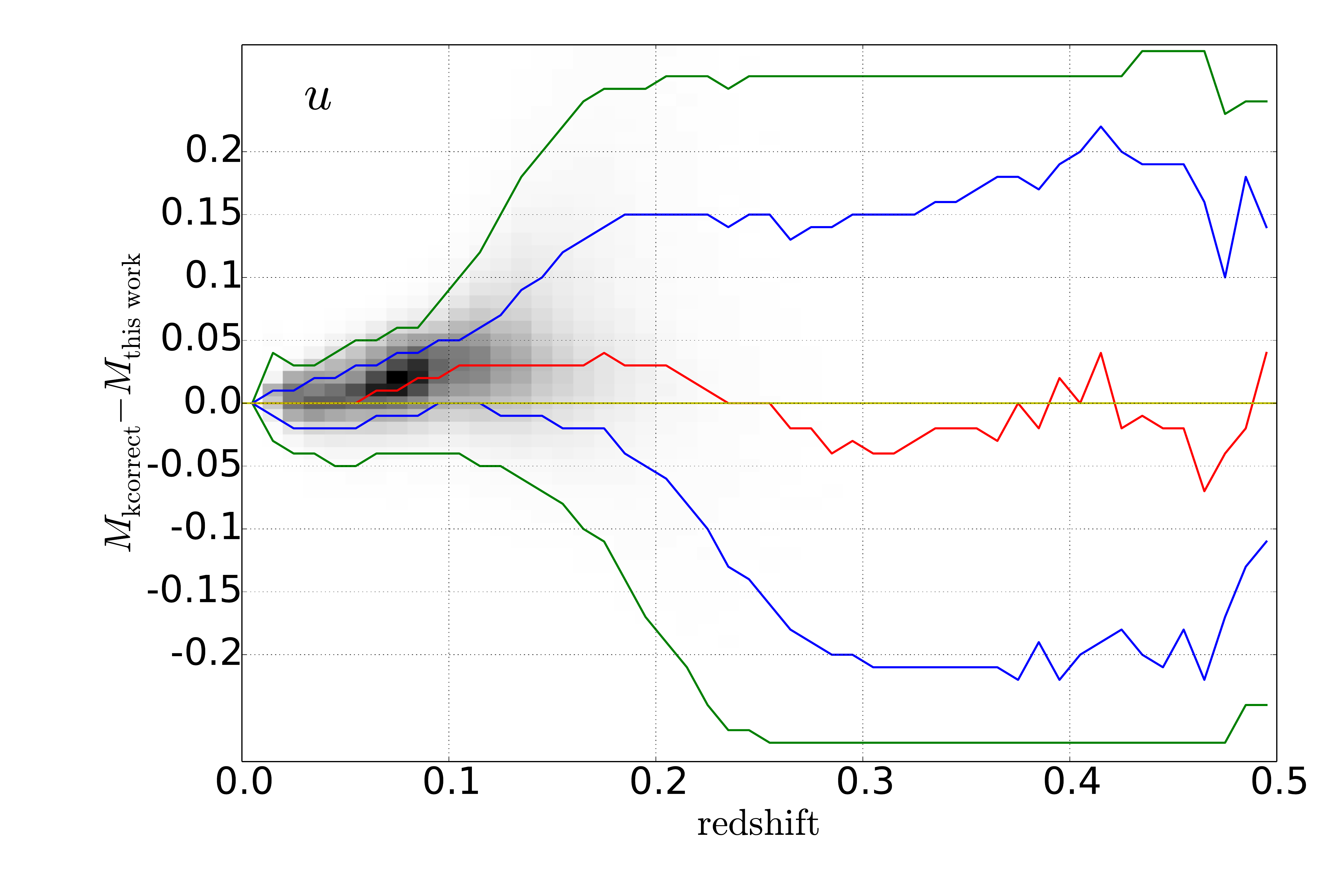}
		\includegraphics[width=0.49\textwidth]{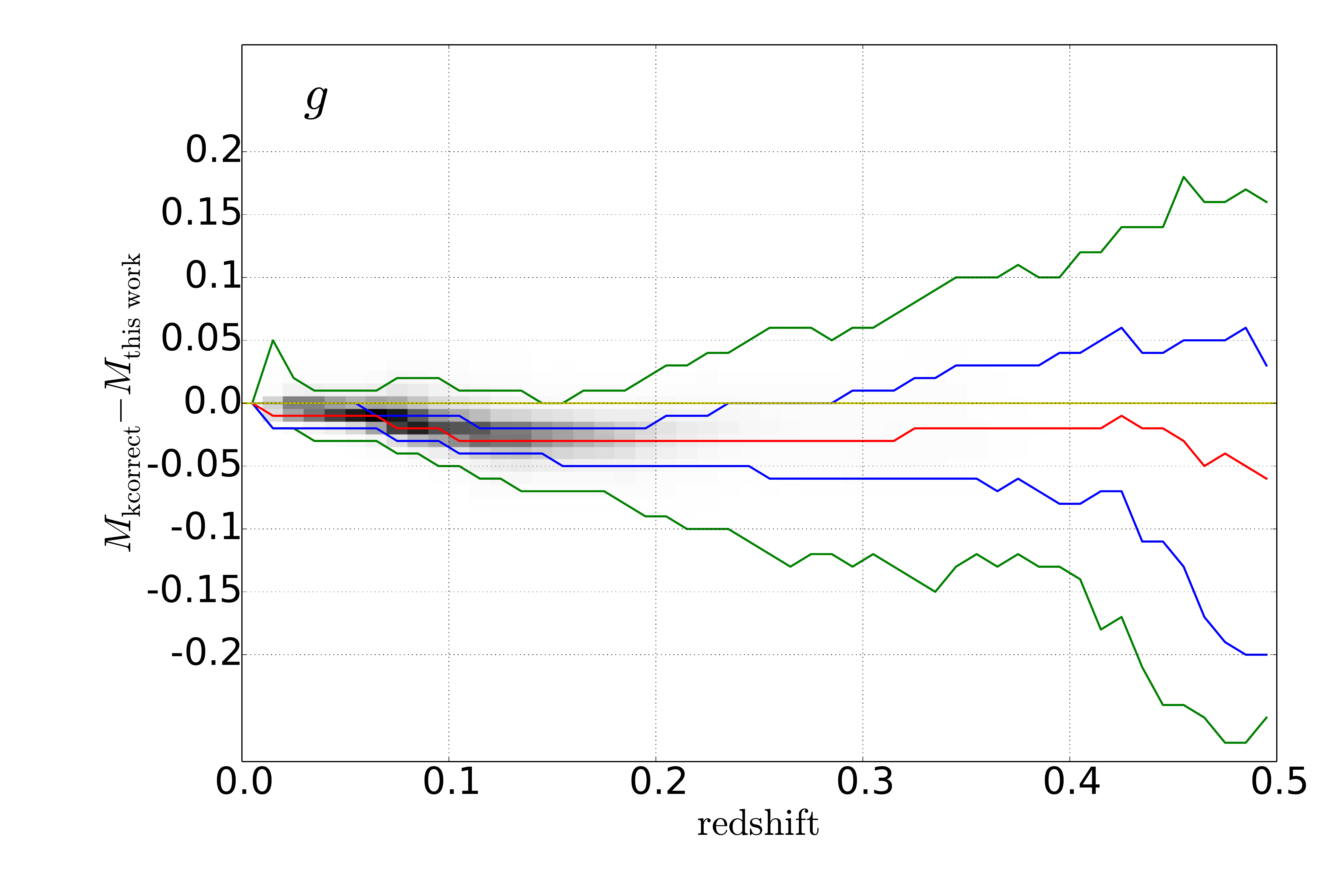}
		\caption{Binned plots comparing absolute magnitudes for the VAGC; \textit{left:} calculated using our method, \textit{right:} calculated using the latest version of \textit{kcorrect} (v\_4\_2). As for the synthetic galaxies in Figure \ref{fig:ug_comparisons_templates}, the $griz$ systematic offsets are less than 0.02. The size of the $u$-band systematic offset is $\sim 0.04$ but varies in sign (unlike that for the synthetic galaxies). This will partly reflect the distribution of galaxy types in the VAGC. In the left hand panel, a cloud of galaxies is visible at $0.12 < z < 0.2$  with $u$-band absolute magnitudes that are much brighter as calculated using \textit{kcorrect}. These galaxies are found to have very faint $u$-band apparent magnitudes (within 1 mag of the SDSS limiting magnitude of $u = 22.0$) .  The red, green and blue lines represent the same percentiles as in Figure \ref{fig:ug_comparisons_templates}.}
		\label {fig:ug_comparisons}
\end{figure*}

In Figure \ref{fig:ug_comparisons} we compare absolute magnitudes determined with our method and  \textit{kcorrect} for VAGC galaxies. The systematic offsets are similar in size to  those for the synthetic galaxies.  However, there are some significant differences for the $u$-band: the systematic offset varies in sign with redshift; the scatter increases markedly beyond $z \sim 0.015$, and there is a substantial population of galaxies that are up to $\sim$0.2 mag fainter in the $u$-band as determined using \textit{kcorrect}. We return again to this latter point below.

\begin{figure*}
	\centering
	\includegraphics[width=0.49\textwidth]{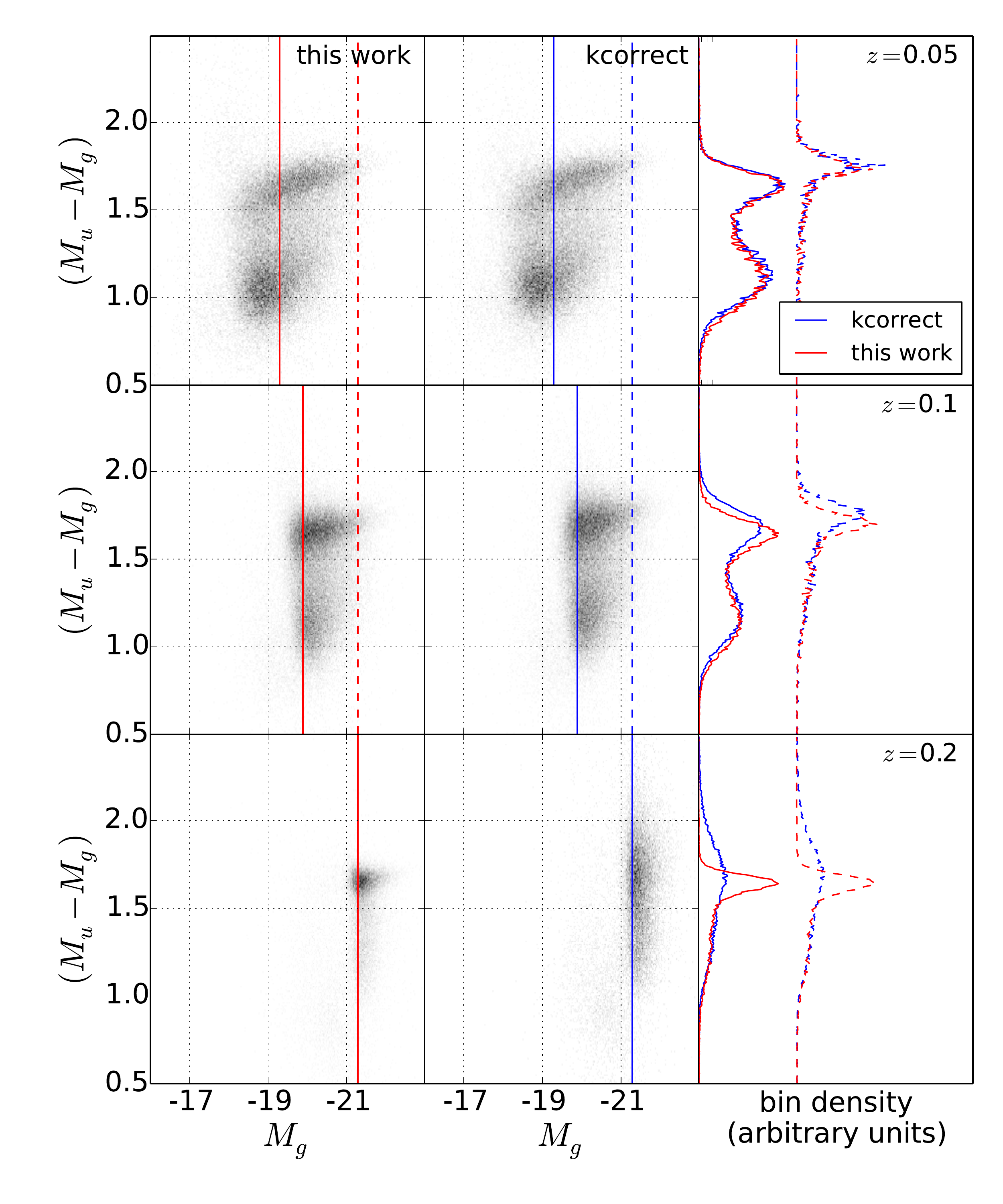}
	\includegraphics[width=0.49\textwidth]{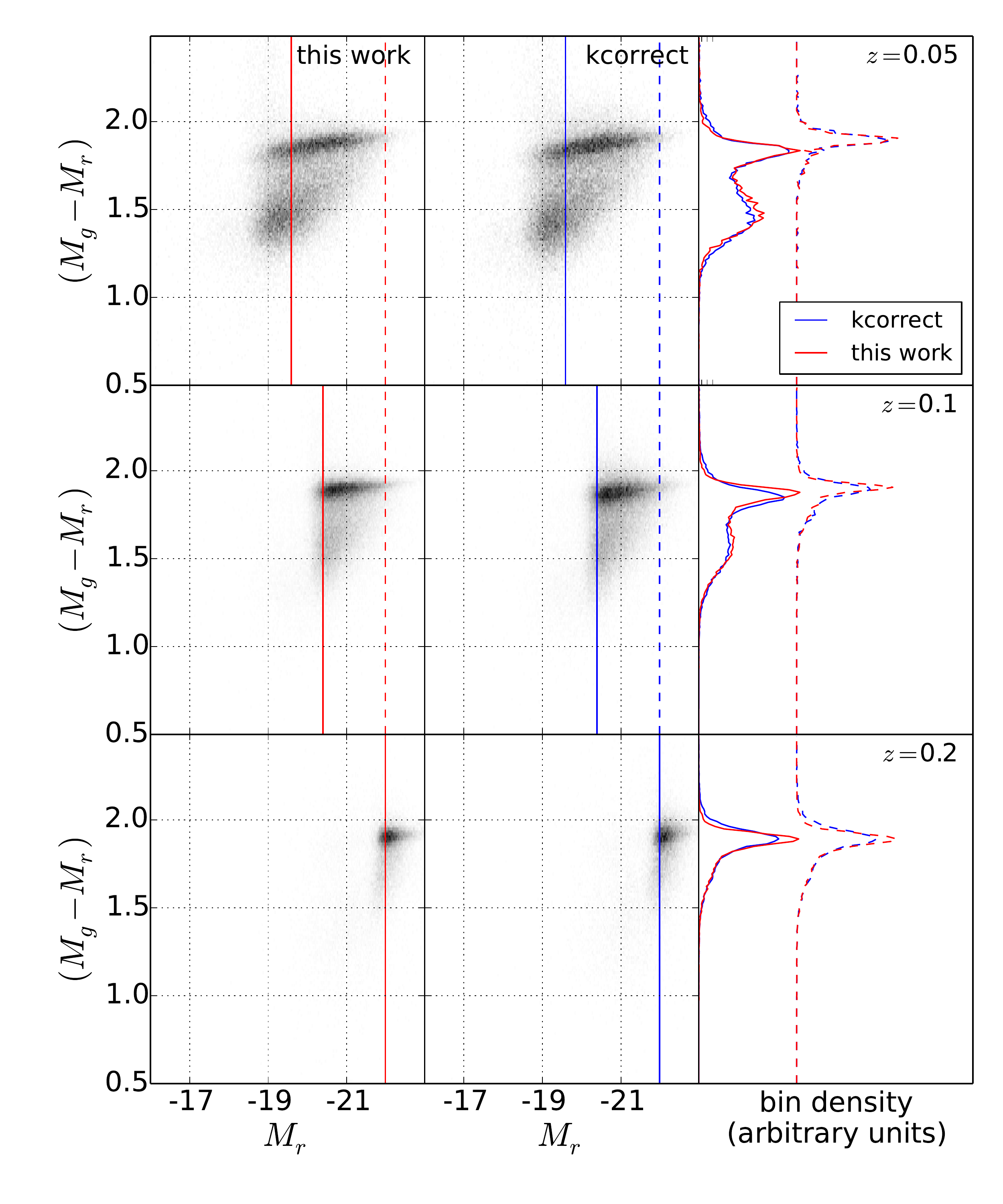}
	\caption{Comparing our restframe color-magnitude plots with those from \textit{kcorrect}  (v\_4\_2) for VAGC galaxies. \textit{left:} $(M_u - M_g)$ versus $M_g$; \textit{right:} $(M_g - M_r)$ versus $M_r$. We see a better defined red sequence, blue cloud and green valley, particularly at higher redshift and in $(M_u-M_g)$ color.  Unlike \textit{kcorrect} we do not see a substantial population of galaxies redder than $(M_u - M_g) \sim 1.8$ at $z  = 0.2$ which is not present in the low redshift Universe.  The dashed histograms correspond to fixed absolute magnitudes which are bright enough to be highly complete at all three redshifts, so ensuring a valid comparison. The right hand panels are histograms of bin density along the solid and dashed lines, the scales being arbitrarily adjusted to give similar maximum densities for our results.}
	\label{fig:CM_comparison}
\end{figure*}

As a further comparison we show color magnitude plots of $(M_u-M_g)$ against $M_g$ and $(M_g-M_r)$ against $M_r$  for our absolute magnitudes and those from \textit{kcorrect} at redshifts of 0.05, 0.1 and 0.2 in Figure \ref{fig:CM_comparison}.  The red sequence, blue cloud and green valley are significantly better defined using our results, particularly at higher redshift and in $(M_u-M_g)$ color, as can be seen from the histograms in the right hand panels. We note that the offset between our $M_u$ values and those from \textit{kcorrect} seen in Figure \ref{fig:ug_comparisons} is still evident at $z = 0.1$ where the majority of galaxies are still relatively bright. The dashed lines provide a direct comparison for the same absolute magnitude at each redshift.

\begin{figure*}
 	\centering
		\includegraphics[width=0.3\textwidth]{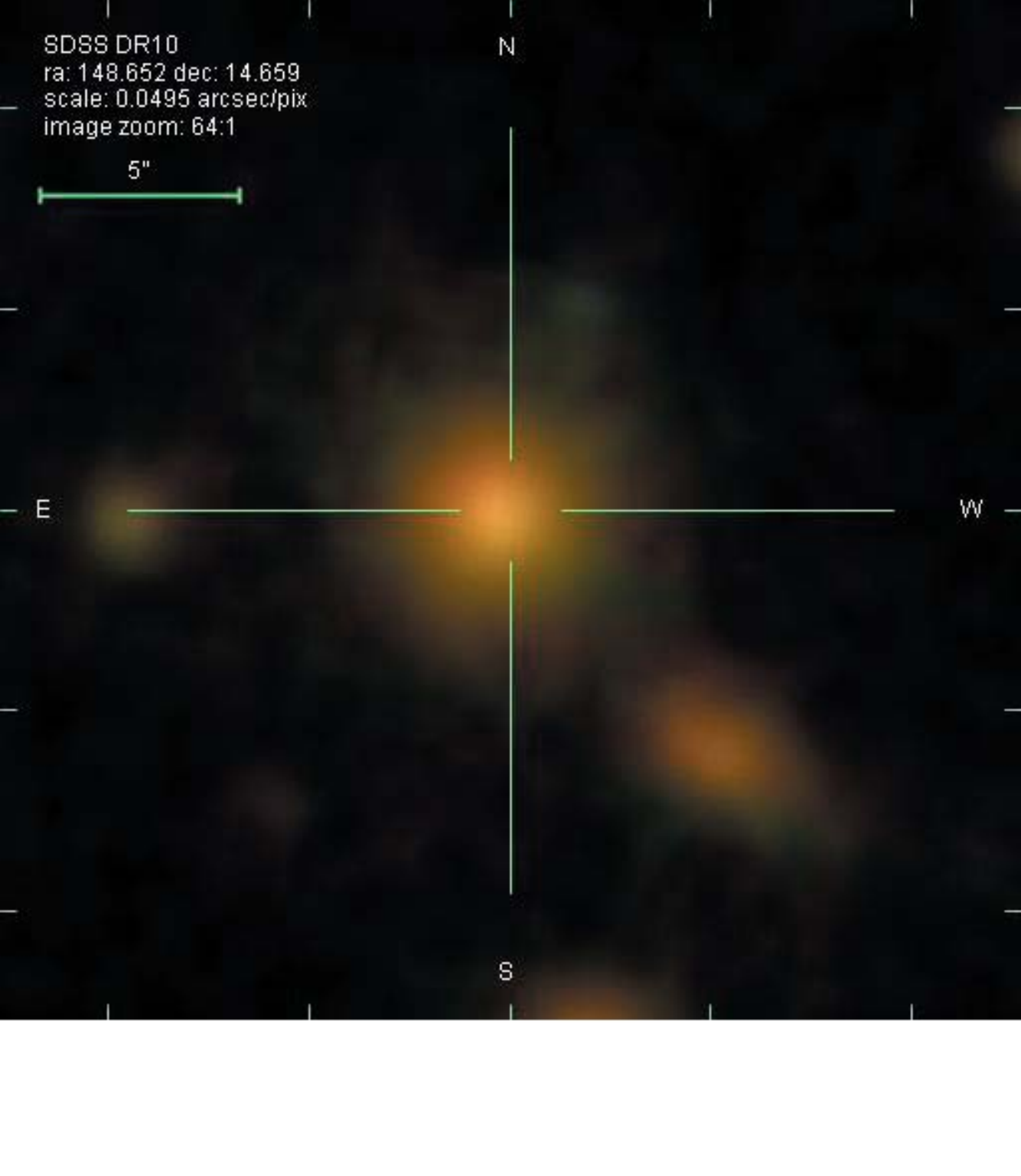}
		\includegraphics[width=0.6\textwidth]{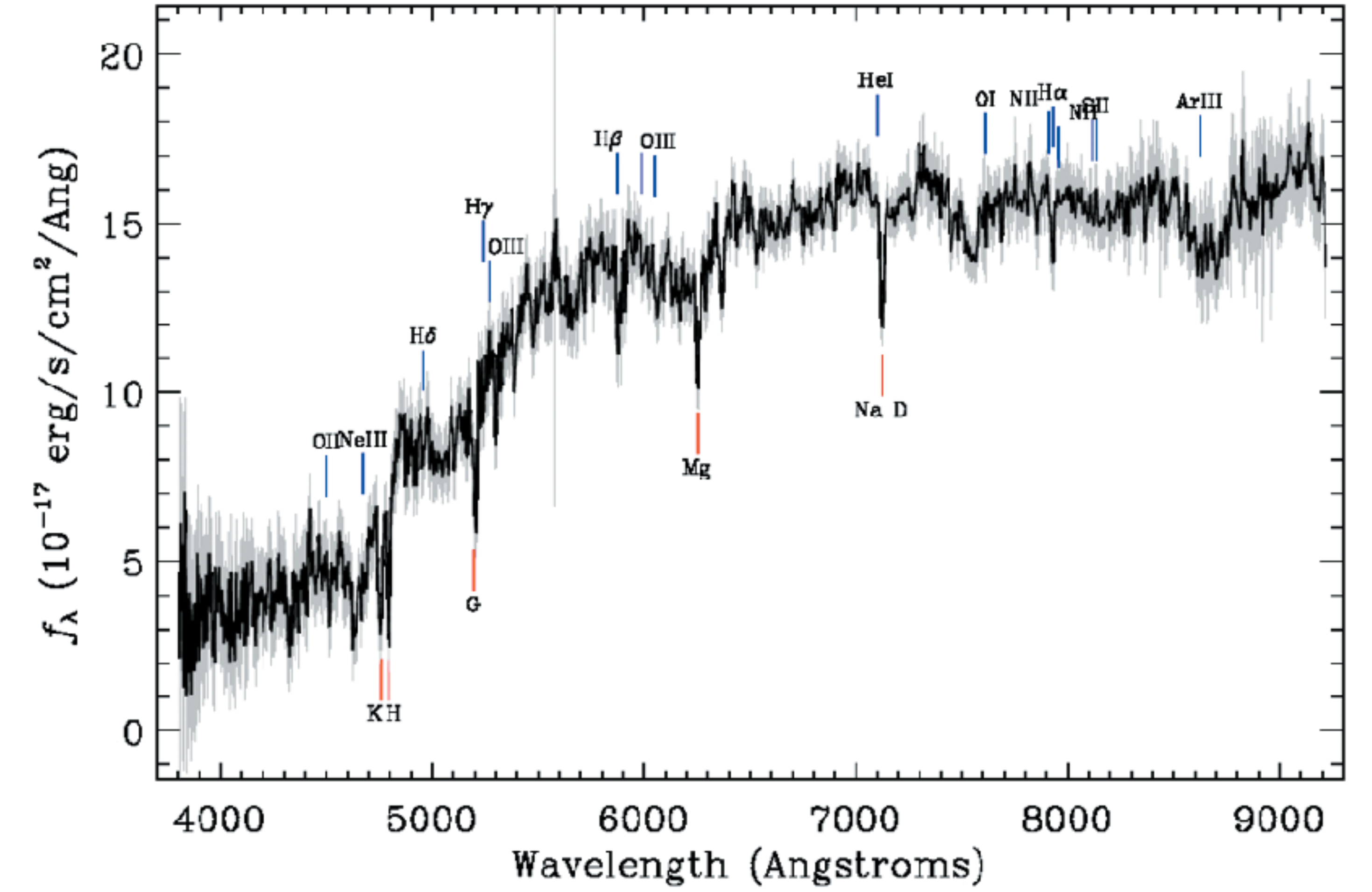}
		\\
		\includegraphics[width=0.3\textwidth]{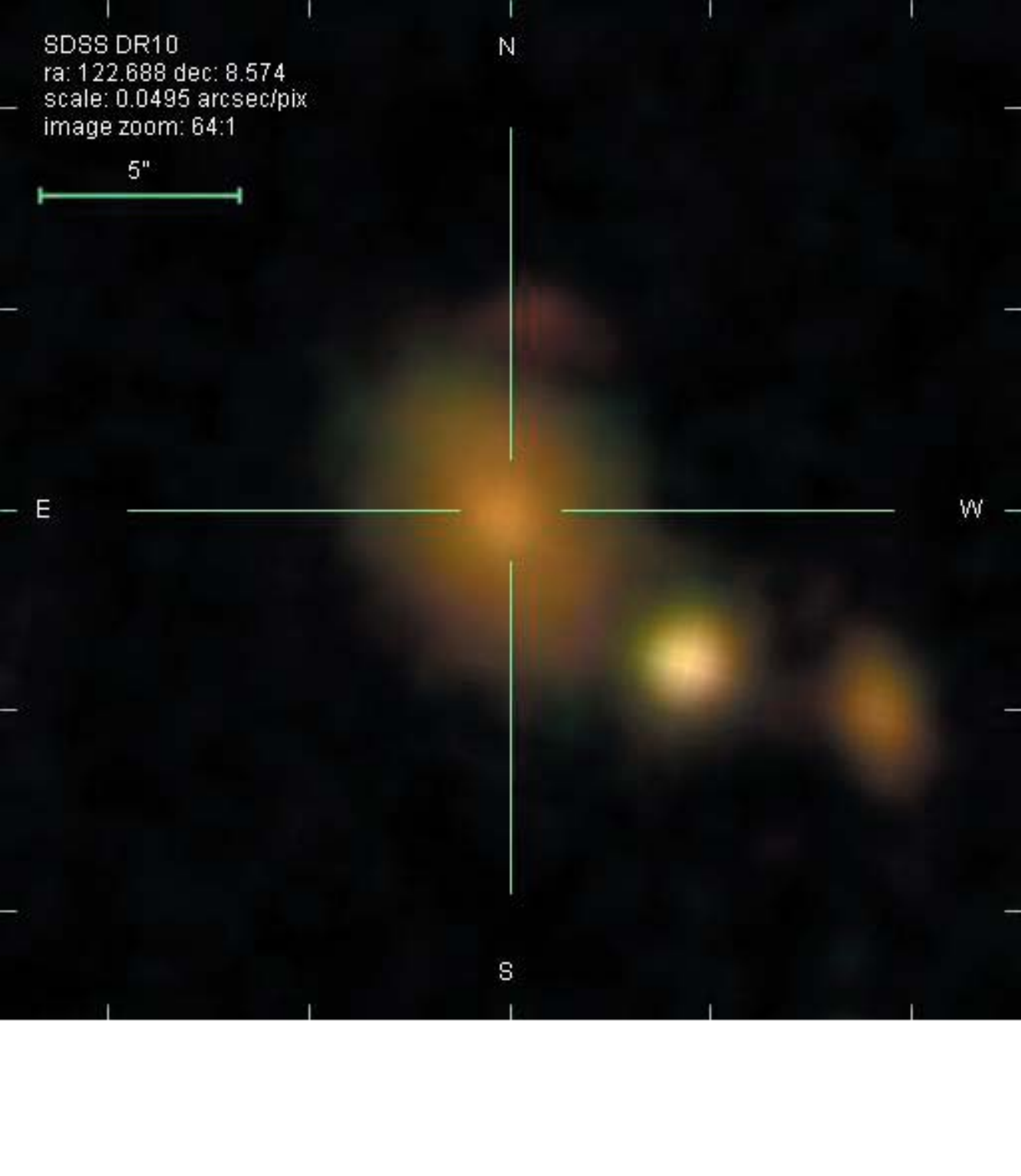}
		\includegraphics[width=0.6\textwidth]{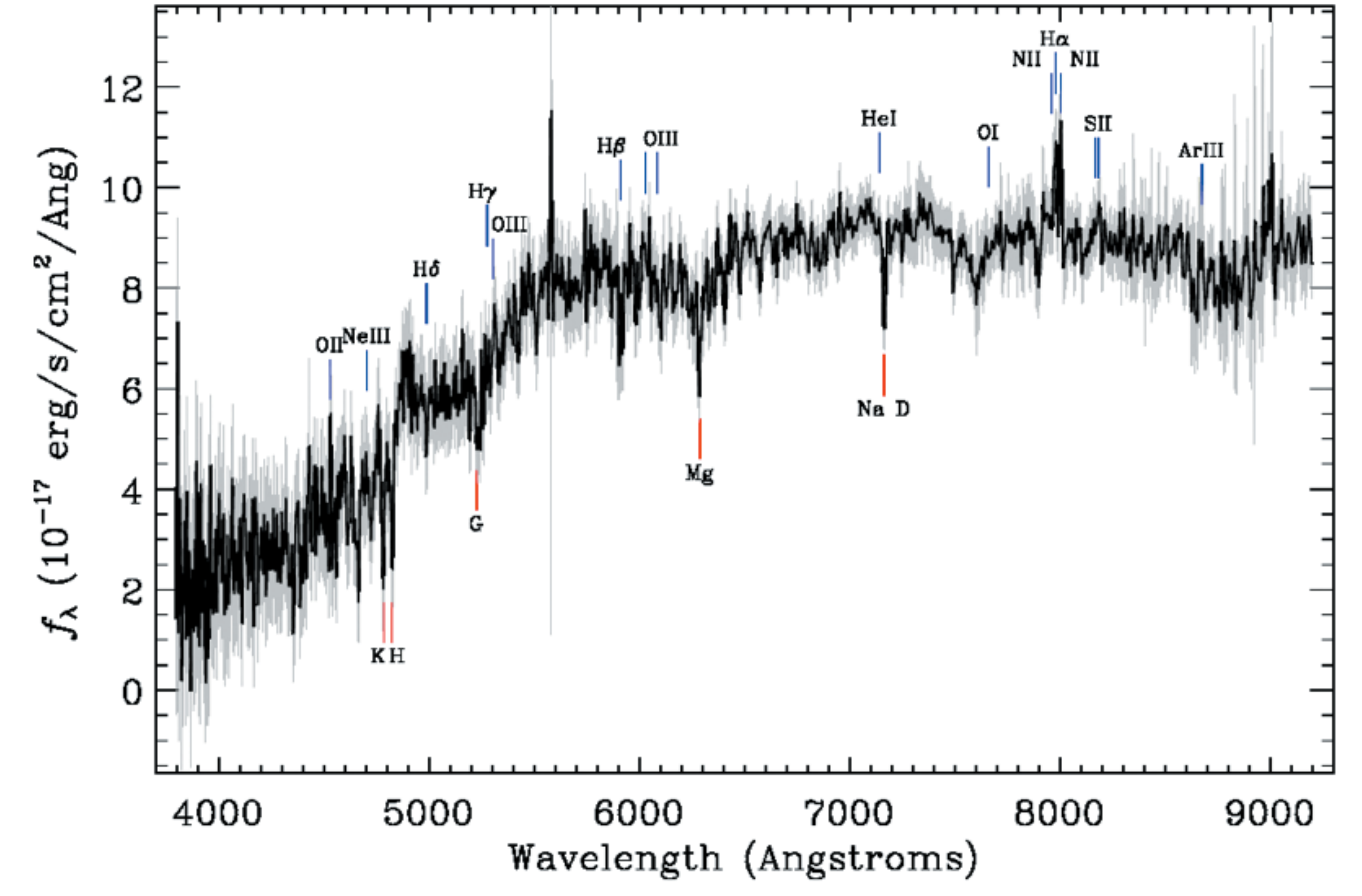}		
		\caption{Two galaxies with anomalously red $(M_u-M_g)$ color as computed by \textit{kcorrect}. \textit{Left panels:} $gri$ color images from SDSS; \textit{right panels:} spectra from SDSS.   The spectra of these objects are typical of normal elliptical galaxies but they have faint $u$-band magnitudes within \s1 mag  of the Sloan 95\% u-band limit of 22.0. (\textit{Top:} SDSS ID 1237671261742235846, \textit{bottom:} SDSS ID 1237671261730832842.)}
		\label{fig:veryred}
\end{figure*}

In addition the \textit{kcorrect}  results for $z  = 0.2$ show a substantial population of galaxies redder than $(M_u-M_g) = 2$. Such galaxies are not seen in the low redshift Universe and they are not present in our results. There are 1402 objects in the VAGC with redshifts between 0.18 and 0.22 that have extremely red \textit{kcorrect} colors $(M_u-M_g) > 2.0$, but normal red sequence colors $(M_u-M_g) < 1.7$ as determined by our method. We visually inspected images and spectra for these objects using the SDSS ImageList tool and found that they are predominantly faint objects whose apparent $u$-band magnitudes are within \s1 mag of the Sloan 95\% u-band limit of 22.0. Their $u$-band uncertainties are \s20 times larger than those for the other four filters, typically \s0.2 mag as opposed to \s0.01 mag. We found that the SDSS spectra of these objects are typical of normal elliptical galaxies (and not star-forming dust-obscured galaxies), and exhibit prominent absorption lines, especially Ca, K, Mg, and (usually) H$_{\alpha}$, with relatively insignificant emission lines. The difference between our $(M_u-M_g)$ colors and those from \textit{kcorrect} is largely due to the difference between the computed $M_u$ values; the $M_g$ values differ very little. We illustrate these points with the example galaxies shown in Figure \ref{fig:veryred}. Figure \ref{fig:CM_comparison} shows that our method handles $u$-band faint galaxies like these with large $u$ magnitude uncertainties more reliably than \textit{kcorrect}.

\section{SUMMARY}
\label{sec:summary}

We have developed an improved method of determining K-corrections whereby the relationship between the absolute magnitude and apparent magnitude is modelled using a second degree polynomial function of just one observed color. The polynomial parameters at any redshift are determined by fitting absolute and apparent magnitudes calculated using the new atlas of 129 template SEDs from \citet{brown14}.  These SEDs form the most extensive and accurate set of galaxy templates yet available, extending from the ultraviolet to the mid-infrared (although we do not make use of the broad wavelength coverage in this paper). They enable us to measure K-corrections with high accuracy, because for suitable observed colors the polynomial fits form a remarkably tight sequence. Uncertainties in computed absolute magnitudes are the sum in quadrate of the uncertainties deriving from inexact photometry, redshift uncertainty and the scatter of templates around the polynomial fits.

The optimum observed color is determined using the RMS deviation of the template SEDs around the best polynomial fit, backed up by visual inspection of the polynomial fits. This minimizes random error in computed absolute magnitudes due to galaxy diversity.  We find that the best pair of apparent magnitudes sometimes does not correspond to the two filters closest to the redshifted restframe filter for which we are seeking an absolute magnitude. We find that the random errors due to inexact photometry are in practice also minimised by choosing observed colors that mimimse the template scatter. We investigate outliers, chiefly faint starburst and compact blue galaxies, and show that these are only significant for these rare and faint galaxy types at certain redshifts for galaxies fainter than $g = 21.5$.

We list our preferred observed colors for determining absolute magnitudes in different redshift ranges between $z =0$ and $z  = 0.5$ for all five SDSS filters ($ugriz$), and we make publicly available the corresponding tables of polynomial coefficients together with the root mean square deviations for the polynomial fits. We also tabulate the coefficients for alternative polynomial fits that avoid the use of $u$ or $z$-band photometry when this is absent or of poor quality.

We have been able to quantify the performance of our method using synthetic galaxy photometry, the SDSS VAGC catalogue, and comparisons with kcorrect absolute magnitudes. All of these tests show that systematic differences between our method and \textit{kcorrect} are less than 0.05 magnitudes across the $ugriz$ filter set, and for some redshifts and filters the residuals are as little as 0.01 magnitudes. The evolution of rest-frame colors as a function of redshift is better behaved using our method than \textit{kcorrect}, with relatively few galaxies being assigned anomalously red colors and a tight red sequence being observed across the whole $0.0 <  z < 0.5$  redshift range.

We make our tables of polynomial parameters and RMS offsets in Tables \ref{tab:u_parameters} to \ref{tab:z_parameters} publicly available on-line in CSV format\footnote{http://dx.doi.org/10.4225/03/54498EC286B08}. These tables use the Sloan filter response curves calculated by J. Gunn which are available from the SDSS website\footnote{http://www.sdss3.org/instruments/camera.php (first two columns)}.

\section{ACKNOWLEDGEMENTS}
\label{sec:acknowledgements}

Richard Beare wishes to thank Monash University for financial support from MGS and MIPRS postgraduate research scholarships. Michael Brown acknowledges financial support from The Australian Research Council (FT100100280), the Monash Research Accelerator Program (MRA).

Funding for the SDSS has been provided by the Alfred P. Sloan Foundation, the Participating Institutions, the National Science Foundation, the U.S. Department of Energy, the National Aeronautics and Space Administration, the Japanese Monbukagakusho, the Max Planck Society, and the Higher Education Funding Council for England. The SDSS Web Site is http://www.sdss.org/. The SDSS is managed by the Astrophysical Research Consortium for the Participating Institutions. The Participating Institutions are the American Museum of Natural History, Astrophysical Institute Potsdam, University of Basel, University of Cambridge, Case Western Reserve University, University of Chicago, Drexel University, Fermilab, the Institute for Advanced Study, the Japan Participation Group, Johns Hopkins University, the Joint Institute for Nuclear Astrophysics, the Kavli Institute for Particle Astrophysics and Cosmology, the Korean Scientist Group, the Chinese Academy of Sciences (LAMOST), Los Alamos National Laboratory, the Max-Planck-Institute for Astronomy (MPIA), the Max-Planck-Institute for Astrophysics (MPA), New Mexico State University, Ohio State University, University of Pittsburgh, University of Portsmouth, Princeton University, the United States Naval Observatory, and the University of Washington.

\clearpage

\appendix

\section{Model parameters for calculating K-corrections for the Sloan filter set}
\label{app:tables} 

\begin{deluxetable}{ccccccccccccc}												
\tablewidth{0pt}												
\tablecolumns{13}												
\tabletypesize{\scriptsize}												
\tablecaption {Parameters for calculating absolute magnitudes $M_u = a(u-g)^2 + b(u-g) + c - D_M + g$ or $M_u = a(g-r)^2 + b(g-r) + c - D_M + r$}												
\tablehead{												
\colhead{red-} & 	\colhead{input} & 	\colhead{} & 	\colhead{parameters} & 	\colhead{} & 	\colhead{rms} & 	\colhead{} & 	\colhead{red-} & 	\colhead{input} & 	\colhead{} & 	\colhead{parameters} & 	\colhead{} & 	\colhead{rms} \\
\colhead{shift} & 	\colhead{colour} & 	\colhead{a} & 	\colhead{b} & 	\colhead{c} & 	\colhead{error} & 	\colhead{} & 	\colhead{shift} & 	\colhead{colour} & 	\colhead{a} & 	\colhead{b} & 	\colhead{c} & 	\colhead{error} \\
}												
 \startdata												
0.00 & 	$(u-g)$* & 	0.0000 & 	1.0000 & 	0.0000 & 	0.0000 & 	 & 	0.00 & 	$(g-r)$ & 	1.2208 & 	1.5209 & 	0.4054 & 	0.0881 \\
0.01 & 	$(u-g)$* & 	0.0245 & 	0.9330 & 	0.0087 & 	0.0084 & 	 & 	0.01 & 	$(g-r)$ & 	1.0792 & 	1.5130 & 	0.4331 & 	0.0876 \\
0.02 & 	$(u-g)$* & 	0.0466 & 	0.8674 & 	0.0199 & 	0.0128 & 	 & 	0.02 & 	$(g-r)$ & 	1.0556 & 	1.4480 & 	0.4690 & 	0.0875 \\
0.03 & 	$(u-g)$* & 	0.0598 & 	0.8196 & 	0.0254 & 	0.0186 & 	 & 	0.03 & 	$(g-r)$ & 	1.0499 & 	1.3174 & 	0.5436 & 	0.0875 \\
0.04 & 	$(u-g)$* & 	0.0646 & 	0.7872 & 	0.0273 & 	0.0242 & 	 & 	0.04 & 	$(g-r)$ & 	0.9470 & 	1.3494 & 	0.5521 & 	0.0894 \\
0.05 & 	$(u-g)$* & 	0.0595 & 	0.7778 & 	0.0189 & 	0.0297 & 	 & 	0.05 & 	$(g-r)$ & 	0.8474 & 	1.3776 & 	0.5491 & 	0.0896 \\
0.06 & 	$(u-g)$* & 	0.0430 & 	0.7895 & 	0.0072 & 	0.0338 & 	 & 	0.06 & 	$(g-r)$ & 	0.5919 & 	1.5813 & 	0.4928 & 	0.0845 \\
0.07 & 	$(u-g)$* & 	0.0235 & 	0.8038 & 	0.0012 & 	0.0363 & 	 & 	0.07 & 	$(g-r)$ & 	0.4257 & 	1.6994 & 	0.4496 & 	0.0816 \\
0.08 & 	$(u-g)$* & 	0.0024 & 	0.8216 & 	-0.0054 & 	0.0384 & 	 & 	0.08 & 	$(g-r)$ & 	0.2828 & 	1.8176 & 	0.3976 & 	0.0766 \\
0.09 & 	$(u-g)$* & 	-0.0157 & 	0.8308 & 	-0.0058 & 	0.0399 & 	 & 	0.09 & 	$(g-r)$ & 	0.1457 & 	1.9509 & 	0.3305 & 	0.0736 \\
0.10 & 	$(u-g)$* & 	-0.0299 & 	0.8290 & 	0.0007 & 	0.0409 & 	 & 	0.10 & 	$(g-r)$ & 	-0.0010 & 	2.0922 & 	0.2617 & 	0.0748 \\
0.11 & 	$(u-g)$* & 	-0.0431 & 	0.8228 & 	0.0096 & 	0.0416 & 	 & 	0.11 & 	$(g-r)$ & 	-0.0992 & 	2.1801 & 	0.2071 & 	0.0759 \\
0.12 & 	$(u-g)$* & 	-0.0553 & 	0.8126 & 	0.0210 & 	0.0420 & 	 & 	0.12 & 	$(g-r)$ & 	-0.1984 & 	2.2869 & 	0.1379 & 	0.0727 \\
0.13 & 	$(u-g)$* & 	-0.0660 & 	0.7975 & 	0.0345 & 	0.0423 & 	 & 	0.13 & 	$(g-r)$ & 	-0.2256 & 	2.2776 & 	0.1138 & 	0.0669 \\
0.14 & 	$(u-g)$* & 	-0.0767 & 	0.7813 & 	0.0480 & 	0.0423 & 	 & 	0.14 & 	$(g-r)$ & 	-0.2663 & 	2.2918 & 	0.0804 & 	0.0655 \\
0.15 & 	$(u-g)$* & 	-0.0866 & 	0.7625 & 	0.0624 & 	0.0422 & 	 & 	0.15 & 	$(g-r)$ & 	-0.2942 & 	2.2782 & 	0.0637 & 	0.0629 \\
0.16 & 	$(u-g)$* & 	-0.0967 & 	0.7436 & 	0.0761 & 	0.0421 & 	 & 	0.16 & 	$(g-r)$ & 	-0.2839 & 	2.2087 & 	0.0688 & 	0.0664 \\
0.17 & 	$(u-g)$* & 	-0.1050 & 	0.7208 & 	0.0910 & 	0.0417 & 	 & 	0.17 & 	$(g-r)$ & 	-0.2712 & 	2.1380 & 	0.0701 & 	0.0646 \\
0.18 & 	$(u-g)$* & 	-0.1128 & 	0.6971 & 	0.1056 & 	0.0413 & 	 & 	0.18 & 	$(g-r)$ & 	-0.2750 & 	2.0905 & 	0.0687 & 	0.0604 \\
0.19 & 	$(u-g)$* & 	-0.1185 & 	0.6702 & 	0.1206 & 	0.0407 & 	 & 	0.19 & 	$(g-r)$ & 	-0.2677 & 	2.0327 & 	0.0684 & 	0.0586 \\
0.20 & 	$(u-g)$* & 	-0.1220 & 	0.6401 & 	0.1358 & 	0.0402 & 	 & 	0.20 & 	$(g-r)$ & 	-0.2772 & 	2.0095 & 	0.0549 & 	0.0573 \\
0.21 & 	$(u-g)$* & 	-0.1237 & 	0.6084 & 	0.1505 & 	0.0397 & 	 & 	0.21 & 	$(g-r)$ & 	-0.2691 & 	1.9521 & 	0.0612 & 	0.0538 \\
0.22 & 	$(u-g)$* & 	-0.1240 & 	0.5765 & 	0.1633 & 	0.0391 & 	 & 	0.22 & 	$(g-r)$ & 	-0.2624 & 	1.9028 & 	0.0636 & 	0.0507 \\
0.23 & 	$(u-g)$* & 	-0.1222 & 	0.5422 & 	0.1761 & 	0.0384 & 	 & 	0.23 & 	$(g-r)$ & 	-0.2538 & 	1.8534 & 	0.0677 & 	0.0478 \\
0.24 & 	$(u-g)$* & 	-0.1184 & 	0.5052 & 	0.1888 & 	0.0376 & 	 & 	0.24 & 	$(g-r)$ & 	-0.2424 & 	1.8008 & 	0.0734 & 	0.0449 \\
0.25 & 	$(u-g)$* & 	-0.1118 & 	0.4629 & 	0.2026 & 	0.0364 & 	 & 	0.25 & 	$(g-r)$ & 	-0.2254 & 	1.7387 & 	0.0835 & 	0.0414 \\
0.26 & 	$(u-g)$* & 	-0.1027 & 	0.4156 & 	0.2177 & 	0.0351 & 	 & 	0.26 & 	$(g-r)$ & 	-0.1959 & 	1.6530 & 	0.1024 & 	0.0412 \\
0.27 & 	$(u-g)$* & 	-0.0922 & 	0.3646 & 	0.2334 & 	0.0332 & 	 & 	0.27 & 	$(g-r)$ & 	-0.1675 & 	1.5670 & 	0.1227 & 	0.0399 \\
0.28 & 	$(u-g)$* & 	-0.0815 & 	0.3139 & 	0.2476 & 	0.0314 & 	 & 	0.28 & 	$(g-r)$ & 	-0.1491 & 	1.4967 & 	0.1408 & 	0.0355 \\
0.29 & 	$(u-g)$* & 	-0.0717 & 	0.2658 & 	0.2598 & 	0.0295 & 	 & 	0.29 & 	$(g-r)$ & 	-0.1354 & 	1.4345 & 	0.1572 & 	0.0317 \\
0.30 & 	$(u-g)$* & 	-0.0616 & 	0.2170 & 	0.2712 & 	0.0272 & 	 & 	0.30 & 	$(g-r)$ & 	-0.1235 & 	1.3749 & 	0.1736 & 	0.0279 \\
0.31 & 	$(u-g)$* & 	-0.0517 & 	0.1694 & 	0.2806 & 	0.0251 & 	 & 	0.31 & 	$(g-r)$ & 	-0.1127 & 	1.3172 & 	0.1902 & 	0.0249 \\
0.32 & 	$(u-g)$* & 	-0.0407 & 	0.1190 & 	0.2905 & 	0.0224 & 	 & 	0.32 & 	$(g-r)$ & 	-0.1001 & 	1.2546 & 	0.2112 & 	0.0219 \\
0.33 & 	$(u-g)$* & 	-0.0294 & 	0.0703 & 	0.2982 & 	0.0194 & 	 & 	0.33 & 	$(g-r)$ & 	-0.0870 & 	1.1936 & 	0.2320 & 	0.0192 \\
0.34 & 	$(u-g)$* & 	-0.0175 & 	0.0222 & 	0.3062 & 	0.0167 & 	 & 	0.34 & 	$(g-r)$ & 	-0.0722 & 	1.1305 & 	0.2564 & 	0.0175 \\
0.35 & 	$(u-g)$* & 	-0.0056 & 	-0.0226 & 	0.3121 & 	0.0136 & 	 & 	0.35 & 	$(g-r)$ & 	-0.0564 & 	1.0698 & 	0.2789 & 	0.0161 \\
0.36 & 	$(u-g)$* & 	0.0061 & 	-0.0643 & 	0.3157 & 	0.0108 & 	 & 	0.36 & 	$(g-r)$ & 	-0.0415 & 	1.0166 & 	0.2955 & 	0.0153 \\
0.37 & 	$(u-g)$* & 	0.0166 & 	-0.1013 & 	0.3165 & 	0.0088 & 	 & 	0.37 & 	$(g-r)$ & 	-0.0321 & 	0.9788 & 	0.3032 & 	0.0152 \\
0.38 & 	$(u-g)$* & 	0.0277 & 	-0.1394 & 	0.3174 & 	0.0088 & 	 & 	0.38 & 	$(g-r)$ & 	-0.0240 & 	0.9450 & 	0.3090 & 	0.0163 \\
0.39 & 	$(u-g)$* & 	0.0384 & 	-0.1764 & 	0.3176 & 	0.0105 & 	 & 	0.39 & 	$(g-r)$ & 	-0.0187 & 	0.9163 & 	0.3133 & 	0.0182 \\
0.40 & 	$(u-g)$* & 	0.0483 & 	-0.2122 & 	0.3179 & 	0.0127 & 	 & 	0.40 & 	$(g-r)$ & 	-0.0163 & 	0.8928 & 	0.3170 & 	0.0204 \\
0.41 & 	$(u-g)$* & 	0.0564 & 	-0.2417 & 	0.3128 & 	0.0156 & 	 & 	0.41 & 	$(g-r)$ & 	-0.0169 & 	0.8766 & 	0.3156 & 	0.0233 \\
0.42 & 	$(u-g)$* & 	0.0618 & 	-0.2629 & 	0.3008 & 	0.0193 & 	 & 	0.42 & 	$(g-r)$ & 	-0.0201 & 	0.8666 & 	0.3098 & 	0.0267 \\
0.43 & 	$(u-g)$* & 	0.0651 & 	-0.2786 & 	0.2858 & 	0.0235 & 	 & 	0.43 & 	$(g-r)$ & 	-0.0261 & 	0.8614 & 	0.3026 & 	0.0304 \\
0.44 & 	$(u-g)$* & 	0.0665 & 	-0.2905 & 	0.2703 & 	0.0280 & 	 & 	0.44 & 	$(g-r)$ & 	-0.0350 & 	0.8607 & 	0.2951 & 	0.0338 \\
0.45 & 	$(u-g)$* & 	0.0677 & 	-0.3039 & 	0.2590 & 	0.0319 & 	 & 	0.45 & 	$(g-r)$ & 	-0.0465 & 	0.8619 & 	0.2908 & 	0.0366 \\
0.46 & 	$(u-g)$* & 	0.0702 & 	-0.3238 & 	0.2557 & 	0.0347 & 	 & 	0.46 & 	$(g-r)$ & 	-0.0597 & 	0.8619 & 	0.2927 & 	0.0386 \\
0.47 & 	$(u-g)$* & 	0.0743 & 	-0.3499 & 	0.2591 & 	0.0365 & 	 & 	0.47 & 	$(g-r)$ & 	-0.0732 & 	0.8591 & 	0.3001 & 	0.0396 \\
0.48 & 	$(u-g)$* & 	0.0791 & 	-0.3780 & 	0.2645 & 	0.0378 & 	 & 	0.48 & 	$(g-r)$ & 	-0.0860 & 	0.8531 & 	0.3106 & 	0.0401 \\
0.49 & 	$(u-g)$* & 	0.0848 & 	-0.4084 & 	0.2708 & 	0.0389 & 	 & 	0.49 & 	$(g-r)$ & 	-0.0971 & 	0.8423 & 	0.3240 & 	0.0402 \\
0.50 & 	$(u-g)$* & 	0.0913 & 	-0.4409 & 	0.2776 & 	0.0400 & 	 & 	0.50 & 	$(g-r)$ & 	-0.1072 & 	0.8281 & 	0.3392 & 	0.0401 \\
\enddata												
\label{tab:u_parameters}												
\end{deluxetable}																														

\clearpage

\begin{deluxetable}{ccccccccccccc}												
\tablewidth{0pt}												
\tablecolumns{13}												
\tabletypesize{\tiny}												
\tablecaption {Parameters for calculating absolute magnitudes $M_g = a(g-r)^2 + b(g-r) + c - D_M + r$ or $M_g = a(r-i)^2 + b(r-i) + c - D_M + i$}												
\tablehead{												
\colhead{red-} & 	\colhead{input} & 	\colhead{} & 	\colhead{parameters} & 	\colhead{} & 	\colhead{rms} & 	\colhead{} & 	\colhead{red-} & 	\colhead{input} & 	\colhead{} & 	\colhead{parameters} & 	\colhead{} & 	\colhead{rms} \\
\colhead{shift} & 	\colhead{colour} & 	\colhead{a} & 	\colhead{b} & 	\colhead{c} & 	\colhead{error} & 	\colhead{} & 	\colhead{shift} & 	\colhead{colour} & 	\colhead{a} & 	\colhead{b} & 	\colhead{c} & 	\colhead{error} \\
}												
 \startdata												
0.00 & 	$(g-r)$* & 	0.0000 & 	1.0000 & 	0.0000 & 	0.0000 & 	 & 	0.00 & 	$(r-i)$ & 	0.7930 & 	2.0554 & 	0.2547 & 	0.0662 \\
0.01 & 	$(g-r)$* & 	-0.0161 & 	0.9609 & 	0.0146 & 	0.0032 & 	 & 	0.01 & 	$(r-i)$ & 	0.8446 & 	2.0698 & 	0.2323 & 	0.0660 \\
0.02 & 	$(g-r)$* & 	-0.0334 & 	0.9318 & 	0.0243 & 	0.0054 & 	 & 	0.02 & 	$(r-i)$ & 	0.9119 & 	2.1994 & 	0.1560 & 	0.0633 \\
0.03 & 	$(g-r)$* & 	-0.0505 & 	0.9121 & 	0.0279 & 	0.0064 & 	 & 	0.03 & 	$(r-i)$ & 	-1.0026 & 	3.5191 & 	-0.1028 & 	0.0583 \\
0.04 & 	$(g-r)$* & 	-0.0642 & 	0.8932 & 	0.0292 & 	0.0073 & 	 & 	0.04 & 	$(r-i)$ & 	-3.5452 & 	5.5473 & 	-0.5453 & 	0.0733 \\
0.05 & 	$(g-r)$* & 	-0.0930 & 	0.8968 & 	0.0227 & 	0.0093 & 	 & 	0.05 & 	$(r-i)$ & 	-3.9089 & 	6.0560 & 	-0.7350 & 	0.0797 \\
0.06 & 	$(g-r)$* & 	-0.1483 & 	0.9413 & 	0.0009 & 	0.0152 & 	 & 	0.06 & 	$(r-i)$ & 	-3.6353 & 	5.7361 & 	-0.7001 & 	0.0858 \\
0.07 & 	$(g-r)$* & 	-0.2178 & 	1.0152 & 	-0.0347 & 	0.0254 & 	 & 	0.07 & 	$(r-i)$ & 	-11.1145 & 	10.9029 & 	-1.5882 & 	0.0877 \\
0.08 & 	$(g-r)$* & 	-0.2863 & 	1.0950 & 	-0.0753 & 	0.0371 & 	 & 	0.08 & 	$(r-i)$ & 	-11.4425 & 	10.9730 & 	-1.5798 & 	0.0915 \\
0.09 & 	$(g-r)$* & 	-0.2469 & 	1.0078 & 	-0.0544 & 	0.0263 & 	 & 	0.09 & 	$(r-i)$ & 	-9.6241 & 	9.6094 & 	-1.3375 & 	0.0896 \\
0.10 & 	$(g-r)$* & 	-0.2673 & 	1.0099 & 	-0.0641 & 	0.0277 & 	 & 	0.10 & 	$(r-i)$ & 	-4.0420 & 	5.6986 & 	-0.6840 & 	0.0818 \\
0.11 & 	$(g-r)$* & 	-0.2686 & 	0.9839 & 	-0.0638 & 	0.0275 & 	 & 	0.11 & 	$(r-i)$ & 	-3.3843 & 	5.3143 & 	-0.6443 & 	0.0811 \\
0.12 & 	$(g-r)$* & 	-0.2478 & 	0.9220 & 	-0.0493 & 	0.0260 & 	 & 	0.12 & 	$(r-i)$ & 	-1.4090 & 	3.6091 & 	-0.2981 & 	0.0746 \\
0.13 & 	$(g-r)$* & 	-0.2101 & 	0.8327 & 	-0.0229 & 	0.0232 & 	 & 	0.13 & 	$(r-i)$ & 	-0.0820 & 	2.3499 & 	-0.0216 & 	0.0779 \\
0.14 & 	$(g-r)$* & 	-0.1466 & 	0.7121 & 	0.0087 & 	0.0309 & 	 & 	0.14 & 	$(r-i)$ & 	0.6666 & 	1.6507 & 	0.1224 & 	0.0686 \\
0.15 & 	$(g-r)$* & 	-0.0999 & 	0.6052 & 	0.0436 & 	0.0308 & 	 & 	0.15 & 	$(r-i)$ & 	0.8826 & 	1.3627 & 	0.1872 & 	0.0635 \\
0.16 & 	$(g-r)$* & 	-0.0810 & 	0.5368 & 	0.0697 & 	0.0259 & 	 & 	0.16 & 	$(r-i)$ & 	0.9476 & 	1.2571 & 	0.2216 & 	0.0468 \\
0.17 & 	$(g-r)$* & 	-0.0662 & 	0.4767 & 	0.0928 & 	0.0221 & 	 & 	0.17 & 	$(r-i)$ & 	0.9809 & 	1.1414 & 	0.2353 & 	0.0513 \\
0.18 & 	$(g-r)$* & 	-0.0553 & 	0.4237 & 	0.1142 & 	0.0188 & 	 & 	0.18 & 	$(r-i)$ & 	0.9257 & 	1.0943 & 	0.2511 & 	0.0466 \\
0.19 & 	$(g-r)$* & 	-0.0450 & 	0.3743 & 	0.1345 & 	0.0159 & 	 & 	0.19 & 	$(r-i)$ & 	0.8771 & 	1.0544 & 	0.2638 & 	0.0418 \\
0.20 & 	$(g-r)$* & 	-0.0346 & 	0.3263 & 	0.1551 & 	0.0135 & 	 & 	0.20 & 	$(r-i)$ & 	0.8135 & 	1.0250 & 	0.2752 & 	0.0376 \\
0.21 & 	$(g-r)$* & 	-0.0253 & 	0.2819 & 	0.1749 & 	0.0115 & 	 & 	0.21 & 	$(r-i)$ & 	0.7460 & 	1.0027 & 	0.2853 & 	0.0339 \\
0.22 & 	$(g-r)$* & 	-0.0165 & 	0.2402 & 	0.1936 & 	0.0101 & 	 & 	0.22 & 	$(r-i)$ & 	0.6771 & 	0.9849 & 	0.2940 & 	0.0309 \\
0.23 & 	$(g-r)$* & 	-0.0073 & 	0.1981 & 	0.2134 & 	0.0093 & 	 & 	0.23 & 	$(r-i)$ & 	0.6100 & 	0.9748 & 	0.3015 & 	0.0275 \\
0.24 & 	$(g-r)$* & 	0.0008 & 	0.1578 & 	0.2334 & 	0.0088 & 	 & 	0.24 & 	$(r-i)$ & 	0.5224 & 	0.9802 & 	0.3072 & 	0.0239 \\
0.25 & 	$(g-r)$* & 	0.0103 & 	0.1139 & 	0.2559 & 	0.0096 & 	 & 	0.25 & 	$(r-i)$ & 	0.4020 & 	1.0077 & 	0.3107 & 	0.0203 \\
0.26 & 	$(g-r)$* & 	0.0195 & 	0.0695 & 	0.2795 & 	0.0111 & 	 & 	0.26 & 	$(r-i)$ & 	0.3033 & 	1.0167 & 	0.3152 & 	0.0181 \\
0.27 & 	$(g-r)$* & 	0.0272 & 	0.0270 & 	0.3032 & 	0.0132 & 	 & 	0.27 & 	$(r-i)$ & 	0.2427 & 	1.0023 & 	0.3203 & 	0.0164 \\
0.28 & 	$(g-r)$* & 	0.0352 & 	-0.0163 & 	0.3283 & 	0.0155 & 	 & 	0.28 & 	$(r-i)$ & 	0.2057 & 	0.9769 & 	0.3255 & 	0.0146 \\
0.29 & 	$(g-r)$* & 	0.0433 & 	-0.0588 & 	0.3538 & 	0.0180 & 	 & 	0.29 & 	$(r-i)$ & 	0.1760 & 	0.9513 & 	0.3303 & 	0.0132 \\
0.30 & 	$(g-r)$* & 	0.0498 & 	-0.0972 & 	0.3784 & 	0.0202 & 	 & 	0.30 & 	$(r-i)$ & 	0.1495 & 	0.9283 & 	0.3347 & 	0.0117 \\
0.31 & 	$(g-r)$* & 	0.0544 & 	-0.1296 & 	0.3999 & 	0.0218 & 	 & 	0.31 & 	$(r-i)$ & 	0.1242 & 	0.9076 & 	0.3385 & 	0.0104 \\
0.32 & 	$(g-r)$* & 	0.0576 & 	-0.1585 & 	0.4195 & 	0.0229 & 	 & 	0.32 & 	$(r-i)$ & 	0.0983 & 	0.8889 & 	0.3413 & 	0.0089 \\
0.33 & 	$(g-r)$* & 	0.0530 & 	-0.1635 & 	0.4210 & 	0.0196 & 	 & 	0.33 & 	$(r-i)$ & 	0.0558 & 	0.8881 & 	0.3387 & 	0.0066 \\
0.34 & 	$(g-r)$* & 	0.0336 & 	-0.1229 & 	0.3873 & 	0.0107 & 	 & 	0.34 & 	$(r-i)$* & 	-0.0226 & 	0.9246 & 	0.3259 & 	0.0048 \\
0.35 & 	$(g-r)$ & 	-0.0038 & 	-0.0301 & 	0.3164 & 	0.0134 & 	 & 	0.35 & 	$(r-i)$* & 	-0.1596 & 	1.0195 & 	0.2986 & 	0.0108 \\
0.36 & 	$(g-r)$ & 	-0.0553 & 	0.0987 & 	0.2241 & 	0.0296 & 	 & 	0.36 & 	$(r-i)$* & 	-0.3490 & 	1.1784 & 	0.2528 & 	0.0253 \\
0.37 & 	$(g-r)$ & 	-0.0871 & 	0.1580 & 	0.1904 & 	0.0268 & 	 & 	0.37 & 	$(r-i)$* & 	-0.3862 & 	1.1818 & 	0.2437 & 	0.0255 \\
0.38 & 	$(g-r)$ & 	-0.1149 & 	0.2123 & 	0.1544 & 	0.0311 & 	 & 	0.38 & 	$(r-i)$* & 	-0.4522 & 	1.2146 & 	0.2282 & 	0.0312 \\
0.39 & 	$(g-r)$ & 	-0.1353 & 	0.2441 & 	0.1339 & 	0.0339 & 	 & 	0.39 & 	$(r-i)$* & 	-0.4112 & 	1.1259 & 	0.2466 & 	0.0275 \\
0.40 & 	$(g-r)$ & 	-0.1521 & 	0.2648 & 	0.1203 & 	0.0361 & 	 & 	0.40 & 	$(r-i)$* & 	-0.2755 & 	0.9552 & 	0.2786 & 	0.0310 \\
0.41 & 	$(g-r)$ & 	-0.1664 & 	0.2758 & 	0.1134 & 	0.0383 & 	 & 	0.41 & 	$(r-i)$* & 	-0.2729 & 	0.8673 & 	0.3045 & 	0.0356 \\
0.42 & 	$(g-r)$ & 	-0.1774 & 	0.2751 & 	0.1147 & 	0.0405 & 	 & 	0.42 & 	$(r-i)$* & 	-0.4159 & 	0.9411 & 	0.2938 & 	0.0306 \\
0.43 & 	$(g-r)$ & 	-0.1848 & 	0.2636 & 	0.1227 & 	0.0427 & 	 & 	0.43 & 	$(r-i)$* & 	-0.2909 & 	0.7830 & 	0.3268 & 	0.0254 \\
0.44 & 	$(g-r)$ & 	-0.1919 & 	0.2502 & 	0.1316 & 	0.0451 & 	 & 	0.44 & 	$(r-i)$* & 	-0.2102 & 	0.6659 & 	0.3523 & 	0.0201 \\
0.45 & 	$(g-r)$ & 	-0.1978 & 	0.2325 & 	0.1426 & 	0.0477 & 	 & 	0.45 & 	$(r-i)$* & 	-0.1557 & 	0.5740 & 	0.3734 & 	0.0149 \\
0.46 & 	$(g-r)$ & 	-0.2017 & 	0.2071 & 	0.1583 & 	0.0503 & 	 & 	0.46 & 	$(r-i)$* & 	-0.1158 & 	0.4960 & 	0.3927 & 	0.0097 \\
0.47 & 	$(g-r)$ & 	-0.2055 & 	0.1781 & 	0.1770 & 	0.0529 & 	 & 	0.47 & 	$(r-i)$* & 	-0.0963 & 	0.4440 & 	0.4057 & 	0.0063 \\
0.48 & 	$(g-r)$ & 	-0.2083 & 	0.1439 & 	0.1993 & 	0.0556 & 	 & 	0.48 & 	$(r-i)$* & 	-0.0869 & 	0.4000 & 	0.4169 & 	0.0050 \\
0.49 & 	$(g-r)$ & 	-0.2083 & 	0.1014 & 	0.2265 & 	0.0585 & 	 & 	0.49 & 	$(r-i)$* & 	-0.0689 & 	0.3441 & 	0.4330 & 	0.0062 \\
0.50 & 	$(g-r)$ & 	-0.2070 & 	0.0556 & 	0.2547 & 	0.0615 & 	 & 	0.50 & 	$(r-i)$* & 	-0.0515 & 	0.2927 & 	0.4475 & 	0.0075 \\
\enddata												
\label{tab:g_parameters}												
\end{deluxetable}																																			

\clearpage

\begin{deluxetable}{ccccccccccccc}												
\tablewidth{0pt}												
\tablecolumns{13}												
\tabletypesize{\scriptsize}												
\tablecaption {Parameters for calculating absolute magnitudes $M_r = a(g-i)^2 + b(g-i) + c - D_M + r$ or $M_r = a(r-z)^2 + b(r-z) + c - D_M + z$}												
\tablehead{												
\colhead{red-} & 	\colhead{input} & 	\colhead{} & 	\colhead{parameters} & 	\colhead{} & 	\colhead{rms} & 	\colhead{} & 	\colhead{red-} & 	\colhead{input} & 	\colhead{} & 	\colhead{parameters} & 	\colhead{} & 	\colhead{rms} \\
\colhead{shift} & 	\colhead{colour} & 	\colhead{a} & 	\colhead{b} & 	\colhead{c} & 	\colhead{error} & 	\colhead{} & 	\colhead{shift} & 	\colhead{colour} & 	\colhead{a} & 	\colhead{b} & 	\colhead{c} & 	\colhead{error} \\
}												
 \startdata												
0.00 & 	$(g-i)$* & 	-0.0910 & 	0.5390 & 	-0.1256 & 	0.0303 & 	 & 	0.00 & 	$(r-z)$ & 	0.0000 & 	1.0000 & 	0.0000 & 	0.0000 \\
0.01 & 	$(g-i)$* & 	-0.0938 & 	0.5250 & 	-0.1223 & 	0.0302 & 	 & 	0.01 & 	$(r-z)$ & 	-0.0092 & 	1.0028 & 	-0.0076 & 	0.0038 \\
0.02 & 	$(g-i)$* & 	-0.0946 & 	0.5045 & 	-0.1114 & 	0.0295 & 	 & 	0.02 & 	$(r-z)$ & 	-0.0449 & 	1.0643 & 	-0.0460 & 	0.0106 \\
0.03 & 	$(g-i)$* & 	-0.0911 & 	0.4742 & 	-0.0927 & 	0.0281 & 	 & 	0.03 & 	$(r-z)$ & 	-0.1879 & 	1.2562 & 	-0.1262 & 	0.0215 \\
0.04 & 	$(g-i)$* & 	-0.0826 & 	0.4325 & 	-0.0652 & 	0.0258 & 	 & 	0.04 & 	$(r-z)$ & 	-0.4428 & 	1.5838 & 	-0.2468 & 	0.0339 \\
0.05 & 	$(g-i)$* & 	-0.0609 & 	0.3587 & 	-0.0155 & 	0.0212 & 	 & 	0.05 & 	$(r-z)$ & 	-0.6180 & 	1.7891 & 	-0.3174 & 	0.0410 \\
0.06 & 	$(g-i)$* & 	-0.0361 & 	0.2782 & 	0.0381 & 	0.0174 & 	 & 	0.06 & 	$(r-z)$ & 	-0.6245 & 	1.7770 & 	-0.3166 & 	0.0427 \\
0.07 & 	$(g-i)$* & 	-0.0090 & 	0.1915 & 	0.0949 & 	0.0158 & 	 & 	0.07 & 	$(r-z)$ & 	-0.6208 & 	1.7518 & 	-0.3115 & 	0.0435 \\
0.08 & 	$(g-i)$* & 	-0.0103 & 	0.1717 & 	0.1103 & 	0.0155 & 	 & 	0.08 & 	$(r-z)$ & 	-0.6075 & 	1.7115 & 	-0.3003 & 	0.0439 \\
0.09 & 	$(g-i)$* & 	-0.0112 & 	0.1553 & 	0.1200 & 	0.0151 & 	 & 	0.09 & 	$(r-z)$ & 	-0.5452 & 	1.5984 & 	-0.2619 & 	0.0414 \\
0.10 & 	$(g-i)$* & 	-0.0148 & 	0.1479 & 	0.1225 & 	0.0146 & 	 & 	0.10 & 	$(r-z)$ & 	-0.4249 & 	1.3992 & 	-0.1920 & 	0.0372 \\
0.11 & 	$(g-i)$* & 	-0.0141 & 	0.1297 & 	0.1324 & 	0.0140 & 	 & 	0.11 & 	$(r-z)$ & 	-0.2743 & 	1.1576 & 	-0.1077 & 	0.0345 \\
0.12 & 	$(g-i)$* & 	-0.0156 & 	0.1183 & 	0.1383 & 	0.0135 & 	 & 	0.12 & 	$(r-z)$ & 	-0.1403 & 	0.9413 & 	-0.0327 & 	0.0349 \\
0.13 & 	$(g-i)$* & 	-0.0193 & 	0.1174 & 	0.1363 & 	0.0124 & 	 & 	0.13 & 	$(r-z)$ & 	-0.0598 & 	0.7990 & 	0.0178 & 	0.0355 \\
0.14 & 	$(g-i)$* & 	-0.0170 & 	0.1024 & 	0.1448 & 	0.0112 & 	 & 	0.14 & 	$(r-z)$ & 	-0.0221 & 	0.7153 & 	0.0491 & 	0.0357 \\
0.15 & 	$(g-i)$* & 	-0.0236 & 	0.1181 & 	0.1279 & 	0.0101 & 	 & 	0.15 & 	$(r-z)$ & 	-0.0078 & 	0.6646 & 	0.0691 & 	0.0355 \\
0.16 & 	$(g-i)$* & 	-0.0558 & 	0.2072 & 	0.0577 & 	0.0137 & 	 & 	0.16 & 	$(r-z)$ & 	-0.0041 & 	0.6300 & 	0.0829 & 	0.0351 \\
0.17 & 	$(g-i)$* & 	-0.0356 & 	0.1366 & 	0.1118 & 	0.0094 & 	 & 	0.17 & 	$(r-z)$ & 	-0.0073 & 	0.6055 & 	0.0930 & 	0.0346 \\
0.18 & 	$(g-i)$* & 	-0.0162 & 	0.0696 & 	0.1616 & 	0.0061 & 	 & 	0.18 & 	$(r-z)$ & 	-0.0137 & 	0.5855 & 	0.1017 & 	0.0341 \\
0.19 & 	$(g-i)$* & 	-0.0146 & 	0.0593 & 	0.1663 & 	0.0046 & 	 & 	0.19 & 	$(r-z)$ & 	-0.0203 & 	0.5677 & 	0.1092 & 	0.0336 \\
0.20 & 	$(g-i)$* & 	-0.0115 & 	0.0465 & 	0.1724 & 	0.0037 & 	 & 	0.20 & 	$(r-z)$ & 	-0.0265 & 	0.5502 & 	0.1169 & 	0.0331 \\
0.21 & 	$(g-i)$* & 	-0.0093 & 	0.0369 & 	0.1763 & 	0.0038 & 	 & 	0.21 & 	$(r-z)$ & 	-0.0336 & 	0.5338 & 	0.1249 & 	0.0326 \\
0.22 & 	$(g-i)$* & 	-0.0074 & 	0.0282 & 	0.1794 & 	0.0048 & 	 & 	0.22 & 	$(r-z)$ & 	-0.0396 & 	0.5151 & 	0.1344 & 	0.0317 \\
0.23 & 	$(g-i)$* & 	-0.0078 & 	0.0277 & 	0.1736 & 	0.0073 & 	 & 	0.23 & 	$(r-z)$ & 	-0.0403 & 	0.4872 & 	0.1486 & 	0.0302 \\
0.24 & 	$(g-i)$* & 	-0.0148 & 	0.0515 & 	0.1437 & 	0.0119 & 	 & 	0.24 & 	$(r-z)$ & 	-0.0330 & 	0.4441 & 	0.1706 & 	0.0279 \\
0.25 & 	$(g-i)$* & 	-0.0360 & 	0.1261 & 	0.0672 & 	0.0206 & 	 & 	0.25 & 	$(r-z)$* & 	-0.0140 & 	0.3807 & 	0.2028 & 	0.0243 \\
0.26 & 	$(g-i)$ & 	-0.0581 & 	0.2043 & 	-0.0126 & 	0.0301 & 	 & 	0.26 & 	$(r-z)$* & 	0.0184 & 	0.2922 & 	0.2477 & 	0.0199 \\
0.27 & 	$(g-i)$ & 	-0.0749 & 	0.2633 & 	-0.0736 & 	0.0376 & 	 & 	0.27 & 	$(r-z)$* & 	0.0653 & 	0.1758 & 	0.3068 & 	0.0158 \\
0.28 & 	$(g-i)$ & 	-0.0737 & 	0.2590 & 	-0.0776 & 	0.0396 & 	 & 	0.28 & 	$(r-z)$* & 	0.0985 & 	0.0850 & 	0.3544 & 	0.0148 \\
0.29 & 	$(g-i)$ & 	-0.0716 & 	0.2514 & 	-0.0779 & 	0.0411 & 	 & 	0.29 & 	$(r-z)$* & 	0.1173 & 	0.0219 & 	0.3891 & 	0.0152 \\
0.30 & 	$(g-i)$ & 	-0.0678 & 	0.2378 & 	-0.0739 & 	0.0423 & 	 & 	0.30 & 	$(r-z)$* & 	0.1263 & 	-0.0214 & 	0.4145 & 	0.0161 \\
0.31 & 	$(g-i)$ & 	-0.0635 & 	0.2212 & 	-0.0664 & 	0.0435 & 	 & 	0.31 & 	$(r-z)$* & 	0.1274 & 	-0.0476 & 	0.4303 & 	0.0169 \\
0.32 & 	$(g-i)$ & 	-0.0588 & 	0.2035 & 	-0.0581 & 	0.0445 & 	 & 	0.32 & 	$(r-z)$* & 	0.1221 & 	-0.0599 & 	0.4376 & 	0.0168 \\
0.33 & 	$(g-i)$ & 	-0.0552 & 	0.1876 & 	-0.0496 & 	0.0453 & 	 & 	0.33 & 	$(r-z)$* & 	0.1106 & 	-0.0574 & 	0.4358 & 	0.0160 \\
0.34 & 	$(g-i)$ & 	-0.0521 & 	0.1712 & 	-0.0398 & 	0.0461 & 	 & 	0.34 & 	$(r-z)$* & 	0.0937 & 	-0.0431 & 	0.4273 & 	0.0145 \\
0.35 & 	$(g-i)$ & 	-0.0441 & 	0.1354 & 	-0.0118 & 	0.0456 & 	 & 	0.35 & 	$(r-z)$* & 	0.0690 & 	-0.0151 & 	0.4136 & 	0.0125 \\
0.36 & 	$(g-i)$ & 	-0.0345 & 	0.0915 & 	0.0249 & 	0.0450 & 	 & 	0.36 & 	$(r-z)$* & 	0.0291 & 	0.0483 & 	0.3800 & 	0.0095 \\
0.37 & 	$(g-i)$ & 	-0.0140 & 	0.0041 & 	0.1046 & 	0.0448 & 	 & 	0.37 & 	$(r-z)$* & 	-0.0337 & 	0.1584 & 	0.3236 & 	0.0086 \\
0.38 & 	$(g-i)$ & 	0.0099 & 	-0.0960 & 	0.1967 & 	0.0466 & 	 & 	0.38 & 	$(r-z)$* & 	-0.0442 & 	0.1685 & 	0.3161 & 	0.0089 \\
0.39 & 	$(g-i)$ & 	0.0340 & 	-0.2009 & 	0.2993 & 	0.0539 & 	 & 	0.39 & 	$(r-z)$* & 	-0.0621 & 	0.1963 & 	0.2983 & 	0.0106 \\
0.40 & 	$(g-i)$ & 	0.0526 & 	-0.2747 & 	0.3617 & 	0.0510 & 	 & 	0.40 & 	$(r-z)$* & 	-0.0626 & 	0.1862 & 	0.3020 & 	0.0105 \\
0.41 & 	$(g-i)$ & 	0.0609 & 	-0.3174 & 	0.4051 & 	0.0507 & 	 & 	0.41 & 	$(r-z)$* & 	-0.0637 & 	0.1849 & 	0.2960 & 	0.0118 \\
0.42 & 	$(g-i)$ & 	0.0748 & 	-0.3829 & 	0.4705 & 	0.0528 & 	 & 	0.42 & 	$(r-z)$* & 	-0.1234 & 	0.3061 & 	0.2260 & 	0.0207 \\
0.43 & 	$(g-i)$ & 	0.0743 & 	-0.3911 & 	0.4812 & 	0.0511 & 	 & 	0.43 & 	$(r-z)$* & 	-0.1298 & 	0.3129 & 	0.2200 & 	0.0220 \\
0.44 & 	$(g-i)$ & 	0.0807 & 	-0.4303 & 	0.5272 & 	0.0507 & 	 & 	0.44 & 	$(r-z)$* & 	-0.1229 & 	0.2957 & 	0.2245 & 	0.0223 \\
0.45 & 	$(g-i)$ & 	0.0847 & 	-0.4564 & 	0.5571 & 	0.0524 & 	 & 	0.45 & 	$(r-z)$* & 	-0.1148 & 	0.2770 & 	0.2294 & 	0.0227 \\
0.46 & 	$(g-i)$ & 	0.0891 & 	-0.4820 & 	0.5844 & 	0.0542 & 	 & 	0.46 & 	$(r-z)$* & 	-0.1050 & 	0.2525 & 	0.2388 & 	0.0224 \\
0.47 & 	$(g-i)$ & 	0.0917 & 	-0.4990 & 	0.6022 & 	0.0559 & 	 & 	0.47 & 	$(r-z)$* & 	-0.0889 & 	0.2133 & 	0.2569 & 	0.0213 \\
0.48 & 	$(g-i)$ & 	0.0912 & 	-0.5045 & 	0.6096 & 	0.0576 & 	 & 	0.48 & 	$(r-z)$* & 	-0.0796 & 	0.1930 & 	0.2615 & 	0.0221 \\
0.49 & 	$(g-i)$ & 	0.0910 & 	-0.5119 & 	0.6181 & 	0.0591 & 	 & 	0.49 & 	$(r-z)$* & 	-0.0852 & 	0.2053 & 	0.2486 & 	0.0247 \\
0.50 & 	$(g-i)$ & 	0.0917 & 	-0.5218 & 	0.6289 & 	0.0607 & 	 & 	0.50 & 	$(r-z)$* & 	-0.0819 & 	0.1973 & 	0.2473 & 	0.0262 \\
\enddata												
\label{tab:r_parameters}												
\end{deluxetable}

\clearpage

\begin{deluxetable}{ccccccccccccc}												
\tablewidth{0pt}												
\tablecolumns{13}												
\tabletypesize{\scriptsize}												
\tablecaption {Parameters for calculating absolute magnitudes $M_i = a(r-z)^2 + b(r-z) + c - D_M + z$ or $M_i = a(g-i)^2 + b(g-i) + c - D_M + r$}												
\tablehead{												
\colhead{red-} & 	\colhead{input} & 	\colhead{} & 	\colhead{parameters} & 	\colhead{} & 	\colhead{rms} & 	\colhead{} & 	\colhead{red-} & 	\colhead{input} & 	\colhead{} & 	\colhead{parameters} & 	\colhead{} & 	\colhead{rms} \\
\colhead{shift} & 	\colhead{colour} & 	\colhead{a} & 	\colhead{b} & 	\colhead{c} & 	\colhead{error} & 	\colhead{} & 	\colhead{shift} & 	\colhead{colour} & 	\colhead{a} & 	\colhead{b} & 	\colhead{c} & 	\colhead{error} \\
}												
 \startdata												
0.00 & 	$(r-z)$* & 	0.0584 & 	0.2701 & 	0.0916 & 	0.0221 & 	 & 	0.00 & 	$(g-i)$ & 	0.0000 & 	0.0000 & 	0.0000 & 	0.0000 \\
0.01 & 	$(r-z)$* & 	0.0531 & 	0.2604 & 	0.0941 & 	0.0209 & 	 & 	0.01 & 	$(g-i)$ & 	-0.0004 & 	-0.0146 & 	0.0071 & 	0.0022 \\
0.02 & 	$(r-z)$* & 	0.0483 & 	0.2614 & 	0.0885 & 	0.0187 & 	 & 	0.02 & 	$(g-i)$ & 	0.0011 & 	-0.0346 & 	0.0194 & 	0.0046 \\
0.03 & 	$(r-z)$* & 	0.0280 & 	0.2841 & 	0.0748 & 	0.0164 & 	 & 	0.03 & 	$(g-i)$ & 	0.0110 & 	-0.0744 & 	0.0455 & 	0.0077 \\
0.04 & 	$(r-z)$* & 	-0.0154 & 	0.3320 & 	0.0549 & 	0.0146 & 	 & 	0.04 & 	$(g-i)$ & 	0.0446 & 	-0.1685 & 	0.1062 & 	0.0129 \\
0.05 & 	$(r-z)$* & 	-0.0483 & 	0.3548 & 	0.0477 & 	0.0140 & 	 & 	0.05 & 	$(g-i)$ & 	0.1053 & 	-0.3284 & 	0.2090 & 	0.0223 \\
0.06 & 	$(r-z)$* & 	-0.0520 & 	0.3359 & 	0.0555 & 	0.0135 & 	 & 	0.06 & 	$(g-i)$ & 	0.1556 & 	-0.4883 & 	0.3288 & 	0.0312 \\
0.07 & 	$(r-z)$* & 	-0.0532 & 	0.3140 & 	0.0646 & 	0.0128 & 	 & 	0.07 & 	$(g-i)$ & 	0.2919 & 	-0.8171 & 	0.5226 & 	0.0398 \\
0.08 & 	$(r-z)$* & 	-0.0518 & 	0.2890 & 	0.0753 & 	0.0121 & 	 & 	0.08 & 	$(g-i)$ & 	0.3353 & 	-0.9575 & 	0.6235 & 	0.0481 \\
0.09 & 	$(r-z)$* & 	-0.0394 & 	0.2501 & 	0.0916 & 	0.0113 & 	 & 	0.09 & 	$(g-i)$ & 	0.3127 & 	-0.9186 & 	0.6020 & 	0.0447 \\
0.10 & 	$(r-z)$* & 	-0.0177 & 	0.1989 & 	0.1124 & 	0.0105 & 	 & 	0.10 & 	$(g-i)$ & 	0.3107 & 	-0.9361 & 	0.6197 & 	0.0454 \\
0.11 & 	$(r-z)$* & 	0.0045 & 	0.1474 & 	0.1330 & 	0.0100 & 	 & 	0.11 & 	$(g-i)$ & 	0.3107 & 	-0.9563 & 	0.6391 & 	0.0468 \\
0.12 & 	$(r-z)$* & 	0.0193 & 	0.1070 & 	0.1495 & 	0.0094 & 	 & 	0.12 & 	$(g-i)$ & 	0.3091 & 	-0.9747 & 	0.6602 & 	0.0478 \\
0.13 & 	$(r-z)$* & 	0.0237 & 	0.0809 & 	0.1609 & 	0.0086 & 	 & 	0.13 & 	$(g-i)$ & 	0.2847 & 	-0.9252 & 	0.6343 & 	0.0456 \\
0.14 & 	$(r-z)$* & 	0.0222 & 	0.0634 & 	0.1688 & 	0.0076 & 	 & 	0.14 & 	$(g-i)$ & 	0.2790 & 	-0.9255 & 	0.6450 & 	0.0450 \\
0.15 & 	$(r-z)$* & 	0.0183 & 	0.0497 & 	0.1744 & 	0.0064 & 	 & 	0.15 & 	$(g-i)$ & 	0.2734 & 	-0.9248 & 	0.6546 & 	0.0437 \\
0.16 & 	$(r-z)$* & 	0.0136 & 	0.0376 & 	0.1789 & 	0.0051 & 	 & 	0.16 & 	$(g-i)$ & 	0.1509 & 	-0.5742 & 	0.3981 & 	0.0330 \\
0.17 & 	$(r-z)$* & 	0.0089 & 	0.0264 & 	0.1831 & 	0.0038 & 	 & 	0.17 & 	$(g-i)$ & 	0.3112 & 	-1.0404 & 	0.7346 & 	0.0459 \\
0.18 & 	$(r-z)$* & 	0.0044 & 	0.0154 & 	0.1867 & 	0.0023 & 	 & 	0.18 & 	$(g-i)$ & 	0.3035 & 	-1.0388 & 	0.7458 & 	0.0465 \\
0.19 & 	$(r-z)$* & 	0.0007 & 	0.0036 & 	0.1906 & 	0.0010 & 	 & 	0.19 & 	$(g-i)$ & 	0.2974 & 	-1.0385 & 	0.7563 & 	0.0478 \\
0.20 & 	$(r-z)$* & 	-0.0027 & 	-0.0088 & 	0.1954 & 	0.0012 & 	 & 	0.20 & 	$(g-i)$ & 	0.2820 & 	-1.0081 & 	0.7446 & 	0.0469 \\
0.21 & 	$(r-z)$* & 	-0.0049 & 	-0.0243 & 	0.2021 & 	0.0029 & 	 & 	0.21 & 	$(g-i)$ & 	0.2725 & 	-0.9925 & 	0.7420 & 	0.0471 \\
0.22 & 	$(r-z)$* & 	-0.0037 & 	-0.0472 & 	0.2134 & 	0.0052 & 	 & 	0.22 & 	$(g-i)$ & 	0.2557 & 	-0.9537 & 	0.7220 & 	0.0451 \\
0.23 & 	$(r-z)$* & 	0.0070 & 	-0.0886 & 	0.2345 & 	0.0085 & 	 & 	0.23 & 	$(g-i)$ & 	0.2417 & 	-0.9185 & 	0.6965 & 	0.0443 \\
0.24 & 	$(r-z)$* & 	0.0325 & 	-0.1582 & 	0.2703 & 	0.0135 & 	 & 	0.24 & 	$(g-i)$ & 	0.2060 & 	-0.8095 & 	0.6078 & 	0.0408 \\
0.25 & 	$(r-z)$* & 	0.0780 & 	-0.2640 & 	0.3249 & 	0.0206 & 	 & 	0.25 & 	$(g-i)$ & 	0.1302 & 	-0.5690 & 	0.4119 & 	0.0415 \\
0.26 & 	$(r-z)$* & 	0.1457 & 	-0.4124 & 	0.4014 & 	0.0306 & 	 & 	0.26 & 	$(g-i)$ & 	0.0827 & 	-0.4143 & 	0.2805 & 	0.0421 \\
0.27 & 	$(r-z)$* & 	0.1765 & 	-0.5175 & 	0.4707 & 	0.0347 & 	 & 	0.27 & 	$(g-i)$ & 	0.0627 & 	-0.3526 & 	0.2299 & 	0.0450 \\
0.28 & 	$(r-z)$* & 	0.2327 & 	-0.6439 & 	0.5386 & 	0.0421 & 	 & 	0.28 & 	$(g-i)$ & 	0.0591 & 	-0.3463 & 	0.2263 & 	0.0471 \\
0.29 & 	$(r-z)$* & 	0.2633 & 	-0.7231 & 	0.5844 & 	0.0470 & 	 & 	0.29 & 	$(g-i)$ & 	0.0610 & 	-0.3511 & 	0.2292 & 	0.0458 \\
0.30 & 	$(r-z)$* & 	0.2812 & 	-0.7765 & 	0.6167 & 	0.0505 & 	 & 	0.30 & 	$(g-i)$ & 	0.0613 & 	-0.3571 & 	0.2341 & 	0.0475 \\
0.31 & 	$(r-z)$* & 	0.2876 & 	-0.8052 & 	0.6346 & 	0.0526 & 	 & 	0.31 & 	$(g-i)$ & 	0.0628 & 	-0.3689 & 	0.2455 & 	0.0493 \\
0.32 & 	$(r-z)$* & 	0.3302 & 	-0.8873 & 	0.6683 & 	0.0515 & 	 & 	0.32 & 	$(g-i)$ & 	0.0644 & 	-0.3796 & 	0.2561 & 	0.0511 \\
0.33 & 	$(r-z)$* & 	0.3467 & 	-0.9233 & 	0.6807 & 	0.0507 & 	 & 	0.33 & 	$(g-i)$ & 	0.0651 & 	-0.3905 & 	0.2680 & 	0.0526 \\
0.34 & 	$(r-z)$* & 	0.4001 & 	-1.0185 & 	0.7154 & 	0.0502 & 	 & 	0.34 & 	$(g-i)$ & 	0.0665 & 	-0.4055 & 	0.2837 & 	0.0539 \\
0.35 & 	$(r-z)$* & 	0.4973 & 	-1.1916 & 	0.7848 & 	0.0511 & 	 & 	0.35 & 	$(g-i)$ & 	0.0711 & 	-0.4306 & 	0.3053 & 	0.0521 \\
0.36 & 	$(r-z)$* & 	0.5611 & 	-1.3055 & 	0.8272 & 	0.0520 & 	 & 	0.36 & 	$(g-i)$ & 	0.0872 & 	-0.5025 & 	0.3731 & 	0.0533 \\
0.37 & 	$(r-z)$* & 	0.4876 & 	-1.1735 & 	0.7628 & 	0.0463 & 	 & 	0.37 & 	$(g-i)$ & 	0.1144 & 	-0.6185 & 	0.4833 & 	0.0583 \\
0.38 & 	$(r-z)$* & 	0.4757 & 	-1.1600 & 	0.7577 & 	0.0443 & 	 & 	0.38 & 	$(g-i)$ & 	0.1404 & 	-0.7272 & 	0.5868 & 	0.0608 \\
0.39 & 	$(r-z)$* & 	0.4459 & 	-1.1105 & 	0.7342 & 	0.0430 & 	 & 	0.39 & 	$(g-i)$ & 	0.1689 & 	-0.8506 & 	0.7111 & 	0.0587 \\
0.40 & 	$(r-z)$* & 	0.4466 & 	-1.1270 & 	0.7478 & 	0.0444 & 	 & 	0.40 & 	$(g-i)$ & 	0.1867 & 	-0.9355 & 	0.8020 & 	0.0577 \\
0.41 & 	$(r-z)$* & 	0.4382 & 	-1.1115 & 	0.7369 & 	0.0433 & 	 & 	0.41 & 	$(g-i)$ & 	0.2101 & 	-1.0447 & 	0.9187 & 	0.0599 \\
0.42 & 	$(r-z)$* & 	0.3434 & 	-0.9172 & 	0.6328 & 	0.0442 & 	 & 	0.42 & 	$(g-i)$ & 	0.2277 & 	-1.1308 & 	1.0118 & 	0.0630 \\
0.43 & 	$(r-z)$* & 	0.3134 & 	-0.8648 & 	0.6104 & 	0.0414 & 	 & 	0.43 & 	$(g-i)$ & 	0.2344 & 	-1.1729 & 	1.0631 & 	0.0649 \\
0.44 & 	$(r-z)$* & 	0.3047 & 	-0.8536 & 	0.6078 & 	0.0410 & 	 & 	0.44 & 	$(g-i)$ & 	0.2374 & 	-1.1914 & 	1.0851 & 	0.0618 \\
0.45 & 	$(r-z)$* & 	0.2978 & 	-0.8452 & 	0.6060 & 	0.0408 & 	 & 	0.45 & 	$(g-i)$ & 	0.2415 & 	-1.2216 & 	1.1254 & 	0.0637 \\
0.46 & 	$(r-z)$* & 	0.2941 & 	-0.8466 & 	0.6116 & 	0.0412 & 	 & 	0.46 & 	$(g-i)$ & 	0.2463 & 	-1.2521 & 	1.1626 & 	0.0660 \\
0.47 & 	$(r-z)$* & 	0.3018 & 	-0.8742 & 	0.6322 & 	0.0428 & 	 & 	0.47 & 	$(g-i)$ & 	0.2449 & 	-1.2546 & 	1.1705 & 	0.0679 \\
0.48 & 	$(r-z)$* & 	0.2944 & 	-0.8621 & 	0.6258 & 	0.0427 & 	 & 	0.48 & 	$(g-i)$ & 	0.2388 & 	-1.2398 & 	1.1604 & 	0.0692 \\
0.49 & 	$(r-z)$* & 	0.2588 & 	-0.7868 & 	0.5852 & 	0.0410 & 	 & 	0.49 & 	$(g-i)$ & 	0.2438 & 	-1.2620 & 	1.1800 & 	0.0683 \\
0.50 & 	$(r-z)$* & 	0.2414 & 	-0.7528 & 	0.5675 & 	0.0410 & 	 & 	0.50 & 	$(g-i)$ & 	0.2401 & 	-1.2554 & 	1.1771 & 	0.0698 \\
\enddata												
\label{tab:i_parameters}												
\end{deluxetable}												
\clearpage

\clearpage

\begin{deluxetable}{ccccccccccccc}												
\tablewidth{0pt}												
\tablecolumns{13}												
\tabletypesize{\scriptsize}												
\tablecaption {Parameters for calculating absolute magnitudes $M_z = a(r-z)^2 + b(r-z) + c - D_M + z$ or $M_z = a(g-i)^2 + b(g-i) + c - D_M + r$}												
\tablehead{												
\colhead{red-} & 	\colhead{input} & 	\colhead{} & 	\colhead{parameters} & 	\colhead{} & 	\colhead{rms} & 	\colhead{} & 	\colhead{red-} & 	\colhead{input} & 	\colhead{} & 	\colhead{parameters} & 	\colhead{} & 	\colhead{rms} \\
\colhead{shift} & 	\colhead{colour} & 	\colhead{a} & 	\colhead{b} & 	\colhead{c} & 	\colhead{error} & 	\colhead{} & 	\colhead{shift} & 	\colhead{colour} & 	\colhead{a} & 	\colhead{b} & 	\colhead{c} & 	\colhead{error} \\
}												
 \startdata												
0.00 & 	$(r-z)$* & 	0.0000 & 	0.0000 & 	0.0000 & 	0.0000 & 	 & 	0.00 & 	$(g-i)$ & 	0.0215 & 	-0.2030 & 	-0.0892 & 	0.0338 \\
0.01 & 	$(r-z)$* & 	-0.0022 & 	-0.0134 & 	0.0057 & 	0.0019 & 	 & 	0.01 & 	$(g-i)$ & 	0.0234 & 	-0.2170 & 	-0.0831 & 	0.0360 \\
0.02 & 	$(r-z)$* & 	-0.0037 & 	-0.0282 & 	0.0124 & 	0.0038 & 	 & 	0.02 & 	$(g-i)$ & 	0.0276 & 	-0.2380 & 	-0.0706 & 	0.0382 \\
0.03 & 	$(r-z)$* & 	-0.0031 & 	-0.0487 & 	0.0221 & 	0.0053 & 	 & 	0.03 & 	$(g-i)$ & 	0.0418 & 	-0.2836 & 	-0.0412 & 	0.0403 \\
0.04 & 	$(r-z)$* & 	0.0045 & 	-0.0795 & 	0.0362 & 	0.0068 & 	 & 	0.04 & 	$(g-i)$ & 	0.0843 & 	-0.3945 & 	0.0298 & 	0.0430 \\
0.05 & 	$(r-z)$* & 	0.0158 & 	-0.1135 & 	0.0501 & 	0.0084 & 	 & 	0.05 & 	$(g-i)$ & 	0.1623 & 	-0.5913 & 	0.1547 & 	0.0478 \\
0.06 & 	$(r-z)$* & 	0.0230 & 	-0.1408 & 	0.0608 & 	0.0099 & 	 & 	0.06 & 	$(g-i)$ & 	0.2561 & 	-0.8461 & 	0.3279 & 	0.0490 \\
0.07 & 	$(r-z)$* & 	0.0299 & 	-0.1673 & 	0.0718 & 	0.0113 & 	 & 	0.07 & 	$(g-i)$ & 	0.4015 & 	-1.2101 & 	0.5537 & 	0.0547 \\
0.08 & 	$(r-z)$* & 	0.0373 & 	-0.1930 & 	0.0828 & 	0.0128 & 	 & 	0.08 & 	$(g-i)$ & 	0.4650 & 	-1.3932 & 	0.6790 & 	0.0599 \\
0.09 & 	$(r-z)$* & 	0.0396 & 	-0.2078 & 	0.0904 & 	0.0144 & 	 & 	0.09 & 	$(g-i)$ & 	0.4714 & 	-1.4225 & 	0.6986 & 	0.0617 \\
0.10 & 	$(r-z)$* & 	0.0322 & 	-0.2064 & 	0.0923 & 	0.0159 & 	 & 	0.10 & 	$(g-i)$ & 	0.4739 & 	-1.4517 & 	0.7260 & 	0.0634 \\
0.11 & 	$(r-z)$* & 	0.0151 & 	-0.1914 & 	0.0891 & 	0.0175 & 	 & 	0.11 & 	$(g-i)$ & 	0.4350 & 	-1.3818 & 	0.6976 & 	0.0628 \\
0.12 & 	$(r-z)$* & 	-0.0043 & 	-0.1731 & 	0.0846 & 	0.0195 & 	 & 	0.12 & 	$(g-i)$ & 	0.4347 & 	-1.4037 & 	0.7236 & 	0.0640 \\
0.13 & 	$(r-z)$* & 	-0.0178 & 	-0.1634 & 	0.0821 & 	0.0214 & 	 & 	0.13 & 	$(g-i)$ & 	0.4038 & 	-1.3376 & 	0.6898 & 	0.0622 \\
0.14 & 	$(r-z)$* & 	-0.0256 & 	-0.1620 & 	0.0816 & 	0.0233 & 	 & 	0.14 & 	$(g-i)$ & 	0.3945 & 	-1.3300 & 	0.6991 & 	0.0619 \\
0.15 & 	$(r-z)$* & 	-0.0294 & 	-0.1660 & 	0.0819 & 	0.0251 & 	 & 	0.15 & 	$(g-i)$ & 	0.3853 & 	-1.3214 & 	0.7075 & 	0.0614 \\
0.16 & 	$(r-z)$* & 	-0.0318 & 	-0.1727 & 	0.0826 & 	0.0269 & 	 & 	0.16 & 	$(g-i)$ & 	0.2603 & 	-0.9545 & 	0.4343 & 	0.0618 \\
0.17 & 	$(r-z)$* & 	-0.0324 & 	-0.1811 & 	0.0841 & 	0.0287 & 	 & 	0.17 & 	$(g-i)$ & 	0.3456 & 	-1.2355 & 	0.6672 & 	0.0621 \\
0.18 & 	$(r-z)$* & 	-0.0318 & 	-0.1906 & 	0.0853 & 	0.0305 & 	 & 	0.18 & 	$(g-i)$ & 	0.3623 & 	-1.2821 & 	0.7016 & 	0.0599 \\
0.19 & 	$(r-z)$* & 	-0.0310 & 	-0.2015 & 	0.0873 & 	0.0323 & 	 & 	0.19 & 	$(g-i)$ & 	0.3475 & 	-1.2555 & 	0.6949 & 	0.0597 \\
0.20 & 	$(r-z)$* & 	-0.0304 & 	-0.2130 & 	0.0905 & 	0.0342 & 	 & 	0.20 & 	$(g-i)$ & 	0.3350 & 	-1.2328 & 	0.6900 & 	0.0596 \\
0.21 & 	$(r-z)$* & 	-0.0281 & 	-0.2290 & 	0.0963 & 	0.0362 & 	 & 	0.21 & 	$(g-i)$ & 	0.3218 & 	-1.2076 & 	0.6834 & 	0.0599 \\
0.22 & 	$(r-z)$* & 	-0.0223 & 	-0.2538 & 	0.1077 & 	0.0383 & 	 & 	0.22 & 	$(g-i)$ & 	0.3271 & 	-1.2424 & 	0.7252 & 	0.0602 \\
0.23 & 	$(r-z)$* & 	-0.0067 & 	-0.2998 & 	0.1310 & 	0.0407 & 	 & 	0.23 & 	$(g-i)$ & 	0.3185 & 	-1.2270 & 	0.7187 & 	0.0619 \\
0.24 & 	$(r-z)$* & 	0.0254 & 	-0.3781 & 	0.1716 & 	0.0440 & 	 & 	0.24 & 	$(g-i)$ & 	0.3168 & 	-1.2078 & 	0.6825 & 	0.0634 \\
0.25 & 	$(r-z)$* & 	0.0142 & 	-0.4159 & 	0.2116 & 	0.0437 & 	 & 	0.25 & 	$(g-i)$ & 	0.2104 & 	-0.8738 & 	0.4209 & 	0.0608 \\
0.26 & 	$(r-z)$* & 	0.0793 & 	-0.5587 & 	0.2891 & 	0.0481 & 	 & 	0.26 & 	$(g-i)$ & 	0.1539 & 	-0.6957 & 	0.2775 & 	0.0645 \\
0.27 & 	$(r-z)$* & 	0.1746 & 	-0.7536 & 	0.3912 & 	0.0548 & 	 & 	0.27 & 	$(g-i)$ & 	0.1237 & 	-0.6094 & 	0.2171 & 	0.0648 \\
0.28 & 	$(r-z)$* & 	0.2307 & 	-0.8915 & 	0.4724 & 	0.0582 & 	 & 	0.28 & 	$(g-i)$ & 	0.1171 & 	-0.5960 & 	0.2118 & 	0.0663 \\
0.29 & 	$(r-z)$* & 	0.3376 & 	-1.0772 & 	0.5552 & 	0.0614 & 	 & 	0.29 & 	$(g-i)$ & 	0.1135 & 	-0.5906 & 	0.2133 & 	0.0677 \\
0.30 & 	$(r-z)$* & 	0.3667 & 	-1.1461 & 	0.5947 & 	0.0645 & 	 & 	0.30 & 	$(g-i)$ & 	0.1097 & 	-0.5820 & 	0.2068 & 	0.0669 \\
0.31 & 	$(r-z)$* & 	0.4445 & 	-1.2830 & 	0.6529 & 	0.0646 & 	 & 	0.31 & 	$(g-i)$ & 	0.1100 & 	-0.5912 & 	0.2191 & 	0.0682 \\
0.32 & 	$(r-z)$* & 	0.3563 & 	-1.1652 & 	0.6160 & 	0.0670 & 	 & 	0.32 & 	$(g-i)$ & 	0.1103 & 	-0.5992 & 	0.2303 & 	0.0694 \\
0.33 & 	$(r-z)$* & 	0.3882 & 	-1.2202 & 	0.6344 & 	0.0663 & 	 & 	0.33 & 	$(g-i)$ & 	0.1099 & 	-0.6077 & 	0.2433 & 	0.0704 \\
0.34 & 	$(r-z)$* & 	0.4764 & 	-1.3660 & 	0.6875 & 	0.0658 & 	 & 	0.34 & 	$(g-i)$ & 	0.1105 & 	-0.6213 & 	0.2602 & 	0.0711 \\
0.35 & 	$(r-z)$* & 	0.6360 & 	-1.6372 & 	0.7952 & 	0.0669 & 	 & 	0.35 & 	$(g-i)$ & 	0.1168 & 	-0.6569 & 	0.2965 & 	0.0715 \\
0.36 & 	$(r-z)$* & 	0.7203 & 	-1.7760 & 	0.8446 & 	0.0672 & 	 & 	0.36 & 	$(g-i)$ & 	0.1378 & 	-0.7427 & 	0.3758 & 	0.0704 \\
0.37 & 	$(r-z)$* & 	0.6956 & 	-1.7140 & 	0.8033 & 	0.0632 & 	 & 	0.37 & 	$(g-i)$ & 	0.1688 & 	-0.8747 & 	0.5030 & 	0.0751 \\
0.38 & 	$(r-z)$* & 	0.7300 & 	-1.7856 & 	0.8382 & 	0.0655 & 	 & 	0.38 & 	$(g-i)$ & 	0.1990 & 	-1.0074 & 	0.6383 & 	0.0682 \\
0.39 & 	$(r-z)$* & 	0.7033 & 	-1.7380 & 	0.8153 & 	0.0654 & 	 & 	0.39 & 	$(g-i)$ & 	0.2426 & 	-1.1876 & 	0.8147 & 	0.0733 \\
0.40 & 	$(r-z)$* & 	0.7120 & 	-1.7696 & 	0.8378 & 	0.0673 & 	 & 	0.40 & 	$(g-i)$ & 	0.2593 & 	-1.2686 & 	0.9026 & 	0.0707 \\
0.41 & 	$(r-z)$* & 	0.7995 & 	-1.8883 & 	0.8695 & 	0.0647 & 	 & 	0.41 & 	$(g-i)$ & 	0.2868 & 	-1.4002 & 	1.0488 & 	0.0708 \\
0.42 & 	$(r-z)$* & 	0.7383 & 	-1.7210 & 	0.7570 & 	0.0610 & 	 & 	0.42 & 	$(g-i)$ & 	0.2999 & 	-1.4689 & 	1.1266 & 	0.0710 \\
0.43 & 	$(r-z)$* & 	0.6581 & 	-1.5900 & 	0.7078 & 	0.0598 & 	 & 	0.43 & 	$(g-i)$ & 	0.3069 & 	-1.5150 & 	1.1853 & 	0.0727 \\
0.44 & 	$(r-z)$* & 	0.6363 & 	-1.5690 & 	0.7095 & 	0.0619 & 	 & 	0.44 & 	$(g-i)$ & 	0.3117 & 	-1.5520 & 	1.2363 & 	0.0743 \\
0.45 & 	$(r-z)$* & 	0.6142 & 	-1.5461 & 	0.7099 & 	0.0635 & 	 & 	0.45 & 	$(g-i)$ & 	0.3333 & 	-1.6534 & 	1.3461 & 	0.0750 \\
0.46 & 	$(r-z)$* & 	0.5944 & 	-1.5253 & 	0.7109 & 	0.0634 & 	 & 	0.46 & 	$(g-i)$ & 	0.3421 & 	-1.7090 & 	1.4197 & 	0.0745 \\
0.47 & 	$(r-z)$* & 	0.5962 & 	-1.5502 & 	0.7354 & 	0.0641 & 	 & 	0.47 & 	$(g-i)$ & 	0.3397 & 	-1.7096 & 	1.4285 & 	0.0763 \\
0.48 & 	$(r-z)$* & 	0.5671 & 	-1.5021 & 	0.7167 & 	0.0639 & 	 & 	0.48 & 	$(g-i)$ & 	0.3285 & 	-1.6796 & 	1.4079 & 	0.0760 \\
0.49 & 	$(r-z)$* & 	0.4863 & 	-1.3411 & 	0.6413 & 	0.0606 & 	 & 	0.49 & 	$(g-i)$ & 	0.3329 & 	-1.7009 & 	1.4284 & 	0.0752 \\
0.50 & 	$(r-z)$* & 	0.4051 & 	-1.1830 & 	0.5701 & 	0.0635 & 	 & 	0.50 & 	$(g-i)$ & 	0.3281 & 	-1.6877 & 	1.4158 & 	0.0778 \\
\enddata												
\label{tab:z_parameters}												
\end{deluxetable}												

\clearpage

%\bibliography{../../references/AllReferences.bib}

\begin{thebibliography}{36}
\expandafter\ifx\csname natexlab\endcsname\relax\def\natexlab#1{#1}\fi

\bibitem[{{Bell} {et~al.}(2004){Bell}, {Wolf}, {Meisenheimer}, {Rix}, {Borch},
  {Dye}, {Kleinheinrich}, {Wisotzki}, \& {McIntosh}}]{bell04}
{Bell}, E.~F., {et~al.} 2004, \apj, 608, 752

\bibitem[{{Blanton} \& {Roweis}(2007)}]{blant07}
{Blanton}, M.~R., \& {Roweis}, S. 2007, \aj, 133, 734

\bibitem[{{Blanton} {et~al.}(2003){Blanton}, {Brinkmann}, {Csabai}, {Doi},
  {Eisenstein}, {Fukugita}, {Gunn}, {Hogg}, \& {Schlegel}}]{blant03b}
{Blanton}, M.~R., {et~al.} 2003, \aj, 125, 2348

\bibitem[{{Blanton} {et~al.}(2005){Blanton}, {Schlegel}, {Strauss},
  {Brinkmann}, {Finkbeiner}, {Fukugita}, {Gunn}, {Hogg}, {Ivezi{\'c}}, {Knapp},
  {Lupton}, {Munn}, {Schneider}, {Tegmark}, \& {Zehavi}}]{blant05b}
---. 2005, \aj, 129, 2562

\bibitem[{{Brammer} {et~al.}(2008){Brammer}, {van Dokkum}, \&
  {Coppi}}]{bramm08}
{Brammer}, G.~B., {van Dokkum}, P.~G., \& {Coppi}, P. 2008, \apj, 686, 1503

\bibitem[{{Brown} {et~al.}(2007){Brown}, {Dey}, {Jannuzi}, {Brand}, {Benson},
  {Brodwin}, {Croton}, \& {Eisenhardt}}]{brown07}
{Brown}, M.~J.~I., {Dey}, A., {Jannuzi}, B.~T., {Brand}, K., {Benson}, A.~J.,
  {Brodwin}, M., {Croton}, D.~J., \& {Eisenhardt}, P.~R. 2007, \apj, 654, 858

\bibitem[{{Brown} {et~al.}(2014){Brown}, {Moustakas}, {Smith}, {da Cunha},
  {Jarrett}, {Imanishi}, {Armus}, {Brandl}, \& {Peek}}]{brown14}
{Brown}, M.~J.~I., {et~al.} 2014, \apjs, 212, 18

\bibitem[{{Bruzual} \& {Charlot}(2003)}]{bruzu03}
{Bruzual}, G., \& {Charlot}, S. 2003, \mnras, 344, 1000

\bibitem[{{Chilingarian} {et~al.}(2010){Chilingarian}, {Melchior}, \&
  {Zolotukhin}}]{chili10}
{Chilingarian}, I.~V., {Melchior}, A.-L., \& {Zolotukhin}, I.~Y. 2010, \mnras,
  405, 1409

\bibitem[{{Coleman} {et~al.}(1980){Coleman}, {Wu}, \& {Weedman}}]{colem80}
{Coleman}, G.~D., {Wu}, C.-C., \& {Weedman}, D.~W. 1980, \apjs, 43, 393

\bibitem[{{Connolly} {et~al.}(1995){Connolly}, {Szalay}, {Bershady}, {Kinney},
  \& {Calzetti}}]{conno95}
{Connolly}, A.~J., {Szalay}, A.~S., {Bershady}, M.~A., {Kinney}, A.~L., \&
  {Calzetti}, D. 1995, \aj, 110, 1071

\bibitem[{{da Cunha} {et~al.}(2008){da Cunha}, {Charlot}, \& {Elbaz}}]{cunha08}
{da Cunha}, E., {Charlot}, S., \& {Elbaz}, D. 2008, \mnras, 388, 1595

\bibitem[{{Davis} {et~al.}(2003){Davis}, {Faber}, {Newman}, {Phillips},
  {Ellis}, {Steidel}, {Conselice}, {Coil}, {Finkbeiner}, {Koo}, {Guhathakurta},
  {Weiner}, {Schiavon}, {Willmer}, {Kaiser}, {Luppino}, {Wirth}, {Connolly},
  {Eisenhardt}, {Cooper}, \& {Gerke}}]{davis03}
{Davis}, M., {et~al.} 2003, in Society of Photo-Optical Instrumentation
  Engineers (SPIE) Conference Series, Vol. 4834, Society of Photo-Optical
  Instrumentation Engineers (SPIE) Conference Series, ed. {P.~Guhathakurta},
  161--172

\bibitem[{{de Vaucouleurs} {et~al.}(1991){de Vaucouleurs}, {de Vaucouleurs},
  {Corwin}, {Buta}, {Paturel}, \& {Fouqu{\'e}}}]{devau91}
{de Vaucouleurs}, G., {de Vaucouleurs}, A., {Corwin}, Jr., H.~G., {Buta},
  R.~J., {Paturel}, G., \& {Fouqu{\'e}}, P. 1991, {Third Reference Catalogue of
  Bright Galaxies. Volume I: Explanations and references. Volume II: Data for
  galaxies between 0$^{h}$ and 12$^{h}$. Volume III: Data for galaxies between
  12$^{h}$ and 24$^{h}$.}

\bibitem[{{Fioc} \& {Rocca-Volmerange}(1997)}]{fioc97}
{Fioc}, M., \& {Rocca-Volmerange}, B. 1997, \aap, 326, 950

\bibitem[{{Geller} {et~al.}(2005){Geller}, {Dell'Antonio}, {Kurtz}, {Ramella},
  {Fabricant}, {Caldwell}, {Tyson}, \& {Wittman}}]{gelle05}
{Geller}, M.~J., {Dell'Antonio}, I.~P., {Kurtz}, M.~J., {Ramella}, M.,
  {Fabricant}, D.~G., {Caldwell}, N., {Tyson}, J.~A., \& {Wittman}, D. 2005,
  \apjl, 635, L125

\bibitem[{{Giavalisco} {et~al.}(2004){Giavalisco}, {Ferguson}, {Koekemoer},
  {Dickinson}, {Alexander}, {Bauer}, {Bergeron}, {Biagetti}, {Brandt},
  {Casertano}, {Cesarsky}, {Chatzichristou}, {Conselice}, {Cristiani}, {Da
  Costa}, {Dahlen}, {de Mello}, {Eisenhardt}, {Erben}, {Fall}, {Fassnacht},
  {Fosbury}, {Fruchter}, {Gardner}, {Grogin}, {Hook}, {Hornschemeier}, {Idzi},
  {Jogee}, {Kretchmer}, {Laidler}, {Lee}, {Livio}, {Lucas}, {Madau},
  {Mobasher}, {Moustakas}, {Nonino}, {Padovani}, {Papovich}, {Park},
  {Ravindranath}, {Renzini}, {Richardson}, {Riess}, {Rosati}, {Schirmer},
  {Schreier}, {Somerville}, {Spinrad}, {Stern}, {Stiavelli}, {Strolger},
  {Urry}, {Vandame}, {Williams}, \& {Wolf}}]{giava04}
{Giavalisco}, M., {et~al.} 2004, \apjl, 600, L93

\bibitem[{{Hogg} {et~al.}(2002){Hogg}, {Baldry}, {Blanton}, \&
  {Eisenstein}}]{hogg02}
{Hogg}, D.~W., {Baldry}, I.~K., {Blanton}, M.~R., \& {Eisenstein}, D.~J. 2002,
  ArXiv Astrophysics e-prints

\bibitem[{{Humason} {et~al.}(1956){Humason}, {Mayall}, \& {Sandage}}]{humas56}
{Humason}, M.~L., {Mayall}, N.~U., \& {Sandage}, A.~R. 1956, \aj, 61, 97

\bibitem[{{Kewley} {et~al.}(2001){Kewley}, {Dopita}, {Sutherland}, {Heisler},
  \& {Trevena}}]{kewle01}
{Kewley}, L.~J., {Dopita}, M.~A., {Sutherland}, R.~S., {Heisler}, C.~A., \&
  {Trevena}, J. 2001, \apj, 556, 121

\bibitem[{{Kinney} {et~al.}(1996){Kinney}, {Calzetti}, {Bohlin}, {McQuade},
  {Storchi-Bergmann}, \& {Schmitt}}]{kinne96}
{Kinney}, A.~L., {Calzetti}, D., {Bohlin}, R.~C., {McQuade}, K.,
  {Storchi-Bergmann}, T., \& {Schmitt}, H.~R. 1996, \apj, 467, 38

\bibitem[{{Lee} \& {Seung}(1999)}]{lee99}
{Lee}, D.~D., \& {Seung}, H.~S. 1999, \nat, 401, 788

\bibitem[{{Martin} {et~al.}(2005){Martin}, {Fanson}, {Schiminovich},
  {Morrissey}, {Friedman}, {Barlow}, {Conrow}, {Grange}, {Jelinsky},
  {Milliard}, {Siegmund}, {Bianchi}, {Byun}, {Donas}, {Forster}, {Heckman},
  {Lee}, {Madore}, {Malina}, {Neff}, {Rich}, {Small}, {Surber}, {Szalay},
  {Welsh}, \& {Wyder}}]{marti05}
{Martin}, D.~C., {et~al.} 2005, \apjl, 619, L1

\bibitem[{{Moresco} {et~al.}(2013){Moresco}, {Pozzetti}, {Cimatti}, {Zamorani},
  {Bolzonella}, {Lamareille}, {Mignoli}, {Zucca}, {Lilly}, {Carollo},
  {Contini}, {Kneib}, {Le F{\`e}vre}, {Mainieri}, {Renzini}, {Scodeggio},
  {Bardelli}, {Bongiorno}, {Caputi}, {Cucciati}, {de la Torre}, {de Ravel},
  {Franzetti}, {Garilli}, {Iovino}, {Kampczyk}, {Knobel}, {Kova{\v c}}, {Le
  Borgne}, {Le Brun}, {Maier}, {Pell{\'o}}, {Peng}, {Perez-Montero},
  {Presotto}, {Silverman}, {Tanaka}, {Tasca}, {Tresse}, {Vergani}, {Barnes},
  {Bordoloi}, {Cappi}, {Diener}, {Koekemoer}, {Le Floc'h}, {L{\'o}pez-Sanjuan},
  {McCracken}, {Nair}, {Oesch}, {Scarlata}, {Scoville}, \&
  {Welikala}}]{mores13}
{Moresco}, M., {et~al.} 2013, \aap, 558, A61

\bibitem[{{Oke} \& {Sandage}(1968)}]{oke68}
{Oke}, J.~B., \& {Sandage}, A. 1968, \apj, 154, 21

\bibitem[{{O'Mill} {et~al.}(2011){O'Mill}, {Duplancic}, {Garc{\'{\i}}a Lambas},
  \& {Sodr{\'e}}}]{omill11}
{O'Mill}, A.~L., {Duplancic}, F., {Garc{\'{\i}}a Lambas}, D., \& {Sodr{\'e}},
  Jr., L. 2011, \mnras, 413, 1395

\bibitem[{{Roche} {et~al.}(2009){Roche}, {Bernardi}, \& {Hyde}}]{roche09}
{Roche}, N., {Bernardi}, M., \& {Hyde}, J. 2009, \mnras, 398, 1549

\bibitem[{{Rudnick} {et~al.}(2003){Rudnick}, {Rix}, {Franx}, {Labb{\'e}},
  {Blanton}, {Daddi}, {F{\"o}rster Schreiber}, {Moorwood}, {R{\"o}ttgering},
  {Trujillo}, {van der Wel}, {van der Werf}, {van Dokkum}, \& {van
  Starkenburg}}]{rudni03}
{Rudnick}, G., {et~al.} 2003, \apj, 599, 847

\bibitem[{{Simha} {et~al.}(2014){Simha}, {Weinberg}, {Conroy}, {Dave},
  {Fardal}, {Katz}, \& {Oppenheimer}}]{simha14}
{Simha}, V., {Weinberg}, D.~H., {Conroy}, C., {Dave}, R., {Fardal}, M., {Katz},
  N., \& {Oppenheimer}, B.~D. 2014, ArXiv e-prints

\bibitem[{{Skrutskie} {et~al.}(2006){Skrutskie}, {Cutri}, {Stiening},
  {Weinberg}, {Schneider}, {Carpenter}, {Beichman}, {Capps}, {Chester},
  {Elias}, {Huchra}, {Liebert}, {Lonsdale}, {Monet}, {Price}, {Seitzer},
  {Jarrett}, {Kirkpatrick}, {Gizis}, {Howard}, {Evans}, {Fowler}, {Fullmer},
  {Hurt}, {Light}, {Kopan}, {Marsh}, {McCallon}, {Tam}, {Van Dyk}, \&
  {Wheelock}}]{skrut06}
{Skrutskie}, M.~F., {et~al.} 2006, \aj, 131, 1163

\bibitem[{{Taylor} {et~al.}(2009){Taylor}, {Franx}, {van Dokkum}, {Quadri},
  {Gawiser}, {Bell}, {Barrientos}, {Blanc}, {Castander}, {Damen},
  {Gonzalez-Perez}, {Hall}, {Herrera}, {Hildebrandt}, {Kriek}, {Labb{\'e}},
  {Lira}, {Maza}, {Rudnick}, {Treister}, {Urry}, {Willis}, \&
  {Wuyts}}]{taylo09}
{Taylor}, E.~N., {et~al.} 2009, \apjs, 183, 295

\bibitem[{{van Dokkum} \& {Franx}(1996)}]{dokku96}
{van Dokkum}, P.~G., \& {Franx}, M. 1996, \mnras, 281, 985

\bibitem[{{van Dokkum} {et~al.}(2000){van Dokkum}, {Franx}, {Fabricant},
  {Illingworth}, \& {Kelson}}]{dokku00}
{van Dokkum}, P.~G., {Franx}, M., {Fabricant}, D., {Illingworth}, G.~D., \&
  {Kelson}, D.~D. 2000, \apj, 541, 95

\bibitem[{{Westra} {et~al.}(2010){Westra}, {Geller}, {Kurtz}, {Fabricant}, \&
  {Dell'Antonio}}]{westr10}
{Westra}, E., {Geller}, M.~J., {Kurtz}, M.~J., {Fabricant}, D.~G., \&
  {Dell'Antonio}, I. 2010, \pasp, 122, 1258

\bibitem[{{Willmer} {et~al.}(2006){Willmer}, {Faber}, {Koo}, {Weiner},
  {Newman}, {Coil}, {Connolly}, {Conroy}, {Cooper}, {Davis}, {Finkbeiner},
  {Gerke}, {Guhathakurta}, {Harker}, {Kaiser}, {Kassin}, {Konidaris}, {Lin},
  {Luppino}, {Madgwick}, {Noeske}, {Phillips}, \& {Yan}}]{willm06}
{Willmer}, C.~N.~A., {et~al.} 2006, \apj, 647, 853

\bibitem[{{York} {et~al.}(2000){York}, {Adelman}, {Anderson}, {Anderson},
  {Annis}, {Bahcall}, {Bakken}, {Barkhouser}, {Bastian}, {Berman}, {Boroski},
  {Bracker}, {Briegel}, {Briggs}, {Brinkmann}, {Brunner}, {Burles}, {Carey},
  {Carr}, {Castander}, {Chen}, {Colestock}, {Connolly}, {Crocker}, {Csabai},
  {Czarapata}, {Davis}, {Doi}, {Dombeck}, {Eisenstein}, {Ellman}, {Elms},
  {Evans}, {Fan}, {Federwitz}, {Fiscelli}, {Friedman}, {Frieman}, {Fukugita},
  {Gillespie}, {Gunn}, {Gurbani}, {de Haas}, {Haldeman}, {Harris}, {Hayes},
  {Heckman}, {Hennessy}, {Hindsley}, {Holm}, {Holmgren}, {Huang}, {Hull},
  {Husby}, {Ichikawa}, {Ichikawa}, {Ivezi{\'c}}, {Kent}, {Kim}, {Kinney},
  {Klaene}, {Kleinman}, {Kleinman}, {Knapp}, {Korienek}, {Kron}, {Kunszt},
  {Lamb}, {Lee}, {Leger}, {Limmongkol}, {Lindenmeyer}, {Long}, {Loomis},
  {Loveday}, {Lucinio}, {Lupton}, {MacKinnon}, {Mannery}, {Mantsch}, {Margon},
  {McGehee}, {McKay}, {Meiksin}, {Merelli}, {Monet}, {Munn}, {Narayanan},
  {Nash}, {Neilsen}, {Neswold}, {Newberg}, {Nichol}, {Nicinski}, {Nonino},
  {Okada}, {Okamura}, {Ostriker}, {Owen}, {Pauls}, {Peoples}, {Peterson},
  {Petravick}, {Pier}, {Pope}, {Pordes}, {Prosapio}, {Rechenmacher}, {Quinn},
  {Richards}, {Richmond}, {Rivetta}, {Rockosi}, {Ruthmansdorfer}, {Sandford},
  {Schlegel}, {Schneider}, {Sekiguchi}, {Sergey}, {Shimasaku}, {Siegmund},
  {Smee}, {Smith}, {Snedden}, {Stone}, {Stoughton}, {Strauss}, {Stubbs},
  {SubbaRao}, {Szalay}, {Szapudi}, {Szokoly}, {Thakar}, {Tremonti}, {Tucker},
  {Uomoto}, {Vanden Berk}, {Vogeley}, {Waddell}, {Wang}, {Watanabe},
  {Weinberg}, {Yanny}, {Yasuda}, \& {SDSS Collaboration}}]{york00}
{York}, D.~G., {et~al.} 2000, \aj, 120, 1579

\end{thebibliography}

\bibliographystyle{apj}

\end{document}